\documentclass[fleqn, 
usenatbib]{mnras}
\usepackage{bookmark}
\usepackage{newtxtext,newtxmath}
\usepackage{graphicx}	
\usepackage{amsmath}	
\usepackage{multicol}        
\usepackage{bm}		
\usepackage{pdflscape}	
\usepackage{booktabs}
\usepackage{makecell}
\usepackage{diagbox}
\usepackage{soul} 
\usepackage{hyperref}
\usepackage{threeparttable}
\usepackage[T1]{fontenc}

\usepackage{eso-pic}

\DeclareRobustCommand{\VAN}[3]{#2}
\let\VANthebibliography\thebibliography
\def\thebibliography{\DeclareRobustCommand{\VAN}[3]{##3}\VANthebibliography}

\usepackage{graphicx}	
\usepackage{amsmath}	

\title[Impact of mass transfer on WD-IMBH orbit]{Impact of mass transfer on the orbital evolution  of a white dwarf close to an intermediate-mass black hole}
 
\author[Yang Yang et al.]{
Yang Yang,$^{1,2}$\thanks{E-mail: yangyang24@stu.pku.edu.cn}
Jie Yang,$^{3,4,5,6}$\thanks{E-mail: yangjiev@lzu.edu.cn}
Xian Chen,$^{1,2}$\thanks{E-mail: xian.chen@pku.edu.cn}
and Zi-Han Zhang$^{7,8,9}$\thanks{E-mail: zhangzihan242@mails.ucas.ac.cn}
\\
$^{1}$Astronomy Department, School of Physics, Peking University, 100871 Beijing, China\\
$^{2}$Kavli Institute for Astronomy and Astrophysics, Peking University, 100871 Beijing, China\\
$^{3}$School of Physical Science and Technology, Lanzhou University, Lanzhou 730000, China\\
$^{4}$Institute of Theoretical Physics $\&$ Research Center of Gravitation, Lanzhou University, Lanzhou 730000, China\\
$^{5}$Key Laboratory of Quantum Theory and Applications of MoE, Lanzhou University, Lanzhou 730000, China\\
$^{6}$Lanzhou Center for Theoretical Physics $\&$ Key Laboratory of Theoretical Physics of Gansu Province, Lanzhou University, Lanzhou 730000, China\\
$^{7}$International Centre for Theoretical Physics Asia-Pacific, University of Chinese Academy of Sciences, 100190 Beijing, China\\
$^{8}$School of Physical Sciences, University of Chinese Academy of Sciences, Beijing 100049, China\\
$^{9}$Taiji Laboratory for Gravitational Wave Universe, University of Chinese Academy of Sciences, Beijing 100049, China
}
\date{Accepted XXX. Received YYY; in original form ZZZ}

\pubyear{2025}

\begin{document}
\label{firstpage}
\pagerange{\pageref{firstpage}--\pageref{lastpage}}
\maketitle

\begin{abstract}
Extreme mass-ratio inspirals (EMRIs) of low-mass white dwarfs (WDs, $0.1$\,$-$\,$0.3$\,$\text{M}_{\sun}$) around spinning intermediate-mass black holes (IMBHs, $10^{3}$\,$-$\,$10^{5}$\,$\text{M}_{\sun}$) offer unique opportunities for multi-messenger astronomy, emitting both gravitational waves (GWs) and electromagnetic (EM) signals. 
Yet, despite their astrophysical relevance, theoretical models often omit key interactions between relativistic dynamics and phase-dependent mass transfer (MT).
In this study, we integrate a perturbed Keplerian formalism with post-Newtonian (PN) corrections to simulate the relativistic orbit of a WD around a rotating IMBH, explicitly resolving the narrow phase near pericentre where Roche-lobe overflow initiates MT. 
We find that GW emission and MT exert competing influences on the orbit: MT episodes can increase both orbital period and eccentricity, potentially enabling the WD to avoid complete tidal disruption and even escape. 
We further quantify the GW phase evolution induced by MT, identifying parameter regimes in which GW detectors could observe a one–radian phase shift over observational timescales. 
Finally, we propose that the orbital expansion driven by MT may lead to the disappearance of quasi-periodic eruptions (QPEs).
Our results underscore the necessity of jointly modeling relativistic effects and dynamic mass transfer in WD–IMBH systems.
\end{abstract}




\section{Introduction}
An extreme-mass-ratio inspiral (EMRI) involves a massive black hole (MBH) and a compact object (CO), such as a white dwarf (WD), neutron star (NS), or stellar-mass black hole (BH), moving along a tightly bound orbit around the MBH \citep{hils1995gradual,amaro2007intermediate}.  
EMRIs are important sources for future space-based gravitational wave (GW) detectors, such as the Laser Interferometer Space Antenna (LISA) \citep{barack2004lisa,gair2009probing,berry2013expectations,gair2017prospects,amaro2017laser}.
Their GW signals contain rich information about the formation and evolution of stars around MBHs \citep{amaro2007intermediate,amaro2018relativistic}, and they also offer a unique opportunity to test general relativity in the strong-gravity regime \citep{barsanti2024testing,speri2024advancing}.  

Among the  various types of EMRIs, WDs inspiraling towards intermediate-mass black holes (IMBHs, $10^{3} - 10^{5} \text{M}_{\sun}$) stand out as a special type, not only because their formation channel was recognized relatively early  \citep{hils1995gradual,MarcFreitag2001,Bogdanovic2014},  
but also because the end products are luminous electromagnetic (EM) flares from the tidal disruption of the WDs by the IMBHs \citep{1989AandA209103L,ivanov2007orbital,zalamea2010white}.  
Such EM counterparts associated with GW signals could enable precise measurements of the geometry of the universe \citep{menou2008cosmological,sesana2008observing}.  

\citet{king2020gsn} suggested that quasi-periodic eruptions (QPEs), a type of X-ray transient recently discovered  \citep[e.g.][]{miniutti2019nine,giustini2020x,arcodia2021x,chakraborty2021possible,
arcodia2022complex,miniutti2023repeating,miniutti2023alive,
arcodia2024ticking,giustini2024fragments},  are produced by WD-IMBH systems. 
QPEs are hypothesised to result from periodic mass transfer (MT) episodes involving low-mass stars on bound orbits around MBHs, where the stars narrowly avoid full tidal disruption. 
Several lines of reasoning suggest that low-mass WDs are particularly plausible candidates for the donor star in such systems. 
First, low-mass WDs are structurally favourable for producing recurrent QPEs. 
Their relatively large radii, governed by the mass–radius relation   $R_{\mathrm{WD}}$\,$\propto$\,$M_{\mathrm{WD}}^{-1/3}$, render them more susceptible to partial tidal stripping at wider orbital separations, reducing the probability of complete disruption. 
This allows for repeated, stable mass-loss episodes over many orbits, consistent with the observed recurrence of QPEs  \citep{king2022quasi,chen2022milli,zhao2022quasi,krolik2022quasiperiodic,wang2022model,lu2023quasi,linial2023unstable,garain2024partialtidaldisruptionsspinning}. 
Second, main-sequence (MS) stars undergoing Roche-lobe overflow must orbit at much greater separations, leading to lower GW emission and less efficient MT \citep{king2022quasi}. This results in lower accretion luminosities, which may be inconsistent with the energetics inferred from QPE light curves. 
Third, \citet{Cufari_2022} demonstrated that the direct formation of QPE binaries via single scattering events involving MS stars is unlikely, as the energy dissipated in such interactions often exceeds the star’s binding energy. Instead, they suggested that QPE binaries might form through the \citet{hills1988hyper} mechanism, in which a stellar binary is tidally disrupted by an MBH, with one component captured and the other ejected as a hypervelocity star.

To produce a stripped mass consistent with QPE signals, the WD must have a sufficiently large radius—exceeding its corresponding Roche radius—which requires its mass to be unusually low. 
Specifically, the WD mass must lie well below the lower limit ($\sim 0.5 \text{M}_{\sun}$) expected from single-star evolution.
The occurrence of QPEs implies a high formation rate of low-mass WDs, which likely originate from binary evolution \citep{Istrate2014, ChenXue2017, Sun_2018, Li_2019}. 
The plausible formation channels are  binary stable MT and common Envelope Evolution.
Alternatively, \citet{king2020gsn} proposes the possibility that an IMBH directly captures the red giant and is partially stripped away. The remaining core evolves into a low-mass WD.

However, not all the observational properties of QPEs have been explained by the WD-IMBH interaction. 
First, in several QPEs, the time between consecutive eruptions alternates between long and short durations \citep{miniutti2019nine}.  
It has been suggested that this "long-short behavior" may result from variations in the amount of mass stripped from the WD  \citep{king2022quasi}, but a detailed model is still lacking. 
Second, in the case of GSN 069, QPEs disappeared for about two years before reappearing \citep{ miniutti2023repeating,miniutti2023alive}. 
This behavior is not predicted by the WD-IMBH model either.

The model for QPEs drive orbits gradually inwards under the emission of GWs, however, a key prediction of this model—a secular increase in QPE luminosity due to the intensifying MT rate—appears inconsistent with long-term monitoring of sources such as GSN 069, where no significant brightening has been observed.
To address this discrepancy, \citet{WangDi-1} proposed a hybrid scenario in which tidal stripping is coupled with dynamical interactions between the WD and a remnant accretion disc formed by a preceding TDE. In this framework, the WD intersects the fallback disc during its orbit, and the resulting hydrodynamical drag moderates the MT rate. This interaction not only prevents the WD from undergoing rapid disruption, but also enables the system to maintain a relatively stable QPE luminosity.
Despite its promise, the detailed impact of WD–disc interactions remains poorly constrained, particularly across a wide range of orbital eccentricities and disc configurations. Furthermore, not all QPE-hosting systems are expected to retain a disc, and mutual inclination between the disc plane and the binary orbit is likely. Given the inherent complexity and transient nature of post-TDE accretion structures, we focus in this work on exploring the long-term evolution of WD–IMBH binaries in the absence of a disc, with the aim of systematically evaluating the effectiveness and generality of the MT mechanism across a broad parameter space.

Our study is motivated by key limitations in current models of QPEs involving WD–IMBH systems, particularly the incomplete treatment of relativistic orbital dynamics and the MT process. 
These prevents direct comparisons of model predictions with observed QPE light curves.

Firstly, the effects of general relativity (GR) have not been fully 
accounted for. 
What has been included in the  model \citep[e.g.][]{zalamea2010white,king2020gsn} is the secular decay of the orbital semimajor axis and eccentricity due to GW radiation, but the  derivation of the formulae is based on the Keplerian orbit approximation  \citep{peters1964gravitational}.
Such a model fails to capture the relativistic precession of the orbital pericenter or the variation in the orientation of the orbital angular momentum due to frame dragging if the central IMBH is spinning.  
These effects could introduce new characteristic frequencies into the orbit of the WD, as well as change our viewing angle of the system. 
Hence, they may be important for explaining the occurrence times or luminosities of the eruptions. 
The relevance of these effects is corroborated by the high orbital eccentricities (i.e., small pericenter distances) inferred for the WD orbits in QPEs \citep{chen2022milli, king2022quasi, king2023angular}.

We need to capture relativistic corrections over each orbital cycle, and can   attach MT in the form of force together. 
The post-Newtonian (PN) form and the perturbed Kepler method  \citep{efroimsky2005gauge, poisson2014gravity} provide a good way to do this.
The PN expansion introduces relativistic acceleration terms that modify key orbital elements—including the semi-major axis, eccentricity, and orbital phase. 
These modifications are naturally embedded within the perturbed Keplerian formalism, which yields a set of coupled differential equations describing the time evolution of orbital parameters. 
Crucially, differential equations allows us to incorporate external perturbative forces, such as those arising from asymmetric MT.
\citet{WangDi-1}

Second, MT from the WD to the IMBH is inevitable and is suspected to be responsible for stabilizing the long-term evolution of the binary  \citep{king2022quasi}. 
However, the detailed mechanism remains unclear.   
Recent study  \citet{wang2022model} and \citet{king2020gsn, king2022quasi, king2023angular}  use the orbital-averaged dynamical equations derived for stellar binaries  \citep{sepinsky2007interacting} to model the secular evolution of a WD-IMBH binary. 
But the analyses from \citet{sepinsky2007equipotential, sepinsky2007interacting} did not focus on the EMRI system, and the discussion of angular momentum transfer is not sufficient.
Since MT in the WD-IMBH scenario occurs mainly at pericenter due to the aforementioned high eccentricity, the validity of the orbital average remains unverified.

Because high-eccentricity WD-IMBH systems are highly dependent on rapid phase changes,   the Roche lobe overflow of MT occurs only in a narrow phase window near pericenter. 
We need a dynamics treatment that is suitable for handling changes within the concentrated phase range.
\citet{sepinsky2007interacting} provides a method to treat high-eccentricity accretion dynamics, modeling this behavior by treating the MT rate as a spike function near pericenter. 
The MT rate is expressed as a power-law expansion of the difference between the WD radius and the instantaneous Roche lobe radius, thus achieving a continuous and differentiable dynamical description.

\begin{figure}
   \quad \includegraphics[width=0.85\columnwidth]{Illustrations_in_the_introduction.pdf}
    \caption{Simulation result of semi-major axis in high eccentricity WD-IMBH system. The mass of the WD is $0.15\text{M}_{\sun}$ and the mass of the IMBH is $4\times 10^5\text{M}_{\sun}$, the initial eccentricity is $0.99$. A more detailed diagram can be seen in Figure \ref{5.2 Variation of the semi-major axis, eccentricity, WD period}.
    }
	\label{Illus intro}
\end{figure}
 
MT process  is expected to exert a dynamical influence on the orbital evolution of a WD–IMBH system. 
In particular, the anisotropic ejection of matter from the donor can impart a recoil force, which may become comparable to—or under certain conditions even exceed—the backreaction from gravitational wave (GW) emission.
Whereas GW radiation typically drives orbital circularization and inspiral, the MT-induced force can act in the opposite direction, potentially increasing the semi-major axis or enhancing orbital eccentricity. 
These considerations suggest a potentially rich interplay between GW-driven contraction and MT-driven expansion in high-eccentricity EMRIs.
The interaction between these competing effects is likely to affect the orbital evolution as shown in Figure \ref{Illus intro}.
In the late stages of binary evolution, the MT force is likely to exceed the gravitational reaction force, which can be expected to lead to the dynamical ejection of the WD from the gravitational potential of the IMBH.

This paper is structured as follows.  
In Section \ref{sec:Dynamics under post-Newton method}, using the perturbed Kepler method, we describe the motion of the WD in the IMBH’s relative coordinate system. 
We assume that all acceleration terms can be expressed in terms of the required parameters.   
In Section \ref{sec:Mass transfer model of WH-BH system under Hills mechanism}, we introduce the WD state formulae \citep{zalamea2010white}, the MT Roche lobe accretion model  \citep{kolb1990comparative, sepinsky2007equipotential,cehula2023theory}, and the dynamics of eccentric binaries under MT \citep{sepinsky2007interacting}.  
In Section \ref{sec:Analyzing orbital results without MT in PN method}, we introduce the initial angular parameter representation, aligning the spin angular momentum of the IMBH with the Z axis in a fixed Cartesian coordinate system, and then numerically compute the short-term orbital evolution using the PN method and the perturbed Kepler method with MT.     
In Section \ref{sec:Analyzing orbit with MT}, we analyze the combined effects of MT and gravitational radiation on the long-term orbital evolution, using the time-averaged rate of change over one period. 
We describe the orbital evolution starting from the point where the Roche lobe radius at pericenter equals the WD radius.  
In Section \ref{sec:Impact on GW sigal and detection}, for the GW signal, we compute the time required for the orbital phase to change by 1 radian due to the MT effect, and identify the initial parameters that enable the detection of phase differences within 5 years.    
In Section \ref{sec:Discussion} and \ref{sec:Conclusion}, we discuss and summarize our results.

\section{Theory of orbital dynamics}\label{sec:Dynamics under post-Newton method}
\subsection{The general  form of relative acceleration}\label{sec:Relative acceleration}
In this study, we model the orbital dynamics of a WD–IMBH binary system by treating both relativistic corrections and MT effects as perturbations to the classical Newtonian two-body problem. 
The relative acceleration between the WD and the IMBH is thus expressed as: 
\begin{equation}
\begin{aligned}
 \boldsymbol{a}=\frac{\mathrm{d} \boldsymbol{v}}{\mathrm{d} t}=-\frac{G m}{r^{2}}\boldsymbol{n}+ \boldsymbol{f},
\end{aligned}
\label{sec1 5 a}
\end{equation}
where $\boldsymbol{a}=\boldsymbol{a}_1-\boldsymbol{a}_2$ denotes the relative acceleration of the WD with respect to the IMBH, $m=m_1+m_2$   is the total mass of the system, and $m_1$ and $m_2$ are the masses of the WD and the IMBH, respectively.
The vector $\boldsymbol{f}$ represents  the perturbative accelerations arising from relativistic corrections and MT effects.  
To facilitate a systematic perturbative analysis, $\boldsymbol{f}$ is decomposed into components along a natural orthonormal triad:  $\boldsymbol{f}=\mathcal{R } \boldsymbol{n}+ \mathcal{ S } \boldsymbol{\lambda}+\mathcal{ W} \boldsymbol{\ell}$. 
Here, $\boldsymbol{n}=\boldsymbol{r}/r$ is the unit vector of the binary star relative position vector, 
$\boldsymbol{\ell}=\boldsymbol{n}\times \boldsymbol{v}/\left | \boldsymbol{n}\times \boldsymbol{v} \right |$ is the unit vector normal to the orbital plane, and $\boldsymbol{\lambda }=\boldsymbol{\ell}\times\boldsymbol{n}$  lies within the orbital plane.  
This decomposition provides a convenient basis for quantifying the secular evolution of orbital elements under the influence of the perturbing forces. 
 
The relativistic motion of point particles in the system is described using a modified geodesic equation, following the post-Newtonian (PN) formalism outlined by \citet{blanchet2024post}. The Newtonian-like equations of motion (EOM) for individual bodies are provided in their Equation (339), while the EOM governing the relative dynamics of the two-body system is given in their Equation (357).  

To incorporate spin effects, we adopt the treatment of \citet{tagoshi2001gravitational} for the case in which body A possesses spin angular momentum $\boldsymbol{S}_{\text{A}}$. In their formalism, the relative acceleration $\boldsymbol{a}$ includes contributions from the spin–orbit coupling term $\boldsymbol{a}_{\mathrm{SO}}$, the spin–spin interaction term $\boldsymbol{a}_{\mathrm{SS}}$, and the post-Newtonian spin–orbit correction $\boldsymbol{a}_{\mathrm{SOPN}}$, as given in their Equation (5.15). The evolution of the spin vector is governed by the spin precession equation (their Equation  2.17).

In addition to the relativistic and spin effects, we include the influence of MT through an additional acceleration term $\boldsymbol{f}_{\mathrm{MT}}$, provided in Equation (14) of \citet{sepinsky2007interacting}. A detailed formulation and discussion of this term will be presented in Section~\ref{sec:Dynamics under mass transfer}.
 
Combining the above considerations, the total perturbative acceleration $\boldsymbol{f}$ is given by:
\begin{align}\label{eq:relative acceleration}
\boldsymbol{f}= -\frac{G m}{r^{2}}\left (  \mathcal{A} \,  \boldsymbol{n}+\mathcal{B}\, \boldsymbol{v} \right ) +\boldsymbol{a}_{\mathrm{SO}}+\boldsymbol{a}_{\mathrm{SS}}+\boldsymbol{a}_{\mathrm{SOPN}}+\boldsymbol{f}_{\mathrm{MT}}+\mathcal{O}(c^{-8}),
\end{align}
where $\mathcal{A}$  and $\mathcal{B}$ are functions of $\boldsymbol{r}$ and $\boldsymbol{v}$, given in Equations (358) and (359) of \citet{blanchet2024post}.
To fully describe the system's dynamics, we also consider the spin precession equations (Equation (2.17) of \citet{tagoshi2001gravitational}) and the evolution of mass $m_1$ and $m_2$ due to MT. 
A comprehensive set of calculation parameters and further details will be provided in the subsequent section. 
This framework enables a detailed analysis of the orbital dynamics of WD–IMBH systems, accounting for both relativistic effects and MT processes.

\subsection{Perturbed Kepler method}\label{sec:Perturbed Kepler method}
\begin{figure}
    \includegraphics[width=\columnwidth]{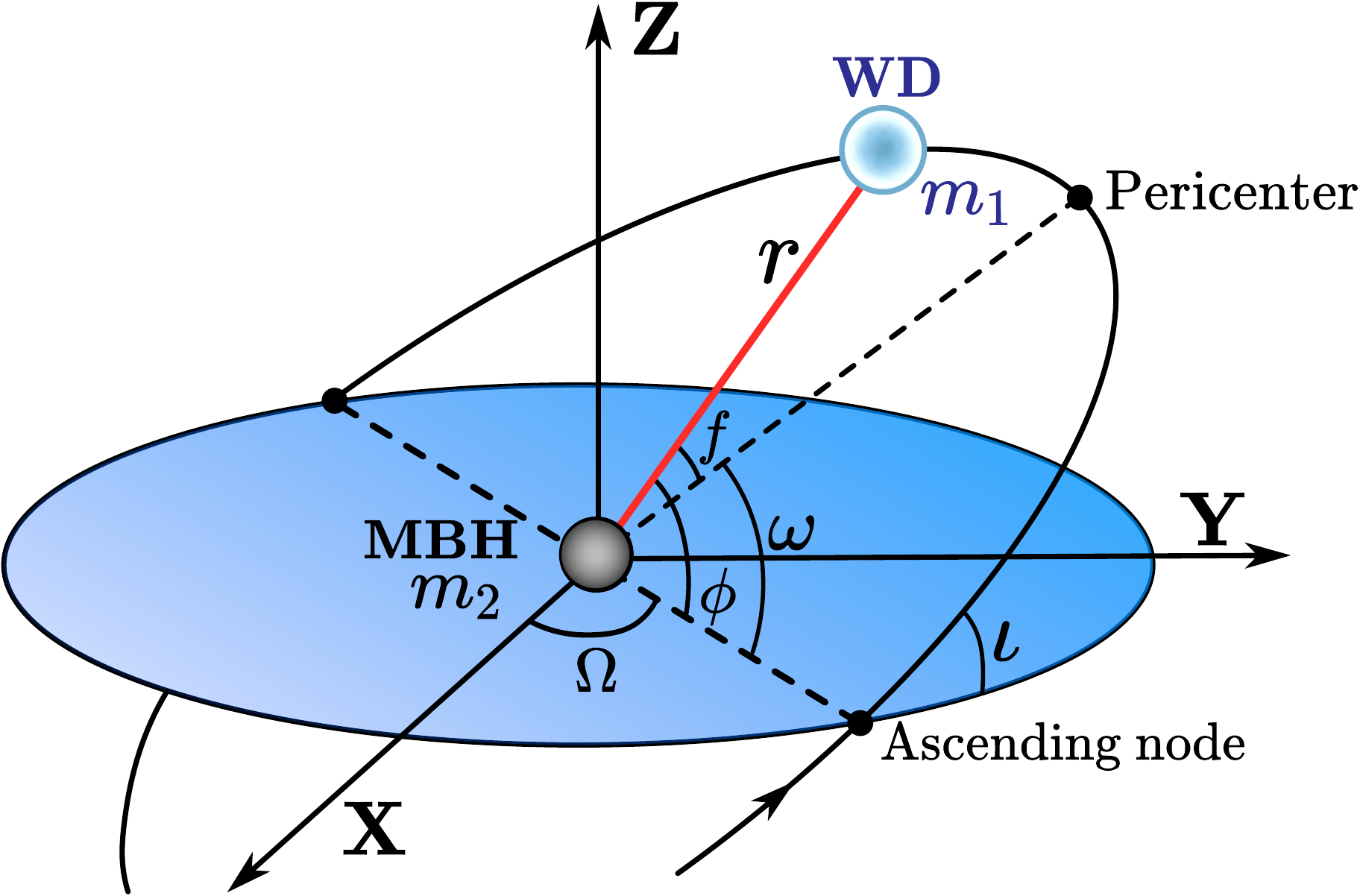}
    \caption{Definition of the ralative coordinate and the angle parameters. 
The origin of the relative coordinate system is placed at   mass $m_2$, while mass $m_1$ follows an orbit defined by  $\boldsymbol{r}$\,$=$\,$r^{X}\boldsymbol{e}_{X}$\,$+$\,$r^{Y}\boldsymbol{e}_{Y}$\,$+$\,$r^{Z}\boldsymbol{e}_{Z}$. 
For a quasi-Keplerian orbit, the system has pericenter and an ascending node, where the ascending node is the point of intersection of the orbit with the $OXY$ plane.  
The parameters are as follows: $\iota$ is the inclination of the orbital plane relative to the  $OXY$ plane; 
$\Omega$ is  the longitude of the ascending node,  the angle between the ascending node and the positive X-axis; 
$\omega$ is the argument of pericenter, the angle between the ascending node and the pericenter; $f$ is the phase angle of $m_1$ relative to the pericenter. 
The angle $\phi$ is defined as $\phi$\,$=$\,$\omega+f$.  
The basis vectors $\boldsymbol{e}_{X}$, $\boldsymbol{e}_{Y}$, $\boldsymbol{e}_{Z}$ define the Cartesian coordinate system with coordinates $(X,Y,Z)$.  }
	\label{Kepler orbit}
\end{figure}

To characterise the relative motion between the WD and the IMBH, we adopt a coordinate system as illustrated in Figure \ref{Kepler orbit}, utilizing the orbital angles ($\phi,f,\omega,\Omega ,\iota$) as defined by  \citet{poisson2014gravity}, to describe the relative motion between the WD and the IMBH. 

For the unperturbed Keplerian orbit ($\boldsymbol{f}$\,$=$\,$\boldsymbol{0}$),  the orbital plane remains fixed without precession, implying that $\dot{\Omega}=\dot{\iota} = \dot{\omega}=0$ and $\dot{\phi}$\,$=$\,$\dot{f}$. The classical Keplerian EOM are then given by: 
\begin{equation} 
\begin{aligned}
& r= \frac{p}{1+e\cos f},\\
&\dot{r}= \sqrt{\frac{G m}{p}} e\sin f, \quad 
\dot{\phi}=\sqrt{\frac{G m}{p^{3}}}\left (  1+e \cos f\right ) ^{2},\\
& v^{2}= \frac{G m}{p}\left [ e^2 \sin ^{2} f +(1+e \cos f)^{2}\right ], 
\end{aligned}
\label{eq:standard Kepler elliptic orbit11}
\end{equation}
where $p$ is the semi-latus rectum and $e$ is the eccentricity. 

In the presence of perturbative accelerations ($\boldsymbol{f}$\,$\ne$\,$ \mathbf{0}$),  
we propose that the Keplerian EOM (\ref{eq:standard Kepler elliptic orbit11}) remains unchanged,  but the orbital elements   ($p$,  $e$, $f$, $\omega$, $\Omega$, $\iota$) evolve over time.  
The evolution of these parameters is governed by the components ($\mathcal{R }$, $\mathcal{S }$, $\mathcal{W }$) in the perturbation acceleration  $\boldsymbol{f}$. 
The differential equations for the orbital elements follow the formalism presented in \citet{poisson2014gravity}:
\begin{subequations}
\begin{align}
\frac{d p}{d t} =&  \sqrt{\frac{p }{G m}}  \frac{2p}{1+e \cos f} \mathcal{S},  \label{eq:elements differential equation11 a}\\
\frac{d e}{d t} =& \sqrt{\frac{p}{G m}}\left[\sin f \mathcal{R}+\frac{2 \cos f+e\left(1+\cos ^{2} f\right)}{1+e \cos f} \mathcal{S}\right],  \label{eq:elements differential equation11 b} \\
\frac{d f}{d t}  =&\sqrt{\frac{G m}{p^{3}}}(1+e \cos f)^{2}\notag\\
+&\sqrt{\frac{p}{G m}} \frac{1}{e}\left[\cos f \mathcal{R}  -\frac{2+e \cos f}{1+e \cos f} \sin f \mathcal{S}\right],   \label{eq:elements differential equation11 c} \\
\frac{d \omega}{d t}=& \sqrt{\frac{p}{G m}}\frac{1}{e} \left[-\cos f \mathcal{R}+\frac{2+e \cos f}{1+e \cos f} \sin f \mathcal{S}  \right. \notag\\
&\quad \quad \quad \quad \left. -\cot \iota \frac{\, e \sin (\omega+f)\, }{1+e \cos f} \mathcal{W}\right],    \label{eq:elements differential equation11 d}  \\
\frac{d \Omega}{d t}= & \sqrt{\frac{p}{G m}} \frac{\sin (\omega+f)}{1+e \cos f} \csc\iota \ \, \mathcal{W},    \label{eq:elements differential equation11 e} \\
\frac{d \iota}{d t} =& \sqrt{\frac{p}{G m}} \frac{\cos (\omega+f)}{1+e \cos f} \mathcal{W}.  \label{eq:elements differential equation11 f}   
\end{align}
\label{eq:elements differential equation11}
\end{subequations}

To express the evolution of the spin angular momentum vector within the natural orthonormal triad ($\boldsymbol{n}$, $\boldsymbol{\lambda}$, $\boldsymbol{\ell}$), we derive the time derivatives of these basis vectors, incorporating the effects of relativistic precession. 
This approach facilitates the projection of spin dynamics onto the orbital frame, which is particularly advantageous for analyzing systems with eccentric orbits and spin-orbit coupling.
The time derivatives of the basis vectors are given by:
\begin{equation} 
\dot{\boldsymbol{n}}=  \dot{f} _\mathrm{N} \boldsymbol{\lambda } ,  \quad   
 \dot{\boldsymbol{\lambda }} =  -\dot{f} _\mathrm{N} \boldsymbol{n} +\dot{f} _\mathrm{N}  \mathcal{W } _{f} \boldsymbol{\ell},  \quad  
 \dot{\boldsymbol{\ell}}= -\dot{f} _\mathrm{N}  \mathcal{W } _{f} \boldsymbol{\lambda  },  
\label{eq:elements differential equation22}
\end{equation}
where       
 \begin{equation}
 \begin{aligned}
& \dot{f} _\mathrm{N} := \sqrt{\frac{G m}{p^{3}}}\left (  1+e \cos f\right ) ^{2},\quad \mathcal{W } _{f}  := \frac{p^2\mathcal{W } }{G m\left (  1+e \cos f\right ) ^{3}}.
 \end{aligned}
 \label{fN Wf}
\end{equation}
Here, $\dot{f} _\mathrm{N}$ has the same form as $\dot{\phi}$ in Equation (\ref{eq:standard Kepler elliptic orbit11}). 
Introducing the dimensionless parameter $\mathcal{W}_f$ facilitates the simplification of Equation (\ref{eq:elements differential equation22}) in scenarios where $\mathcal{W} \ne 0$.  
Equation (\ref{eq:elements differential equation22}) enables the decomposition of the spin precession equation into components aligned with the natural basis, thereby allowing for a detailed analysis of the spin dynamics influenced by relativistic corrections.

The spin angular momentum $\boldsymbol{S}_\mathrm{A}$ ($\mathrm{A}=1,2$) can be expressed as $S_{\mathrm{A}n} \boldsymbol{n}$\,$+$\,$S_{\mathrm{A}\lambda} \boldsymbol{\lambda}$\,$+$\,$S_{\mathrm{A}\ell} \boldsymbol{\ell}$.  
We obtain the differential equations for the spin components $S_{\mathrm{A}n}$, $S_{\mathrm{A}\lambda }$, and $S_{\mathrm{A}\ell}$: 
\begin{subequations} 
  \begin{align}
&  \frac{d {S}_{\mathrm{A}n}}{d t} =  \left (   \boldsymbol{\Omega}_{\mathrm{A}} \times \boldsymbol{S}_{\mathrm{A}} \right )_{n} +  \dot{f} _\mathrm{N} {S}_{\mathrm{A}\lambda } , \\
&\frac{d {S}_{\mathrm{A}\lambda }}{d t} = \left (   \boldsymbol{\Omega}_{\mathrm{A}} \times \boldsymbol{S}_{\mathrm{A}} \right )_{\lambda}-\dot{f} _\mathrm{N}{S}_{\mathrm{A}n}+\dot{f} _\mathrm{N} \mathcal{W } _{f} {S}_{\mathrm{A}\ell},\\
&\frac{d {S}_{\mathrm{A}\ell}}{d t}= \left (   \boldsymbol{\Omega}_{\mathrm{A}} \times \boldsymbol{S}_{\mathrm{A}} \right )_{\ell}-\dot{f} _\mathrm{N} \mathcal{W } _{f} {S}_{\mathrm{A}\lambda },
\end{align}
\label{eq:differential equation representation of the spin component} 
\end{subequations}
where $\boldsymbol{\Omega}_{\mathrm{A}}$ denotes the rates at which the spin angular momentum rotates around the total angular momentum.

Therefore, a comprehensive set of differential equations governing the dynamics of the  inspiralling binary system with the MT process can be derived. This includes the following variables:
$$\left \{ 
\dot{p}, \dot{e}, \dot{f },\dot{\omega },  \dot{\Omega },  \dot{\iota},    
\dot{S}_{1n}, \dot{S}_{1\lambda }, \dot{S}_{1\ell}, \dot{S}_{2n}, \dot{S}_{2\lambda }, \dot{S}_{2\ell}, 
\dot{m}_1, \dot{m}_2 \right \}.$$

\section{Mass transfer for eccentric binary}\label{sec:Mass transfer model of WH-BH system under Hills mechanism}
To complete the orbital evolution analysis of the WD–IMBH system, we incorporate the MT rate ($\dot{m}_1, \dot{m}_2$) and the associated perturbative acceleration $\boldsymbol{f}_{\mathrm{MT}}$. 
In our study of the mass stripping of WD, we utilize the WD model 
proposed by \citet{zalamea2010white} to characterize the surface properties of the WD in Section \ref{sec:Model of white dwarf surface}. 
The MT rate $\dot{m}_1$ is introduced in Section \ref{sec:MT rate and the first lagrangian point} following theory outlined by \citet{kolb1990comparative,jackson2017new,cehula2023theory}. 
Our application of the Lagrangian point L1 and its corresponding equivalent Roche radius refers to \citet{sepinsky2007equipotential}.
To calculate $\boldsymbol{f}_{\mathrm{MT}}$, we use the analysis of the conservation of orbital angular momentum derived in Section \ref{sec:Dynamics under mass transfer}.    
We extend this analysis to the case where it may not be conserved.

\subsection{Hydrostatic equilibrium model of white dwarf}\label{sec:Model of white dwarf surface}  

In the follow, we assume that the polytropic equation of state (EOS)  for the surface layer of the WD is $P$\,$=$\,$K \rho^{\Gamma}$, where $\Gamma$\,$=$\,$5/3$ and $K$\,$\equiv$\,$ (3 / \pi)^{2 / 3} h^{2} / 20 m_{\mathrm{ e}}\left(\mu_{\mathrm{ e}} m_{\mathrm{p} }\right)^{5 / 3}$. 
Here, $h$ is the Planck constant, $m_{\mathrm{e}}$ is the electron mass, $m_{\mathrm{p}}$ is the proton mass, and $\mu_{\mathrm{e}}\approx 2$ is the mean molecular weight per electron.   
The WD's radius can be expressed as \citep{zalamea2010white}: 
\begin{equation}
R_{\mathrm{WD}}(M_{\mathrm{WD}})\approx 6120\frac{G \text{M}_{\sun}}{c^2}  \left (\frac{M_{\mathrm{WD}}}{M_{\mathrm{Ch}}}\right )^{-1/3}\left ( 1- \frac{M_{\mathrm{WD}}}{M_{\mathrm{Ch}}}\right )^{0.447},
\label{eq:RWD}
\end{equation}
where $M_{\mathrm{Ch}}$ is the Chandrasekhar mass.

\begin{figure}
    \includegraphics[width=\columnwidth]{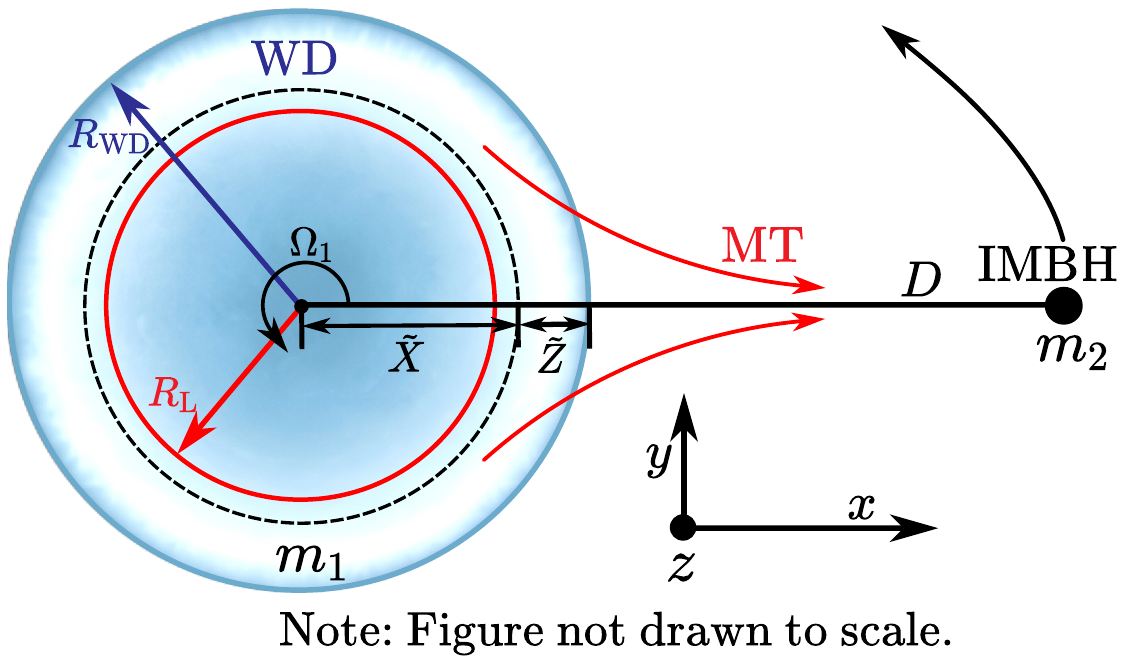}
    \caption{The top view of the instantaneous orbital plane.   
    The directions of $x$, $y$, $z$ are determined by the line connecting the center-of-mass and the rotation axis of WD. 
The donor (WD) has mass $m_1$, and the accretor (IMBH) has mass $m_2$.
The parameters are as follows: 
$R_{\mathrm{WD}}$ is the radius of WD, 
$R_{\mathrm{L}}$ is the Roche-lobe radius.
$\tilde{Z}$ is the depth  beneath the donor surface,  
$\tilde{X}=R_{\mathrm{WD}}-\tilde{Z}$ is the distance from the center-of-mass (CM) of the donor to the mass element, 
$D$ is the distance from the CM  of the donor to the accretor, 
$\Omega _{1}$ is the spin angular velocity of the donor.  
 The Schwarzschild radius of the IMBH is much larger than the WD radius, so it is not drawn to scale.}
	\label{Roche Lobe Tidal Effects}
\end{figure}

Assuming a quasi-static equilibrium and incorporating tidal and rotational potentials, the modified hydrostatic equilibrium equation can be integrated using the polytropic EOS. Applying the boundary condition at the photosphere yields the following density distribution: 
\begin{align} \label{eq:56 rho}
&\rho ^{\Gamma -1} (\tilde{Z})=\rho_{\mathrm{ph}} ^{\Gamma -1} +\frac{\Gamma -1}{K\Gamma } \left \{\frac{G m_1}{R_{\mathrm{WD}}} \left ( \frac{R_{\mathrm{WD}}}{R_{\mathrm{WD}}-\tilde{Z}} -1  \right ) \right.  \\
&-\frac{1}{2} \Omega _{1}^2 \left. (2 R_{\mathrm{WD}}-\tilde{Z}) \tilde{Z}+\frac{G m_2}{D}\left [ \frac{\tilde{Z}}{D} +\frac{D}{D-R_{\mathrm{WD}}+\tilde{Z}} -\frac{D}{D-R_{\mathrm{WD}}} \right ]   \right \}, \nonumber   
\end{align}
where ($m_1$, $m_2$, $R_{\mathrm{WD}}$, $D$, $\tilde{Z}$, $\tilde{X}$, $\Omega _{1}$) are described in Figure \ref{Roche Lobe Tidal Effects}.  

The surface (photosphere) density $\rho_{\mathrm{ph}}$ of a WD is challenging to determine directly due to the presence of an overlying non-degenerate atmosphere. 
This parameter is influenced by factors such as metallicity, evolutionary stage, boundary conditions, and the EOS \citep{cehula2023theory}. 
To facilitate analysis, we introduce a dimensionless surface equivalent depth, $\xi_{0}$, as an adjustable parameter:   
\begin{align}\label{65 fMT}
\rho_{\mathrm{ph}}:= \left [   \frac{\Gamma-1}{K \Gamma}\left(\frac{G M_{\mathrm{WD}}}{R_{\mathrm{WD}}} \frac{\xi_{0}}{1-\xi_{0}}\right)\right ]^{\frac{1}{\Gamma-1}}. 
\end{align}
This formulation equates $\rho_{\mathrm{ph}}$ to the density at a depth $\tilde{Z} = \xi_{0} R_{\mathrm{WD}}$ in a non-rotating WD absent tidal effects, providing a convenient framework for subsequent mathematical treatments. 
Observational data indicate that the outermost layers of WDs possess relatively low densities, typically ranging from $10^{-5}$ to $10^{-3}\,\mathrm{g\,cm^{-3}}$  \citep{ritter1988turning,seymour1984black}. 
We posit that this tenuous atmosphere is stripped away before the WD reaches pericenter and is not replenished within a single orbital cycle. 
Consequently, the exposed surface comprises a denser hydrogen or helium layer, with densities estimated between $10 $ and $10^{4}\,\mathrm{g\,cm^{-3}}$ \citep{seymour1984black}. 
While a minimum density of $10\,\mathrm{g\,cm^{-3}}$ is considered, the actual value is likely substantially higher.
Using Equation (\ref{eq:RWD}), the parameter $\xi_0$ can be estimated as
\begin{align} \label{eq:24 rhoph}
\xi _{0}&= \frac{P_{\mathrm{ph}} /\rho _{\mathrm{ph}}  }{c^2}   \frac{R_{\mathrm{WD} } }{GM_{\mathrm{WD}} /c^2}\frac{ \Gamma }{\Gamma -1}\nonumber \\
&\approx 3.5\times 10^{-3}\left ( \frac{\rho _{\mathrm{ph}} }{10\, \mathrm{g \ cm^{-3}} }  \right )^{2/3} \left (  \frac{M_{\mathrm{WD} }}{0.15\,  \text{M}_{\sun} } \right ) ^{-4/3} \!\!.
\end{align} 
This estimation provides a practical means to relate the surface density to the structural parameters of the WD, facilitating further analysis of MT processes in WD–IMBH systems.

\subsection{Mass transfer model}\label{sec:MT rate and the first lagrangian point} 
With the density distribution of the WD, the transferred mass rate  from the WD to the IMBH can be determined.
Following the framework established by  \citet{cehula2023theory}, we divide the MT process over one orbital period into two phases: isothermal (optically thin) and adiabatic (optically thick) accretion.  
The transition between these two phases is continuous and is related to the Roche radius $R_{\mathrm{L}}$ of the Roche lobe model.

If the donor (WD) is Roche-lobe underfilling ($R_{\mathrm{WD}}$\,$<$\,$R_{\mathrm{L}}$), the MT stream is isothermal \citep{lubow1975gas, ritter1988turning, jackson2017new}.  
The prescription assumes that, even if $R_{\mathrm{WD}}$\,$<$\,$R_{\mathrm{L}}$,  the density profile of donor’s atmosphere extends above its photosphere to the first Lagrange point L1. 
The gas above the photosphere is optically thin to radiation from the donor, but is in equilibrium with the surrounding radiation field  \citep{cehula2023theory}.
After the gas passes through L1, it accretes supersonically on to the accretor \citep{lubow1975gas}. 
Conversely, if the donor is Roche-lobe overflowing ($R_{\mathrm{WD}}>R_{\mathrm{L}}$),  the MT is  optically thick and adiabatic \citep{kolb1990comparative, pavlovskii2015mass, marchant2021role}. 
It is assumed that there is no energy transport within the gas and between the gas and the surrounding radiation field.

According to our considerations, we first give the formula for the MT rate we use. When $R_{\mathrm{WD}}$\,$<$\,$R_{\mathrm{L}}$, the Roche lobe is underfilling, the MT rate is 
\begin{equation} 
\begin{aligned}
\dot{m}_1^{\mathrm{iso}}=-\frac{2 \pi }{\sqrt{B C}}&  \frac{\rho_{\mathrm{ph}}}{\mathrm{ exp}(1/2) } c_{\mathrm{T}}^3  \cdot \mathrm{ exp}\left(-\frac{\phi_{1}-\phi_{\mathrm{ph}}}{c_{\mathrm{T}}^{2}}\right).
\end{aligned}
\label{eq:58 dotm1}
\end{equation}
When $R_{\mathrm{WD}}>R_{\mathrm{L}}$, the MT rate is 
\begin{align}\label{eq:59 dotm1}
\dot{m}_1^{\mathrm{adi}}=-\frac{2 \pi }{\sqrt{B C}}& \left \{ \frac{\rho_{\mathrm{ph}}}{\mathrm{ exp}(1/2) } c_{\mathrm{T}}^3  +F(K,\Gamma) \left ( P_\text{L}^{\frac{3\Gamma -1}{2\Gamma } }-P_{\mathrm{ph}}^{\frac{3\Gamma -1}{2\Gamma } } \right )  \right \}.
\end{align} 

We begin to explain the physical quantities involved in Equations (\ref{eq:58 dotm1}) and (\ref{eq:59 dotm1}), as detailed in  \citet{cehula2023theory}). 
Here, the surface density $\rho_{\mathrm{ph}}$ is determined by Equation (\ref{65 fMT}),  while the sound speed  $c_{\mathrm{T}}$ is given by $c_{\mathrm{T}}^2 = \Gamma P_{\mathrm{ph}}/\rho_{\mathrm{ph}}$.
In Equation (\ref{eq:58 dotm1}), $\phi_{1}$ represents the specific potential at the Lagrange point L1 within the rotating frame, and $\phi_{\mathrm{ph}}$ corresponds to the potential at the WD's surface. 
Both points lie along the line connecting the CM of the binary components. 
The position of L1, accounting for orbital eccentricity and asynchronous rotation, can be determined following the methodology of \citet{sepinsky2007equipotential}. 
Equation (\ref{eq:59 dotm1}) involves the pressures $P_\text{L} = P\left ( \tilde{Z}=R_{\mathrm{WD}}-R_{\mathrm{L}}\right )$ and $P_{\mathrm{ph}} = P\left ( \tilde{Z}=0 \right )$. 
The corresponding density profile is described by Equation (\ref{eq:56 rho}). 
The factor $2\pi/\sqrt{BC}$ arises from the three-dimensional expansion of the effective potential near L1. 
The  coefficients $B$ and $C$ define the effective cross-sectional area through which MT occurs, they  satisfy \citep{cehula2023theory} 
\begin{align}  \label{29 equaiton of phi near X1}
\phi=\phi^{x}(x)+\frac{B}{2}  y^{2}  +\frac{C}{2}  z^{2} + \mathcal{O}\left(y^3, z^3\right),
\end{align}
where $x$ represents the distance from point L1 on the  line connecting the CM, i.e. $x$\,$=$\,$\tilde{X}$\,$-$\,$\tilde{X}_1 D$. 
Here, $y$ and $z$ are shown in Figure \ref{Roche Lobe Tidal Effects}.
The function $F (K,\Gamma) $, dependent on $K$ and  $\Gamma$, is expressed as 
\begin{align}  \label{eq:58 F(K,Gamma)}
F(K,\Gamma)=K^{\frac{1}{2\Gamma } } \left ( \frac{2}{\Gamma+1}  \right ) ^{\frac{\Gamma+1}{2(\Gamma-1)} }  \left ( \frac{2\Gamma^{3/2}}{3\Gamma-1} \right ). 
\end{align} 

While we delineate the MT rate into isothermal and adiabatic phases, the transition between these regimes is inherently continuous. 
This continuity ensures that $\dot{m}_1^{\mathrm{iso}} = \dot{m}_1^{\mathrm{adi}}$ precisely when the WD's radius equals its Roche lobe radius ($R_{\mathrm{WD}} = R_{\mathrm{L}}$), thereby maintaining a smooth MT rate across the transition. 
Notably, near pericenter in highly eccentric orbits, the adiabatic MT rate significantly exceeds the isothermal rate at the point where $R_{\mathrm{WD}} = R_{\mathrm{L}}$, i.e. $\dot{m}_1^{\mathrm{adi}}(r_{\mathrm{p}}) \gg \dot{m}_1^{\mathrm{iso}}(R_{\mathrm{WD}} = R_{\mathrm{L}})$. 
Conversely, at orbital phases far from pericenter, where $(\phi_{1} - \phi_{\mathrm{ph}})/c_{\mathrm{T}}^{2} \gg 1$, the exponential term in Equation (\ref{eq:58 dotm1}) diminishes towards zero, rendering the MT rate negligible.
 
So far, we have given complete expression for the MT rate. 
However, to accurately determine the onset of MT, it is essential to evaluate the Roche lobe radius $R_{\mathrm{L}}$ of WD. 
The Roche lobe radius can be written as $R_{\mathrm{L}}$\,$=$\,$f\left ( A,q\right )D $, where $f(A,q)$ is a dimensionless function of $A$\,$=$\,$\Omega _{1}^2  D^3 / G m$ and the mass ratio is $q$\,$=$\,$m_1/m_2$.  
The factor $A$ represents the ratio of the centrifugal potential by self-rotation to the gravitational potential. 
In the specific context of a WD donor in a highly eccentric orbit around an IMBH, the spin angular velocity  $\Omega  _{1}$ of the WD is typically negligible compared to the orbital angular velocity at pericenter. 
This condition implies $A$\,$\to$\,$0$ at pericenter, satisfying $\Omega_{1} r_{\mathrm{p}}/c$\,$ \ll$\,$\sqrt{\mathrm{R}_\mathrm{s} /r_{\mathrm{p}}} $. 
Under this approximation, $f(A,q)$ can be expanded to zeroth order in  $A$, yielding: 
 \begin{align}\label{coefficient f A,q}
f(A,q)= \frac{0.5439-0.0245  \exp \left[-(0.5+\log_{10} q)^{2}\right] }{0.6  +\ln \left(1+q^{1 / 3}\right)q^{-2 / 3}}+ \mathcal{O}\left(A\right).
\end{align}
 The fitting formula used to expand to get Equation (\ref{coefficient f A,q}) is the Equation (48) of \citet{sepinsky2007equipotential}.  
This expression offers a refined estimate of the Roche lobe radius, accounting for the effects of asynchronism and orbital eccentricity, and is particularly applicable to systems where the donor's rotation is significantly slower than the orbital motion.

\subsection{Dynamics under mass transfer }\label{sec:Dynamics under mass transfer}
Now, only $\boldsymbol{f}_{\mathrm{MT}}$ remains undecided. 
We analyze from the perspective of the conservation and redistribution of orbital angular momentum $J_{\text{orb }}$ \citep{king2020gsn,king2022quasi}. 
MT results in the formation of an accretion disk, or the angular momentum might be lost from the system. 
Orbital angular momentum loss also occurs when binary total mass conservation. 
Part of the angular momentum of the ejected/accreted mass may be transferred to the donor/accretor's spin angular momentum.  

To account for both conservative and non-conservative  MT processes, we introduce two functions, $\tilde{f}$ and $\tilde{g}$, representing   these distinct influences. 
We define the orbital angular momentum loss due to MT as 
\begin{align} \label{38 Conservation of angular momentum 1} 
\frac{\dot{J}_{\text {orb }}}{J_{\text {orb }}}:=\gamma \tilde{f}(\alpha, q, \gamma) \frac{\ \dot{m}_{1}}{\ m_{1}}+(1-\gamma) \tilde{g}(\beta, q, \gamma) \frac{\dot{m}}{m}.
\end{align} 
The functions $\tilde{f}(\alpha,q,\gamma)$ and $\tilde{g}(\beta,q,\gamma)$ represent $J_{\text{orb }}$ loss for  conservative and non-conservative MT processes, respectively.  
The constants $\alpha$ and $\beta$  control the $J_{\text{orb }}$ loss in these functions. 
These functions  satisfy the conditions 
$\tilde{f}(0,q,\gamma)$\,$=$\,$\tilde{g}(0,q,\gamma)$\,$=$\,$0$,  indicating no $J_{\text{orb }}$ loss when $\alpha$ and $\beta$ vanish.

Now, we start to derive the component of the $\boldsymbol{f}_{\mathrm{MT}}$. 
Using the definition $J_{\text{orb }}$\,$=$\,$m \eta\sqrt{Gm p} $, where $\eta $\,$=$\,$ m_1 m_2 / m^2$. 
The time derivative of $J_{\text{orb }}$ can be expressed  as:
\begin{align}\label{38 Conservation of angular momentum} 
\frac{\dot{J}_{\text {orb }}}{J_{\text {orb }}} =\frac{\ \dot{m}_{1}}{\ m_{1}}+\frac{\ \dot{m}_{2}}{\ m_{2}}-\frac{1}{2} \frac{\dot{m}}{m}+\sqrt{\frac{r}{G m (1+e \cos f)} }  \mathcal{ S },
\end{align}
where $\mathcal{S}$ arises from $dp/dt$ in Equation (\ref{eq:elements differential equation11 a}).  
Here, $\dot{m}_{1}$ and $\dot{m}_{2}$ are from $d\eta/dt$. 
In addition, based on the previous assumptions for $\dot{m}_{2}$ $=$ $-\gamma \dot{m}_{1}$, we can rewrite these mass derivative using $\gamma$ and $q$, 
\begin{equation} 
\begin{aligned}
\frac{\ \dot{m}_{1}}{\ m_{1}}+\frac{\ \dot{m}_{2}}{\ m_{2}}-\frac{1}{2} \frac{\dot{m}}{m}=\frac{\ \dot{m}_{1}}{\ m_{1}} \frac{2(1- \gamma  q^2)+q(1-\gamma )}{2(1+q)}.
\end{aligned}
\end{equation}

Equation (\ref{38 Conservation of angular momentum}) represents a generalized expression determined by the system's intrinsic parameters, while Equation (\ref{38 Conservation of angular momentum 1}) serves as an additional relation capturing the possible non-conservation of $J_{\text{orb}}$ induced by MT. 
These two equations are inherently independent. 
By equating their right-hand sides and expressing $\dot{m}_{2}$ and $\dot{m}$ in terms of $\gamma$, $q$, and $\dot{m}_{1}$, we can formulate the function $\mathcal{S}$. 
This function represents the component of the MT-induced force $\boldsymbol{f}_{\text{MT}}$ along the direction $\boldsymbol{\lambda}$ and can be expressed as a function of $q$, $\gamma$, $\tilde{f}$, and $\tilde{g}$: 
\begin{align}\label{39 Conservation of angular momentum for S} 
\mathcal{ S }=&\left (-\frac{\ \dot{m}_{1}}{\ m_{1}}  \right ) r \dot{f} _\mathrm{N} \left \{   \frac{2(1- \gamma  q^2)+q(1-\gamma )[1-2(1-\gamma )\tilde{g}]}{2(1+q)} -\gamma \tilde{f}\right \} \nonumber \\
:=&\left (-\frac{\ \dot{m}_{1}}{\ m_{1}}  \right ) r \dot{f} _\mathrm{N}(1-q)\tilde{h}(q,\gamma,\alpha,\beta) .
\end{align}  
For  $\gamma $\,$= $\,$1$ and $\alpha$\,$ =$\,$ 0$,  corresponding to mass conservation  $(\dot{m} = 0)$ and no $J_{\text {orb }}$ loss, (i.e., the right-hand side of Equation (\ref{38 Conservation of angular momentum 1}) is zero), we find $\tilde{h}\left (  q,1,0,\beta \right )$\,$=$\,$1$,  consistent with previous results.
Therefore, Equation (\ref{39 Conservation of angular momentum for S}) gives the dynamics of the MT effect.

However,  using the analysis of orbital angular momentum only provides the acceleration in the direction of $\boldsymbol{\lambda}$ for the MT process, 
because  $\dot{J}_{\text {orb }}/J_{\text {orb }}$ in Equation (\ref{38 Conservation of angular momentum}) only have the function $\mathcal{S}$.  
The form of $\mathcal{ R }$ in the radial direction $\boldsymbol{n}$ cannot be derived from orbital angular momentum (Equation \ref{38 Conservation of angular momentum}).

For the following reasons, we can assume that the MT process does not induce acceleration in the radial direction.  
Because the orbit has high eccentricity, the MT process occurs exclusively at pericenter, i.e. $\dot{m}_{1 }$\,$ \propto $\,$ \delta\left (  f \right )$, then $\mathcal{R}$ does not need to be determined. 
This is because the coefficient of  $\mathcal{R}$  in   $de/dt$ in Equation  (\ref{eq:elements differential equation11 b}) contains  $\sin f$, which vanishes at $f$\,$=$\,$0$, and is an odd function.  
In addition, according to  Equation (\ref{eq:elements differential equation11}), for the pericenter distance variation, $\dot{ r}_{\text{p}}=\left ( \dot{ p}-\dot{ e}\, r_{\text{p}} \right )\left (  r_{\text{p}}/p \right )$, it is
\begin{align}\label{5.2 long-term orbital effects of MT discrete method drp}
\frac{d r_{\text{p}}}{d t} = \sqrt{\frac{r_{\text{p}}^3}{G m (1+e)}}\left[-\sin f \, \mathcal{R}+\frac{2(2+e+e \cos f) }{1+e \cos f}  \sin^2 \! \frac{f}{2}\, \mathcal{S}\right].  
\end{align}  
The presence of ($\sin$\,$f$) and ($\sin^2$$f/2$) in this equation,  result from $r_{\text{p}}$ will hardly change compared with $p$.
Therefore,  assuming $\mathcal{R} $\,$= $\,$0$ is a reasonable  simplification for high-eccentricity systems. 

Several important aspects warrant further discussion. 
First, the additional acceleration term $\boldsymbol{f}_{\text{MT}}$ given in Equation (\ref{39 Conservation of angular momentum for S}) is derived within a classical Galilean framework. 
Here, the orbital angular momentum $J_{\text {orb }}$ is approximated to Newtonian (0PN) as $\boldsymbol{J} _{\text {orb }}$\,$=$\,$\sum  m_i\boldsymbol{y}_i\times \boldsymbol{v}_i$. However, near pericentre, where $r_{\text{p}}$\,$<$\,$10 G m/c^2$ and characteristic velocities reach $0.1 c$—this approximation may introduce non-negligible errors. 
Such limitations highlight the necessity for a fully relativistic treatment of MT-induced perturbations, which should be pursued in future work. 
Second, it is notable that some previous studies, particularly those not focused on orbital angular momentum evolution (e.g., \citealt{zalamea2010white}), assume that MT does not induce any perturbative acceleration. 
This corresponds to setting $\mathcal{S}$\,$=$\,$0$ in Equation (\ref{38 Conservation of angular momentum}). 
Under such assumptions, $J_{\text {orb}}$ is not conserved. 
Assuming mass conservation ($\gamma$\,$=$\,$1$) and an extreme mass ratio ($q$\,$\ll$\,$1$), Equation (\ref{38 Conservation of angular momentum}) is  $\dot{J}_{\text {orb}}/J_{\text {orb}}$\,$\approx$\,$\dot{m}_{1}/m_{1}$, implying that the angular momentum associated with the transferred mass is entirely lost from the system.  
This is the maximum limit of $J_{\text {orb}}$ loss when mass conservation.

In Sections \ref{sec:Analyzing orbital results without MT in PN method}, \ref{sec:Analyzing orbit with MT}, and \ref{sec:Impact on GW sigal and detection}, we restrict our analysis to scenarios where total mass and $J_{\text {orb}}$ are conserved, corresponding to $\tilde{h}$\,$=$\,$1$. 
Consequently, we do not specify explicit forms for the functions $\tilde{f}(\alpha,q,\gamma)$ and $\tilde{g}(\beta,q,\gamma)$ introduced in the generalized conservation laws. 
A more detailed discussion on how deviations from these conservation assumptions could affect the long-term orbital evolution will be presented in Section \ref{sec:Discussion 1 2}.

\section{Short-term orbital evolution using post-Newtonian method}\label{sec:Analyzing orbital results without MT in PN method}
In this section, we analyze the orbital dynamics of an EMRI system  as described in Section \ref{sec:Dynamics under post-Newton method}.    
We adopt the PN method rather than the Kerr spacetime geodesic approach because it provides a specific form of the perturbation acceleration,  as shown in Equation (\ref{eq:relative acceleration}). 
The acceleration $\boldsymbol{f}_{\text{MT}}$  from MT can be directly added to this perturbation.  
Such a simulation is simple. 
In contrast, converting the geodesic equation into an acceleration form and applying the perturbed Kepler method is challenging, as is incorporating the effect of gravitational radiation into geodesics.

Section \ref{sec:Angle relationship in relative coordinate} establishes the relationship between the latitude angle $\psi$, and the orbital elements ($ f, \omega,\phi,\Omega, \iota$). 
Here, $\psi$ is the angle between the instantaneous orbital plane and the equatorial plane, which is perpendicular to the IMBH’s spin axis.  
Section \ref{sec:Mass limit of WD} indicates the upper limit of the mass of a white dwarf in this system.
Section \ref{sec:Orbital evolution with MT process} includes the MT process and compare the effects of MT and gravitational radiation on key orbital parameters.

\subsection{Angular relationship in relative coordinate}\label{sec:Angle relationship in relative coordinate}

We adopt the relative coordinate system illustrated in Figure \ref{Kepler orbit}, where the Cartesian axes (X,Y,Z) correspond to the spherical coordinates $  ( r,\theta ,\varphi ) $, with the transformation $ (r^{X},r^{Y},r^{Z} )$\,$=$\,$r\left (    \sin\theta \cos\varphi,  \sin\theta \sin\varphi,  \cos\theta\right )$. From this definition, the polar angle $\theta$  can be written as $\theta$\,$=$\,$\cos^{-1}\left ( \sin \iota \sin \phi \right )$. Here, $\theta$ is the polar angle, while its complement $\psi$\,$=$\,$\pi/2$\,$-$\,$\theta$ is referred to as the latitude.

We align the spin axis of the IMBH along the Z-axis and place the WD at pericentre at the initial time, setting $f_0 =\phi_0$\,$-$\,$\omega_0 =0$. Additionally, we choose $\varphi_0$\,$=$\,$\pi/2$. Throughout, a subscript $0$ denotes quantities evaluated at the initial time. The orbital angular parameters are then determined as $\Omega_0$\,$=$\,$\pi/2$\,$-$\,$\varphi_0$,  $\iota _0$\,$=$\,$\pi/2$\,$-$\,$\theta_0$,  $\phi_0$\,$=$\,$\varphi_0$,  $\omega_0$\,$=$\,$\phi_0$\,$-$\,$f_0$.

The spin orientation of a body is described using angles $\vartheta$ and $\varphi$ in conjunction with the basis vectors $\boldsymbol{n}, \boldsymbol{\lambda} , \boldsymbol{\ell}$, such that
\begin{align}\label{66 SA} 
&\frac{\boldsymbol{S}_\mathrm{A} }{\left | \boldsymbol{S}_\mathrm{A} \right | }:=\sin \vartheta_{\mathrm{A}} \cos\varphi _{\mathrm{A}} \boldsymbol{n}+ \sin\vartheta_{\mathrm{A}} \sin\varphi_{\mathrm{A}} \boldsymbol{\lambda}+\cos\vartheta_{\mathrm{A}} \boldsymbol{\ell},
\end{align}
where $\left | \boldsymbol{S}_\mathrm{A} \right | $\,$:=$\,$ \chi _\mathrm{A}G m_\mathrm{A}^2/c$, and $\chi _\mathrm{A}$ is the dimensionless spin size parameter, generally ranging from $0$ to $1$. 
The initial spin $\boldsymbol{S}_{20}$ of the IMBH can be expressed  as $\left | \boldsymbol{S}_{20} \right |$$\left(\cos \theta_{0} \boldsymbol{n}+ \sin \theta_{0} \boldsymbol{\iota}\right)$. 
Comparing with Equation (\ref{66 SA}), the initial angles satisfy     $\vartheta_{20}$\,$=$\,$\frac{\pi}{2}-\theta_{0}$\,$=$\,$\iota _0$ and $\varphi_{20}$\,$=$\,$0$. 
In the context of Kerr spacetime, the quantity $\left | \psi_0 \right |$\,$=$\,$\left | \pi/2- \theta_0 \right |$ represents the maximum latitude of the corresponding geodesic motion \citep{schmidt2002celestial,fujita2009analytical}. 

In general, the initial angle conditions are 
\begin{align}\label{66 SA1}
&\Omega_0=0,\quad \psi_0=\iota _0=\frac{\pi}{2}-\theta_0, \quad  \phi_0=\frac{\pi}{2},\quad \omega_0 =\frac{\pi}{2},\quad  f_0=0, \nonumber \\
&\vartheta_{20}=\frac{\pi}{2}-\theta_{0},\quad  \varphi_{20}=0. 
\end{align}

\subsection{Upper mass limit of WD}\label{sec:Mass limit of WD}
\begin{figure}
    \includegraphics[width=1\linewidth]{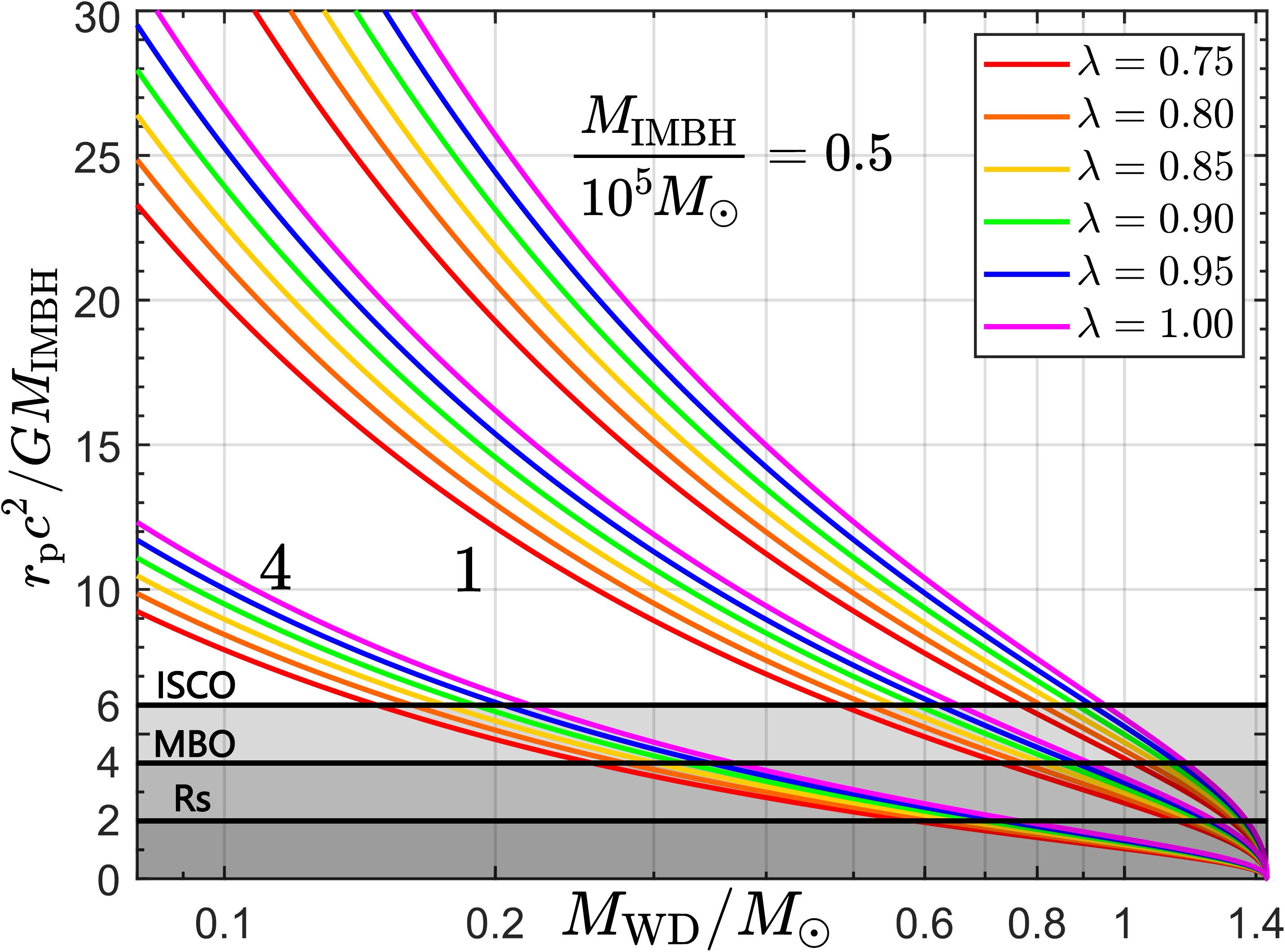} 
    \begin{minipage}{\linewidth}
\caption{The pericenter distance $r_{\text{p}}$ for Roche-lobe overflow  given by $r_{\text{p}}$\,$=$\,$\lambda R_{\mathrm{WD}}(M_{\mathrm{WD}})/f(A,q)$, is shown as a function of $ M_{\mathrm{WD}} $ for fixed values of $ M_{\mathrm{IMBH}}/10^{5} \text{M}_{\sun}$\,$=$\,$0.5$, $1$, $4$. 
Here, $\lambda$\,$\le$\,$1$ ensures $R_{\text{L}}$\,$\le$\,$R_{\mathrm{WD}}$ at pericenter.
The three black lines represent different distances: 
ISCO is the innermost stable circular orbit of a Schwarzschild black hole, $r_{\mathrm{ISCO}}$\,$= $\,$6 GM_{\text{IMBH}}/c^2$;
MBO is the marginally bound orbit, $r_{\mathrm{MBO}}$\,$=$\,$4 GM_{\text{IMBH}}/c^2$;
R$_{\text{s}}$   is the Schwarzschild radius, R$_{\text{s}}$\,$=$\,$2 GM_{\text{IMBH}}/c^2$. }  
	\label{RWD_Roche}
\end{minipage} 
\end{figure}
 
To carry out the simulations, the mass of the companion star must be specified. For a given IMBH mass $M_{\mathrm{IMBH}}$, there exists an upper limit on the WD mass $M_{\mathrm{WD}}$ that permits Roche-lobe overflow without resulting in direct infall into the IMBH. 
This limit arises because the WD radius $R_{\mathrm{WD}}(M_{\mathrm{WD}})$ decreases with increasing mass, allowing Roche-lobe overflow to occur at a smaller pericentre distance, satisfying $r_{\text{p}} \le R_{\mathrm{WD}}/f (A,q )$. 
However, the pericentre distance $r_{\mathrm{p}}$ cannot be reduced indefinitely, as excessively small values would cause the WD to plunge directly into the IMBH.

A minimum pericenter distance, $r_{\mathrm{p}}^{\mathrm{bound}} $\,$\le$\,$ r_{\text{p}}$, must be maintained to ensure a bound orbit. 
The limiting WD mass is determined by the condition $R_{\mathrm{WD}}$\,$=$\,$f  ( A,q )r_{\text{p}}$, evaluated at $r_{\text{p}}=r_{\mathrm{p}}^{\mathrm{bound}}$.

In Figure \ref{RWD_Roche}, we illustrate the relationship between the pericentre distance $r_{\mathrm{p}}$ and the WD mass $M_{\mathrm{WD}}$ required for Roche-lobe overflow, where $\lambda$\,$:=$\,$ f( A,q ) r_{\text{p}}/R_{\mathrm{WD}} $\,$\le $\,$1$, for a fixed IMBH mass. 
Imposing the bound orbit condition $r_{\mathrm{p}}^{\mathrm{bound}} = r_{\mathrm{ISCO}}$, we find that the maximum WD mass compatible with Roche-lobe overflow is constrained to $M_{\mathrm{WD}} $\,$< 0.2$\,$\text{M}_{\sun}$ when $M_{\mathrm{IMBH}}$\,$=$\,$4\times 10^5\, \text{M}_{\sun}$.
This constraint underscores the necessity of a highly eccentric orbit, finely tuned to match the QPE period, for MT to occur—implying that the donor star should be a low-mass WD.

\subsection{Orbital evolution with mass transfer}\label{sec:Orbital evolution with MT process} 
In this subsection, we analyze the orbits using the PN method with the MT processes. 
The pericenter distance $r_{\text{p}}$ and radial eccentricity $e_{\text{r}}$\,$:=$\,$\left ( r_{\mathrm{a} }  -r_{\mathrm{p} } \right ) /\left ( r_{\mathrm{a} }  +r_{\mathrm{p} } \right )$ can be roughly determined.  
First,  MT requires  the Roche lobe radius, $R_{\text{Lp}}$\,$=$\,$f ( A,q )r_{\text{p}}$\,$\lesssim$\,$R_{\mathrm{WD}} ( M_{\mathrm{WD}} ) $, as outlined in Section \ref{sec:Mass limit of WD}, with $M_{\mathrm{WD}}$\,$<$\,$M_{\mathrm{WD}}^{\mathrm{limit}}(M_{\text{IMBH}}, r_{\mathrm{p}}^{\mathrm{bound}})$\,$ \ll $\,$M_{\text{ch}}$,  which directly constrains the range of $r_{\text{p}}$.
Second, due to the Hills mechanism, and to match the periods observed in QPE light curves \citep{miniutti2019nine, miniutti2023repeating}, the radial eccentricity $e_{\text{r}}> 0.95$ \citep{wang2022model, chen2022milli,WangDi-1}. 

We assume that the spin of the low-mass WD  can be ignored, setting $\chi_1$\,$=$\,$0$. We also set the mass $m_{1}$\,$=$\,$M_{\mathrm{WD}}$\,$=$\,$0.15$\,$\text{M}_{\sun}$,  $m_2$\,$=$\,$M_{\mathrm{IMBH}}$\,$=$\,$4 \times 10^{5} \, \text{M}_{\sun}$, and $q$\,$=$\,$m_{1}/m_{2}$\,$=$\,$3.75 \times 10^{-7}$.

\begin{figure}
\includegraphics[width=\linewidth]{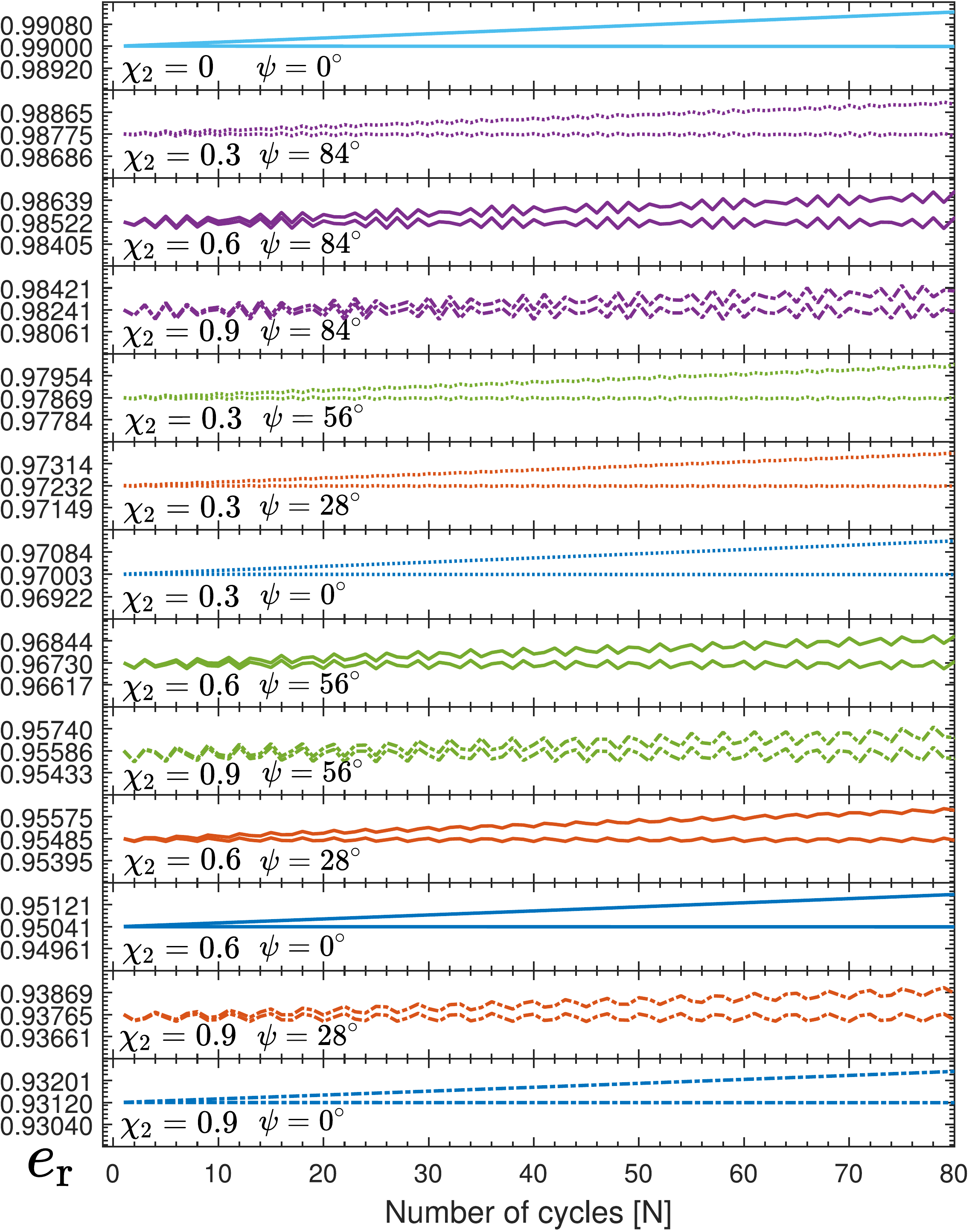}
\begin{minipage}{\linewidth}
\caption{The evolution of the radial eccentricity $e_{\mathrm{r}}$ with and without MT.  
In each panel, the higher curve represents the evolution of $e_{\mathrm{r}}^{\mathrm{MT}}$ with MT, and the low one represents  the evolution of $e_{\mathrm{r}}^{*}$ without MT. 
Different line colors correspond to the same initial latitude $\psi_0$,   
and different line styles denote the same IMBH spin parameters  $\chi_2$.   }
\label{Variation of eccentricity er in PN with MT}
\end{minipage}
\end{figure} 

The initial pericenter distance $r_{\mathrm{p}0}$ is set based on the relationship between $R_{\mathrm{Lp}}$ and $R_{\mathrm{WD}}$, i.e.   $R_{\mathrm{Lp}}$\,$=$\,$f(A,q) r_{\mathrm{p}}$\,$:=$\,$(1-\epsilon )R_{\mathrm{WD}}$\,$<$\,$R_{\mathrm{WD}}$. 
here, to quantify the degree of overflow, we define a dimensionless overflow  parameter, $\epsilon(r_{\mathrm{p} },M_{\mathrm{WD}},q)$\,$=$\,$1-r_{\mathrm{p} }C_{\text{L}}(q)/ R_{\mathrm{WD}}(M_{\mathrm{WD}})$, where $C_{\text{L}}$\,$=$\,$f$\,$\left ( A = 0,q \right ) $ characterises the Roche geometry. 
We set the initial overflow parameter to $\epsilon_0$\,$=$\,$5\times10^{-3}$, yielding: $r_{\mathrm{p}0}$\,$\simeq $\,$3.1502\times 10^{6} G \text{M}_{\sun} /c^2$\,$\simeq$\,$ 7.875\ G m_2 /c^2$\,$\simeq$\,$ 253.87 R_{\mathrm{WD} }  ( M_{\mathrm{WD}}  ) $.
The value of $r_{\mathrm{p}0}$ is fixed in all subsequent calculations.
For the initial radial eccentricity, we adopt $e_{\mathrm{r0}}$\,$=$\,$0.99$, consistent with the orbit of the QPE, when assume zero MBH spin ($\chi_2$\,$=$\,$0$). 
Because lower $\chi_2$ and higher initial latitude $\psi_0$ lead to higher initial $e_{\mathrm{r}}$.

For initial angular configurations, we use the angular relationships described in Section \ref{sec:Angle relationship in relative coordinate}. 
We only need to adjust initial latitude $\psi_0$ and  IMBH spin parameter $\chi_2$.  
We set $\psi_{0}$\,$=$\,$0\degr$, $28\degr$, $56\degr$, $84\degr$, and   $\chi_2$\,$=$\,$0$, $0.3$, $0.6$, $0.9$. 
The only remaining parameter introduced is $\xi_0$, which regulates the   WD surface density $\rho_{\text{ph}}$   in Equation (\ref{65 fMT}).  
According to the result of Equation (\ref{eq:24 rhoph}), we set $\xi_0$\,$=$\,$3.98\times10^{-3}$. 
To summarize, all the initial parameters that need to be determined independently include $\left \{ e_{\mathrm{r0}}, M_{\text{WD}}, M_{\text{IMBH}}, \epsilon_0, \psi_{0}, \chi_2, \xi_0  \right \}$.

To illustrate the impact of MT on orbital dynamics, we examine the evolution of the radial eccentricity $e_{\text{r}}$, under scenarios with and without MT.
As shown in Figure \ref{Variation of eccentricity er in PN with MT}, GW emission alone leads to a gradual decrease in $e_{\text{r}}$, with a rate of change $\delta e_{\mathrm{r}}^{\mathrm{GW}}$\,$ \approx $\,$-10^{-7}$ per cycle.
In contrast, when MT is included, the eccentricity increases, with $\delta e_{\mathrm{r}}^{\text{GW+MT}}$\,$ \approx$\,$ 10^{-5}$ per cycle.

 \begin{figure}   
\includegraphics[width=\linewidth]
{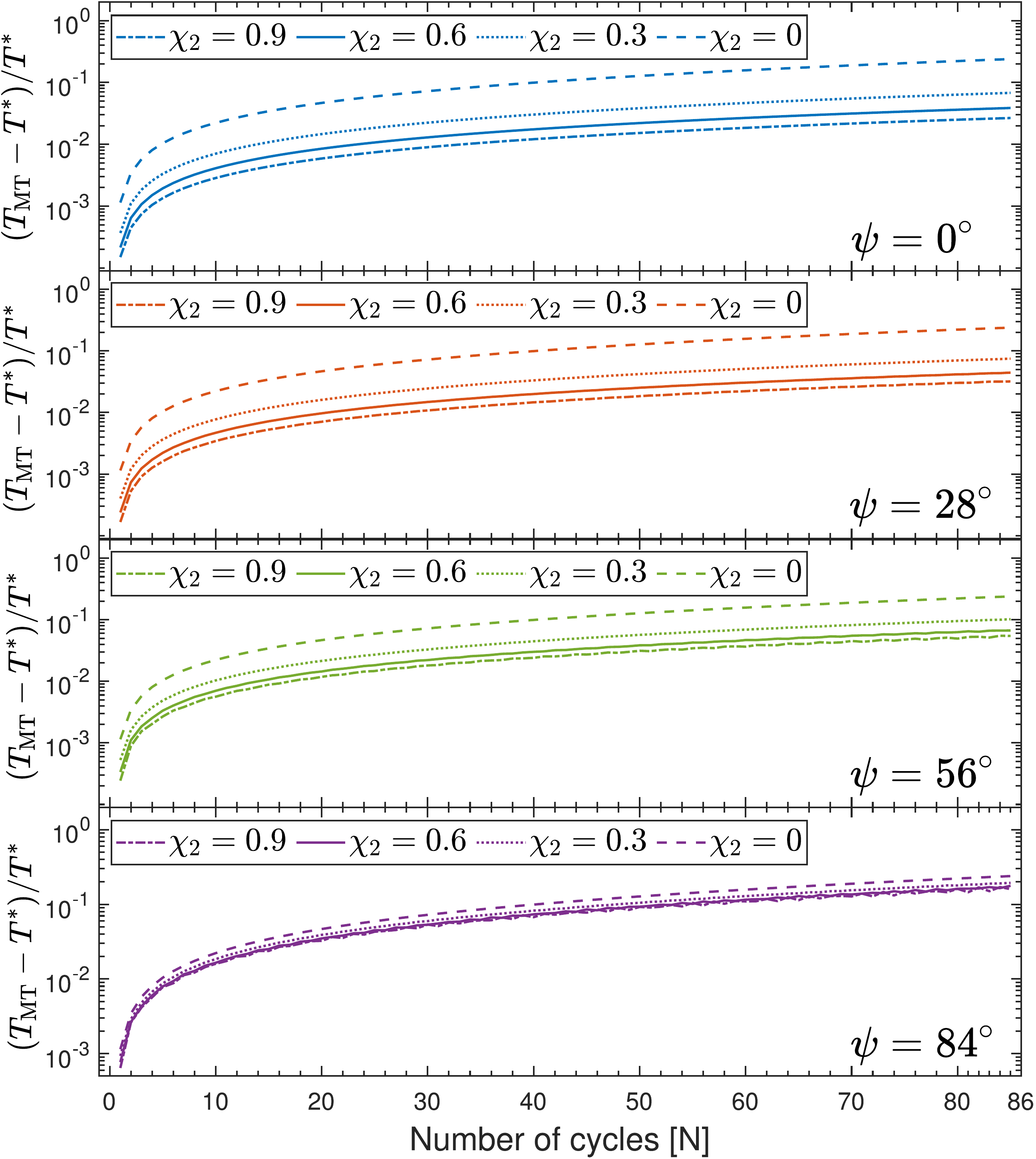}
\begin{minipage}{\linewidth}
\caption{The evolution of $ (T_{\text{MT}}-T^*) /T^*$  due to the MT process, where $T^*(\mathrm{N})$ denotes the period without MT process. 
In contrast, $T_{\text{MT}}(\mathrm{N})$ represents the period with  MT ($\xi_0=3.98\times10^{-3}$).   
We give the influence of different $\chi_2$ under identical initial $\psi_0$.  
}
\label{Variation of period T in PN with MT}
\end{minipage} 
\end{figure} 

This pronounced eccentricity growth driven by MT highlights its dominant influence on the orbital evolution under the current parameter regime.
The perturbative acceleration associated with MT, as described by Equation (\ref{eq:elements differential equation11 b}), exerts a significantly stronger effect on the orbital parameters than the eccentricity damping caused by GW radiation.
These results underscore the necessity of incorporating MT effects in models aiming to accurately capture the dynamical behaviour of such systems.

In addition, there are some relativistic effects brought by IMBH spin.
In Figure \ref{Variation of eccentricity er in PN with MT}, lower $\chi_2$ and higher initial $\psi_0$ lead to higher initial $e_{\mathrm{r}}$.
This relationship arises because, at a fixed pericenter distance, the IMBH's spin can transfer orbital energy to the companion star.
For prograde orbits—those aligned with the IMBH's spin—the enhanced frame-dragging effect leads to increased orbital velocities, reduced semi-major axes, and consequently, lower orbital periods and eccentricities.
Conversely, orbits with higher inclinations, deviating from the equatorial plane, reduce this spin-induced effect, resulting in eccentricities that more closely resemble scenarios without spin.
This behavior aligns with post-Newtonian predictions, where spin-orbit coupling induces variations in orbital elements, contingent on the spin magnitude and orbital orientation.

The change of eccentricity in Figure \ref{Variation of eccentricity er in PN with MT} is not monotonous, but has obvious fluctuations. 
Higher $\chi_2$  and  initial $\psi_0$  enhance the jitter phenomenon. 
This fluctuation is more pronounced in the first few weeks of the orbit, but as the simulation time increases, the effects of mass transfer and gravitational radiation allow us to identify long-term evolutionary features from the fluctuations.
However, the distinctive periodic oscillation effect  does not occur in the geodesic of Kerr spacetimes, nor is it present in non-spinning binary systems using the PN method. 
We speculate that this effect arises from the spin terms $\boldsymbol{a}_{\mathrm{SO}}$, $\boldsymbol{a}_{\mathrm{SS}}$, $\boldsymbol{a}_{\mathrm{SOPN}}$ in the PN orbital dynamics described in Equation (\ref{eq:relative acceleration}), derived from \citet{tagoshi2001gravitational} using  the harmonic gauge condition. 
Furthermore, we find that the changes in $e_{\mathrm{r}}$ does not decrease in higher PN order relative acceleration terms (below  4 PN), when using equations from different PN orders to reproduce the same $r_{\mathrm{p}}$ and $T$.
This suggests that higher PN order terms involving the spin parameter $\chi_2$ should be considered, or it may indicate a subtle connection between harmonic coordinates and Boyer-Lindquist coordinates.

Figure \ref{Variation of period T in PN with MT} shows that, the orbital period becomes longer when MT is introduced, and the increasing is larger for higher $\psi_0$ and lower $\chi_2$.
This result can be inferred from the increase of $e_\mathrm{r}$. 
We give the average values and the maximum jitter difference of $e_{\text{r}}$ and $T$ in Table \ref{tab:average eccentricity}. 
Lower $\chi_2$ and higher initial  $\psi_0$ lead to higher  $\bar{e}_{\mathrm{r} }$ and $\overline{T}$. However, higher $\chi_2$  and  initial $\psi_0$  enhance the jitter difference of the $\Delta e_{\mathrm{r}}$ and $\Delta T$.

\begin{table}
\caption{The average radial eccentricity $\bar{e}_{\mathrm{r} }$ and the difference $e_{\mathrm{rmax} }-\bar{e}_{\mathrm{r} }$, and the average period $\overline{T}$ and the ratio of the difference $ (T_{\mathrm{max} }-\overline{T} )/\overline{T}$. }
 \label{tab:average eccentricity}
\begin{tabular*}{0.9445\columnwidth}{|@{\hspace*{3.5pt}}c@{\hspace*{3.5pt}}|@{\hspace*{7.5pt}}c@{\hspace*{15pt}}c@{\hspace*{15pt}}c@{\hspace*{15pt}}c@{\hspace*{7.5pt}}|}
  \hline
  $\bar{e}_{\mathrm{r} }-0.99$ & $\chi_2=0$ & $\chi_2=0.3$ & $\chi_2=0.6$ & $\chi_2=0.9$  \\ \hline
        $\psi_0=0^\circ\ \  $ & $0$   & $-0.0200 $ & $-0.0396 $ & $-0.0588 $   \\
        $\psi_0=28^\circ$ & $0$ & $-0.0177 $ & $-0.0352 $ & $-0.0525 $   \\ 
        $\psi_0=56^\circ$ & $0$ & $-0.0113 $ & $-0.0228 $ & $-0.0344 $   \\ 
        $\psi_0=84^\circ$ & $0$ & $-0.0023 $ & $-0.0048 $ & $-0.0077 $  \\ \hline
  \hline
  $e_{\mathrm{rmax} }-\bar{e}_{\mathrm{r} }$ & $\chi_2=0$ & $\chi_2=0.3$ & $\chi_2=0.6$ & $\chi_2=0.9$  \\ \hline
        $\psi_0=0^\circ\ \  $ & $5.358\text{e-6}$ & $5.019\text{e-6}$ & $4.984\text{e-6}$ & $4.954\text{e-6}$  \\
        $\psi_0=28^\circ$ & $5.358\text{e-6}$                 & $2.146\text{e-5}$ & $7.786\text{e-5}$ & $1.959\text{e-4}$   \\ 
        $\psi_0=56^\circ$ & $5.358\text{e-6}$                 & $5.164\text{e-5}$ & $2.168\text{e-4}$ & $5.223\text{e-4}$   \\ 
        $\psi_0=84^\circ$ & $5.358\text{e-6}$                 & $7.477\text{e-5}$ & $2.904\text{e-4}$ & $6.508\text{e-4}$  \\ \hline
        \hline
  $\overline{T}/\text{hr}$ & $\chi_2=0$ & $\chi_2=0.3$ & $\chi_2=0.6$ & $\chi_2=0.9$  \\ \hline
        $\psi_0=0^\circ\ \  $  & $76.460$ & $14.911$ & $7.0774$ & $4.3667$   \\
        $\psi_0=28^\circ$ & $76.460$     & $16.772$ & $8.1180$ & $5.0305$   \\ 
        $\psi_0=56^\circ$ & $76.460$     & $24.730$ & $13.055$ & $8.3270$   \\ 
        $\psi_0=84^\circ$ & $76.460$     & $56.415$ & $42.455$ & $32.573$  \\ \hline
        \hline
  $  (T_{\mathrm{max} }-\overline{T} )/\overline{T}$ & $\chi_2=0$ & $\chi_2=0.3$ & $\chi_2=0.6$ & $\chi_2=0.9$  \\ \hline
        $\psi_0=0^\circ\ \  $ & $7.992\text{e-4}$ & $2.511\text{e-4}$ & $1.501\text{e-4}$ & $1.080\text{e-4}$  \\
        $\psi_0=28^\circ$ & $7.992\text{e-4}$                 & $1.158\text{e-3}$ & $2.602\text{e-3}$ & $4.850\text{e-3}$   \\ 
        $\psi_0=56^\circ$ & $7.992\text{e-4}$                 & $3.588\text{e-3}$ & $9.925\text{e-3}$ & $1.789\text{e-2}$   \\ 
        $\psi_0=84^\circ$ & $7.992\text{e-4}$                 & $9.162\text{e-3}$ & $2.961\text{e-2}$ & $5.620\text{e-2}$  \\ \hline
 \end{tabular*}
\end{table}

We note that the relationship between $\bar{e}_{\mathrm{r}}$ and $\overline{T}$ and the relationship between $\Delta e_{\mathrm{r}}$ and $\Delta T$, can be explained by Kepler's law even though the EOM include PN corrections. 
Using Kepler's period expression $T$\,$\propto$\,$a^{3/2}$, where semi-major axis $a$\,$=$\,$(r_{\mathrm{a}}+r_{\mathrm{p}})/2$, and applying a first-order expansion for $\Delta T\ll \overline{T}$, we can get 
\begin{align}\label{46 T - er} 
 \frac{\overline{T} _1}{\overline{T}_2}&=\left ( \frac{\bar{r}_{\mathrm{p1} } }{\bar{r}_{\mathrm{p2} } }  \frac{1-\bar{e} _\mathrm{r2} }{1-\bar{e} _\mathrm{r1}}\right ) ^{3/2}\! =\left ( \frac{\bar{r}_{\mathrm{a1} } }{\bar{r}_{\mathrm{a2} } }  \frac{1+\bar{e} _\mathrm{r2} }{1+\bar{e} _\mathrm{r1}}\right ) ^{3/2} \! \simeq \left ( \frac{1-\bar{e} _\mathrm{r2} }{1-\bar{e} _\mathrm{r1}} \right ) ^{3/2}, \nonumber\\ 
\frac{\Delta T}{\overline{T} }  &  \simeq\frac{3}{2} \frac{\Delta r_{\mathrm{a} }+\Delta r_{\mathrm{p} }}{\bar{r}_{\mathrm{a} }+\bar{r}_{\mathrm{p} }} \simeq \frac{3}{2} \frac{\Delta e_{\mathrm{r} }}{1-\bar{e}_{\mathrm{r} }}. 
\end{align}
The above two equations provide good fit to the results  presented in Table \ref{tab:average eccentricity}.   
These relationships can also help us calculate the effects of MT and gravitational radiation.

As described in Equation (\ref{46 T - er}). 
We get $\delta \overline{T}_{\text{GW+MT}}/\overline{T} $\,$\approx $\,$10^{-3} $, and $\delta \overline{T}_{\text{GW}}/\overline{T}$\,$\approx$\,$10^{-5}$. 
By comparing with $(T_{\mathrm{max} }-\overline{T} )/\overline{T}$ in Table \ref{tab:average eccentricity}, we find that within  $10 $ to $10^2$ cycles, the change of period by the MT process completely masks the PN jitter effect, so no obvious jitter effect is found in Figure \ref{Variation of period T in PN with MT}.

\begin{figure} 
\includegraphics[width=\linewidth]{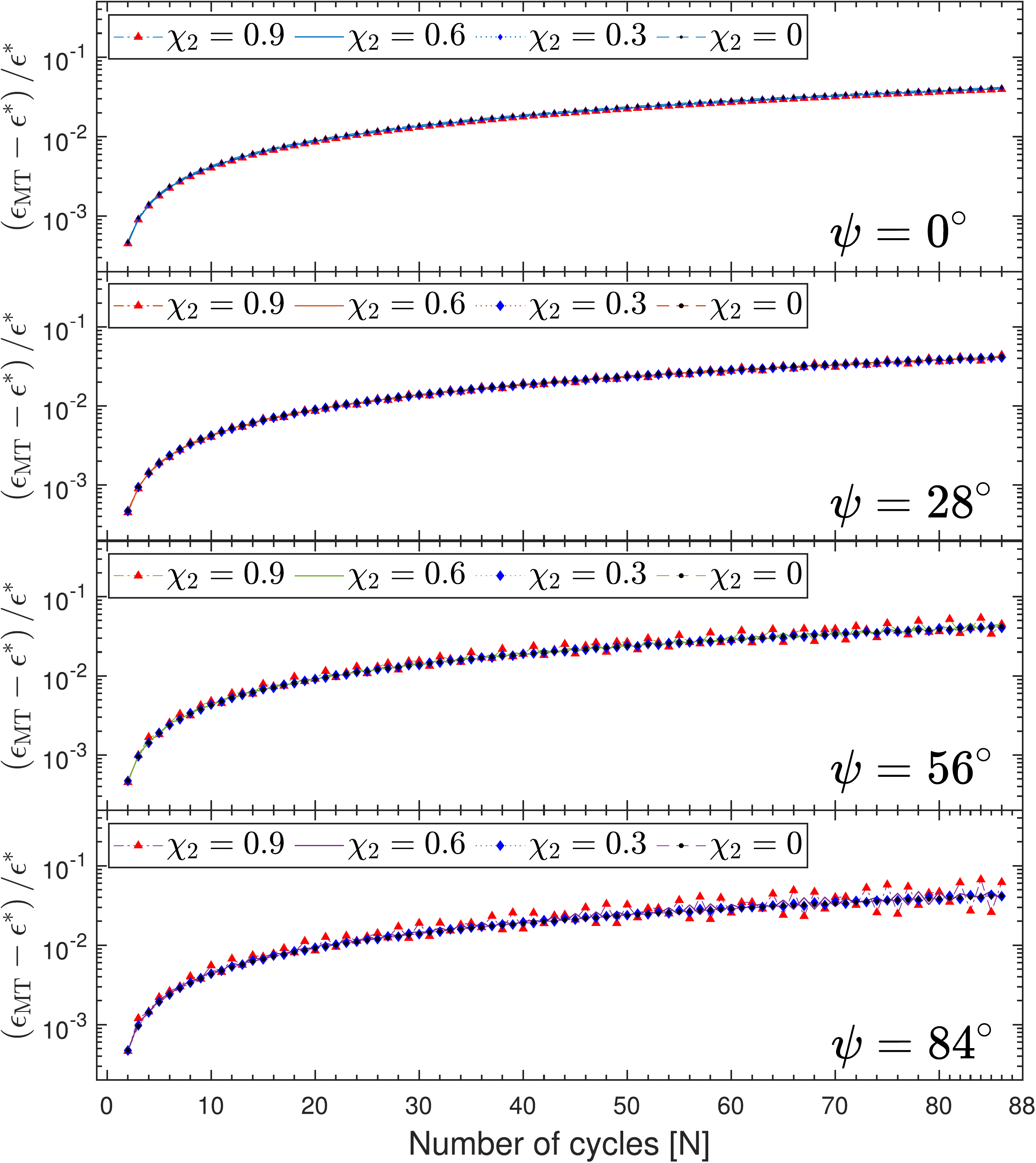}     
\begin{minipage}{\linewidth}
\caption{The evolution of $\left (  \epsilon_{\text{MT}}  -\epsilon^*  \right )$\,$ /$\,$\epsilon^*$ due to the MT process, where  $\epsilon_{\text{MT}} $ denotes  $\epsilon $ with  MT ($\xi_0$\,$=$\,$3.98\times10^{-3}$), 
and $\epsilon^* $ denotes $\epsilon $ without  MT.  
For the same initial $\psi_0$,  the results for different $\chi_2$ are shown in a single subfigure. 
In the top panel, where $\psi_0=0^\circ$,  the four lines do not overlap; the lower the $\chi_2$, the greater the value.
} 
\label{Variation of epsilon in PN with MT}
\end{minipage}  
\end{figure}   

Finally, since the MT rate depends sensitively on the ratio of the WD’s radius to the Roche lobe radius near pericenter, we use the dimensionless overflow parameter 
\begin{align}\label{44 epsilon - rp defintion} 
& \epsilon:=\frac{R_{\mathrm{WD}}-R_{\mathrm{Lp}}}{R_{\mathrm{WD}}}=1-\frac{r_{\mathrm{p}}\, C_{\mathrm{L}}(q)}{R_{\mathrm{WD}}\left(M_{\mathrm{WD}}\right)},
\end{align} 
to assess how PN orbital effects influence MT and its rate. 
In Figure \ref{Variation of epsilon in PN with MT}, different values of $\chi_2$ and  $\psi_0$  have almost no effect on $\left (  \epsilon_{\text{MT}}  -\epsilon^*  \right )  /\epsilon^*$,  
the periodic oscillations caused by the PN method are tightly concentrated around a single curve on the logarithmic plot. 
This indicates that spin is not strongly related to the growth of $ \epsilon_{\text{MT}}$ under MT.

For the change of pericenter distance $r_{\text{p}}$, we do not need to draw its figure separately, because it is closely related to parameter $\epsilon$.
The relationship between $\Delta \epsilon$ and $\Delta r_{\text{p}}$  can be expressed  using a first-order Taylor expansion ($\Delta \epsilon$\,$\ll$\,$\epsilon$, $\Delta r_{\text{p}}$\,$\ll $\,$r_{\text{p}}$) :
\begin{align}\label{45 epsilon - rp} 
& \frac{\Delta \epsilon }{1-\epsilon} \simeq -\frac{\Delta r_{\mathrm{p} }}{r_{\mathrm{p} }}-\frac{\Delta \left (  C_{\mathrm{L} }/R_{\mathrm{WD} }  \right )}{C_{\mathrm{L} }/R_{\mathrm{WD} }} ,
\end{align}
where $\Delta \left (  C_{\mathrm{L} }/R_{\mathrm{WD} }  \right )$ reflects the change of $M_{\text{WD}}$ and $M_{\text{IMBH}}$. 
First, if without the MT process, the mass of WD and the mass ratio $q$ will not change, i.e. $\Delta \left (  C_{\mathrm{L} }/R_{\mathrm{WD} }  \right )$ is zero.  Therefore,  $\epsilon$ only changes with $r_{\mathrm{p}}$.
Because $\epsilon$\,$ \ll$\,$  1$, the relative change $\Delta \epsilon/\epsilon$   is
\begin{align}\label{46 epsilon - rp} 
&\frac{\Delta \epsilon}{\epsilon}   \simeq -\frac{\Delta r_{\mathrm{p}}}{r_{\mathrm{p}} \epsilon}    \gg-\frac{\Delta r_{\mathrm{p}}}{r_{\mathrm{p}} } .
\end{align} 
When $\chi_2$ and $\psi_0$ are  large, a jitter   $\Delta r_{\text{p}}/r_{\text{p}} $\,$\sim$\,$10^{-4}$  can alter $\epsilon$\,$\sim$\,$10^{-3}$ by over  $10\%$, significantly impacting the MT rate upon reconsideration of the MT process. 
This reflects that even a very small change in $r_{\text{p}}$ can have a big impact on MT rate.  
 
Second, if with the MT process, $\Delta \left (  C_{\mathrm{L} }/R_{\mathrm{WD} }\right )$ leads to a gradual increase in the relative deviation of the overflow parameter $\epsilon_{\text{MT}}$ from  the parameter $\epsilon^*$ which is without MT, as shown in Figure \ref{Variation of epsilon in PN with MT}. 
To understand the  importance of $\Delta \left (  C_{\mathrm{L} }/R_{\mathrm{WD} }\right )$, we give the first-order expansion by $\Delta M_{\text{WD}}/  M_{\text{WD}} $, using Equations (\ref{eq:RWD}) and (\ref{coefficient f A,q}):
\begin{equation} 
\begin{aligned}
& -\frac{\Delta \left (  C_{\mathrm{L} }/R_{\mathrm{WD} }  \right )}{C_{\mathrm{L} }/R_{\mathrm{WD} }} \simeq -\frac{2+q\gamma }{3} \frac{\Delta M_{\text{WD}}}{M_{\text{WD}}}>0.
\end{aligned}
\label{47 CLdRWD - MWDdMWD} 
\end{equation}
Here, $\Delta$ represents the difference between the MT process and the process without MT. From the results of numerical calculation, we find  $\left |  \Delta M_{\text{WD}}/M_{\text{WD}}\right | \gg  \left |  \Delta r_{\text{p}}/r_{\text{p}}\right | $. Therefore, by Equation (\ref{45 epsilon - rp}), 
\begin{equation} 
\begin{aligned}
& \frac{\epsilon_{\text{MT}}  -\epsilon^*   }{ \epsilon^*} \simeq -\frac{\Delta \left (  C_{\mathrm{L} }/R_{\mathrm{WD} }  \right )}{C_{\mathrm{L} }/R_{\mathrm{WD} }}\frac{1}{\epsilon^*}  \simeq   \frac{2+q\gamma }{3}\left |    \frac{\Delta M_{\text{WD}}/M_{\text{WD}}}{\epsilon^*} \right | >0.
\end{aligned}
\label{47 CLdRWD - MWDdMWD1} 
\end{equation}
This equation accurately reproduces the trend in Figure \ref{Variation of epsilon in PN with MT}. 
 Since the MT rate remains largely insensitive to the IMBH's spin, the PN-induced periodic oscillations converge tightly around a single curve in the logarithmic plot. 
 
Through numerical simulations, we quantify the per-orbit variations in both pericenter distance $r_{\text{p}}$ and overflow parameter $\epsilon$. For each orbit number N, we define these variations as:
$\delta r_{\text{p}} = r_{\text{p}}(\mathrm{N}) - r_{\text{p}}(\mathrm{N}-1)$ for the pericenter distance, $\delta \epsilon = \epsilon(\mathrm{N}) - \epsilon(\mathrm{N}-1)$ for the outflow parameter.  
After averaging over multiple orbital cycles, we find distinct evolutionary trends depending on the physical processes included:
First, GW emission alone, the per-orbit variations  are $\delta \bar{r}_{\text{p}}^{\text{GW}}/\bar{r}_{\mathrm{p} }$\,$\approx$\,$ -3.3\times 10^{-8}$ and $\delta \bar{\epsilon}^{\text{GW}}/\bar{\epsilon} $\,$\approx $\,$5.5\times 10^{-6}$.
Second,  combined GW emission and MT process, the per-orbit variations  are $\delta \bar{r}_{\text{p}}^{\text{GW+MT}}/\bar{r}_{\mathrm{p} }$\,$\approx $\,$-1.7\times 10^{-8}$ and $\delta \bar{\epsilon}^{\text{GW+MT}}/\bar{\epsilon}$\,$\approx$\,$ 4.6\times 10^{-4}$.

We draw several key conclusions from our analysis. 
First, the orbital decay driven by GW emission is slowed by the MT process. This is evident from the fact that the pericenter shift,
$\delta \bar{r}_{\mathrm{p}}^{\mathrm{GW+MT}}$ is less negative than the pure GW-driven shift, $\delta \bar{r}_{\mathrm{p}}^{\mathrm{GW}}$. 
Second, the outflow evolution is strongly enhanced when MT is active. The change in outflow parameter, $\delta \bar{\epsilon}^{\text{GW+MT}}$, far exceeds that from GWs alone, $ \delta \bar{\epsilon}^{\text{GW}}$. This means the mass stripping process rapidly widens the gap between the WD's radius and its Roche lobe.
The underlying physics can be understood through two competing effects: First, as the WD loses mass, its radius $R_{\text{WD}}$ increases following the mass-radius relation (Equation \ref{eq:RWD}).
Second, the Roche radius $R_{\text{L}}$ scales with mass ratio as $q^{1/3} $ (Equation \ref{coefficient f A,q}). Since MT reduces $q$, the Roche radius $R_{\text{L}}$ shrinks slightly.
The final result is that $R_{\text{WD}}$ expands much faster than $R_{\text{L}}$ contracts, creating a positive feedback loop that dramatically boosts the MT rate. This explains why the relative mass loss rate $\left |  \delta M_{\text{WD}}/M_{\text{WD}}\right | $ dominates over the pericenter variation $ \left |  \delta r_{\text{p}}/r_{\text{p}}\right | $—a hallmark of WD donors that distinguishes them from compact objects whose radius shrink upon mass loss. 
Therefore, for WDs, the orbital dynamics of MT cannot be ignored.

\section{Long-Term evolution with Mass Transfer}\label{sec:Analyzing orbit with MT}
Applying the PN method for calculations is computationally expensive, especially for  long-duration simulations.  
Consequently, we employ a Keplerian orbital approximation and time-average approach to efficiently evaluate the long-term evolution.

First, we analyze the combined effects of the MT process and gravitational radiation on the long-term orbital evolution, as described in Section \ref{sec:Comparison of the effects of MT and gravitational radiation on orbit}.   
To calculate the transferred mass $\delta M_{\text{WD}}$  from the WD over one orbital period, we use the MT rate $\dot{m}_1$,  as detailed in Section \ref{sec:Simplification of mass transfer model} and Appendices \ref{App: Simplification of the mass transfer model} and \ref{App: Calculate transferred mass every period}.
Second, we orbital-average the parameters and analyze the long-term MT effects on a Keplerian elliptical orbit in Section \ref{sec:The long-term orbital effects of MT are calculated using discrete method for classic elliptical orbits}.  
Third, we verify the accuracy of our long-term calculations through short-term numerical calculations in Section \ref{sec:Verification of long-term results}.

\subsection{Relevance of the mass transfer effects}\label{sec:Estimation of mass transfer effects}
\subsubsection{The relative importance of MT and gravitational radiation}\label{sec:Comparison of the effects of MT and gravitational radiation on orbit} 

In this section, we use the second method in Section \ref{sec:Dynamics under mass transfer} to calculate the orbital evolution with MT. 
According to Equation (\ref{39 Conservation of angular momentum for S}), the MT acceleration $\boldsymbol{f}_{\text{MT}}^{\mathcal{B}}$ is 
\begin{align}\label{f_MT change_2 form}
\boldsymbol{f}_{\text{MT}}^{\mathcal{B}}=&\left (-\frac{\ \dot{m}_{1}}{\ m_{1}}  \right ) r \dot{f} _\mathrm{N} (1-q)\tilde{h}\boldsymbol{\lambda }.
\end{align}
We can give the instantaneous change   $\dot{a}$ using $\dot{p}$ and $\dot{e}$ in Equation (\ref{eq:elements differential equation11}). 
In this case, using the periodic integral of Equation (\ref{f_MT change_2 form}), we provide the secular change rate of the orbital semimajor axis $\left \langle   \dot{a}^{\mathcal{B}}_{\mathrm{MT} }\right \rangle_{\text{sec}}$,  which can be expressed as
\begin{equation}
\begin{aligned}
\left \langle   \dot{a}^{\mathcal{B}}_{\mathrm{MT} }\right \rangle_{\text{sec}} \!  =  \left ( -\frac{\delta M_{\mathrm{WD}} }{ M_{\mathrm{WD} }}  \right )\sqrt{\frac{G M_{\mathrm{WD} } }{c^2 a_{\text{r}} } }\frac{c}{\pi }\frac{1+e_{\text{r}}}{1-e_{\text{r}}} (1-q)\sqrt{\frac{1+q}{q} }  \tilde{h} ,
\end{aligned}
\label{48 daBMTdt1}
\end{equation}
where $\delta M_{\mathrm{WD}}$ represents the transferred mass of the WD during each orbital period. Here, $\left \langle \, \cdots  \,  \right \rangle_{\text{sec}}$ is the periodic average operator.

In Section \ref{sec:Dynamics under mass transfer}, we have assumed a MT rate $\dot{m}_1$ without providing a formula to calculate it. 
Based on the assumptions made by \citet{sepinsky2007interacting}, we propose that the MT rate $\dot{m}_1$ at high eccentricity $ e_{\text{r}}\lesssim 1  $, can be expressed as $\dot{M}_{\mathrm{WD} }$\,$:=$\,$\dot{M}_{0} \delta(f) $, where $\dot{M}_{0}$ is a quantity related to the orbit to be determined. 
Therefore, the mass stripped from WD in one cycle is:
\begin{align}\label{42 dMWDdt }
\delta M_{\mathrm{WD}}= \! \int_{-\pi}^{\pi}  \dot{M}_{0} \, \delta(f)  \dot{f}_{\text{N} }^{-1} df \simeq \dot{M}_{0}\sqrt{\frac{a_{\text{r}}^3}{G m}  } \frac{(1-e_{\text{r}})^{3/2}}{(1+e_{\text{r}})^{1/2}}.
\end{align}
Conversly, quantity $\dot{M}_{0} $  can be expressed in terms of $\delta M_{\mathrm{WD}} $:
\begin{align}\label{423 dMWDdt }
\dot{M}_{0}  \simeq \frac{\delta M_{\mathrm{WD}}}{T}  2 \pi \frac{(1+e_{\text{r}})^{1/2}}{(1-e_{\text{r}})^{3/2}}.
\end{align}
In addition, the average change in the semi-major axis due to  quadrupole gravitational radiation  $\left \langle \dot{a}_{\mathrm{GW}} \right \rangle_{\text{sec}}$  is given by  \citet{peters1964gravitational}:
\begin{equation}
\begin{aligned}
\left \langle \dot{a}_{\mathrm{GW}} \right \rangle_{\text{sec}} = - \left (\frac{G M_{\mathrm{WD}}}{c^2 a_{\text{r}}}   \right )^3 \frac{64}{5}c\frac{1+q}{q^2} \frac{ 1+\frac{73}{24} e_{\text{r}}^{2}+\frac{37}{96} e_{\text{r}}^{4}}{(1-e_{\text{r}}^2  )^{7/2} }.
\end{aligned}
\label{Comparing the effects of MT and gravitational radiation on secular change of orbital semimajor axis aGW}
\end{equation}

Because $\delta M_{\mathrm{WD}}$\,$<$\,$0$, MT causes the orbit to expand outward, we have $\left \langle \dot{a}_{\mathrm{GW}} \right \rangle_{\text{sec}}$\,$<0<$\,$\left \langle   \dot{a}^{\mathcal{B}}_{\mathrm{MT} }\right \rangle_{\text{sec}} $.
To identify which is the dominant process that determines the evolution of $a$, we calculate 
 the change ratio $\left |\dot{a}^{\mathcal{B}}_{\mathrm{MT} } / \dot{a}_{\mathrm{GW}}   \right |_{\text{sec}} $, which is given by
\begin{equation}
\begin{aligned}
& \left |\frac{ \dot{a}^{\mathcal{B}}_{\mathrm{MT} } }{ \dot{a}_{\mathrm{GW}} }   \right | _{\text{sec}} = \left ( -\frac{\delta M_{\mathrm{WD}} }{ M_{\mathrm{WD} }}  \right ) \sigma^{-1}, \\
&\sigma= \tilde{\lambda } ^{-5/2} q \frac{64}{5} \pi \frac{1+\frac{73}{24} e_{\text{r}}^{2}+\frac{37}{96} e_{\text{r}}^{4}}{(1+e_{\text{r}})^{9/2} }\frac{\sqrt{1+q} }{(1-q)\tilde{h} } ,
\end{aligned}
\label{44 aMT divided aGW}
\end{equation}
where $\tilde{\lambda }$\,$:=$\,$r_{\mathrm{p}}/(G m_{2}/c^2)$\,$\approx$\,$7.875$, $q$\,$=$\,$3.75\times 10^{-7}$, $e_{\text{r}}$\,$=$\,$0.99$ and $\tilde{h}$\,$=$\,$1$ under the orbital conditions specified in Section \ref{sec:Analyzing orbital results without MT in PN method}.  Following this condition, we find $\sigma$\,$=$\,$1.7 \times 10^{-8}$.  

For gravitational radiation to dominate, the system should  satisfy $\left |\dot{a}^{\mathcal{B}}_{\mathrm{MT} } / \dot{a}_{\mathrm{GW}}  \right |_{\text{sec}}$\,$<$\,$1$, which implies $ \left |\delta M_{\mathrm{WD}}   \right |$\,$<$\,$1.7\times 10^{-8}M_{\mathrm{WD} } $ per period. 
Under the above orbital conditions with $M_{\text{WD}}$\,$=$\,$0.15 $\,$\text{M}_{\sun}$ and $T$\,$\sim$\,$76$\,hr, this leads to $ \left | \dot{M} _{\mathrm{WD}} \right |_{\text{sec}}$\,$<$\,$2.94\times 10^{-7} \text{M}_{\sun} \, \text{yr}^{-1}$. 
This represents an exceptionally low MT rate.  
For $\left \langle \dot{a}^{\mathcal{B}}_{\mathrm{MT} }\right \rangle_{\text{sec}}$\,$=$\,$-\left \langle \dot{a}_{\mathrm{GW} }\right \rangle_{\text{sec}}$, we can get $\left | \delta M_{\mathrm{WD}} \right |$\,$\propto$\,$ - \left \langle \dot{a}_{\mathrm{GW} }\right \rangle_{\text{sec}} (1-e_{\text{r}})$ using Equation (\ref{48 daBMTdt1}).
The factor $(1-e_{\text{r}})$   is a small amount arises from the integral of the MT rate in Equation (\ref{42 dMWDdt }). 
Therefore, the low transferred mass $ \left |\delta M_{\mathrm{WD}} \right |$ is due to the high eccentricity.

Our above conclusion differs from that of \citet{wang2022model}.  
Their $\dot{M}_{\mathrm{WD} }$ corresponds to our $\dot{M}_0$ in Equation (\ref{423 dMWDdt }); 
however, $\dot{M}_0$\,$<$\,$0$ does not represent an instantaneous MT rate. 
They neglected   the factor $(1-e_{\text{r}})^{-3/2}$\,$\gg $\,$1$ for $\dot{M}_0$.  
As a result, they suggest that  $\left|\delta M_{\mathrm{WD}}\right|$\,$>$\,$10^{-3} M_{\mathrm{WD}}$ is required for the MT process to dominate, i.e. $\left |\dot{a}^{\mathcal{B}}_{\mathrm{MT} } / \dot{a}_{\mathrm{GW}}  \right |_{\text{sec}}$\,$>$\,$1$, based on their parameters.

\subsubsection{Simplification of MT model}\label{sec:Simplification of mass transfer model}
To investigate  the orbital evolution induced by $\boldsymbol{f}_{\text{MT}}^{\mathcal{B}}$, it is essential to estimate the $\delta M_{\mathrm{WD}}$.  
While direct numerical integration of the MT rate (Equations \ref{eq:58 dotm1} and \ref{eq:59 dotm1}) is possible, this approach obscures the underlying dependence of $\delta M_{\mathrm{WD}}$ on key system parameters: the pericenter distance $r_{\mathrm{p}}$, orbital eccentricity  $e$, and the masses of the WD ($M_{\mathrm{WD}}$) and IMBH ($M_{\mathrm{IMBH}}$). We therefore develop a simplified analytic treatment.

The MT process depends critically on the relative size of the WD compared to its Roche lobe. We characterize this through the dimensionless parameter $\varepsilon$\,$=$\,$\left ( R_{\text{WD}}-R_{\text{L}} \right )/R_{\text{WD}}$. Note that $\varepsilon$  varies throughout the eccentric orbit as the binary separation changes. This is distinct from the parameter $\epsilon$ introduced earlier, where $\epsilon$\,$=$\,$\varepsilon(D$\,$=$\,$r_{\text{p}})$\,$\ll$\,$1$.

For our $\delta M_{\mathrm{WD}}$ calculation, we focus exclusively on the adiabatic MT component (Equation \ref{eq:59 dotm1}), neglecting the isothermal contribution (Equation \ref{eq:58 dotm1}). 
This simplification is justified because:
Isothermal transfer only becomes significant far from pericenter. 
Our delta-function approximation for the MT rate makes dynamical effects negligible away from pericenter.

To enable integration over the orbital phase and determine the total mass stripped per orbital period, we first develop a simplified expression for the instantaneous MT rate $\dot{m}_1$. This takes the form:
 \begin{align}\label{48 Simplification of mass transfer model 1} 
\dot{m}_1\simeq -&H(\varepsilon)\cdot 2  \pi \cdot W(K,\Gamma) \cdot \left ( G M_{\mathrm{IMBH} } \right )^2   \frac{C_\text{L}^3}{\sqrt{\mathcal{R}_0 } } \times \nonumber \\
&\left ( \sum_{m=0}^{\infty } \mathcal{S}_m \varepsilon ^m    \right ) \left ( \sum_{n=0}^{\infty } \mathcal{U}_{\varepsilon n}\varepsilon ^n     \right ), 
\end{align}
where $W(K,\Gamma)$\,$=$\,$F(K,\Gamma)$\,$\cdot$\,$\left [  K^{-\frac{1}{\Gamma } }(\Gamma-1)/\Gamma \right ]^{\frac{3\Gamma-1}{2(\Gamma-1)}} $.  Detailed definitions of the coefficients  $\mathcal{R}_0 $, $\mathcal{S}_m$,  $ \mathcal{U}_{\varepsilon n}$  provided in Appendix \ref{App: Simplification of the mass transfer model}.

For a Keplerian elliptical orbit, the integral of  $\dot{m}_1$ over a full period can be rearranged as  
\begin{equation} 
\begin{aligned}
&\delta M_{\mathrm{WD}} \simeq \! \int_{-\pi}^{\pi}  \dot{m}_{1} \dot{f}_{\text{N} }^{-1}  df= C_{\dot{m}_1} \sqrt{\frac{p^3}{Gm} } \sum_{n=0}^{\infty } C_{\mathrm{\varepsilon n} }  \int_{-f_0}^{f_0}\frac{\varepsilon(f)^n df}{(1+e \cos f)^{2}}, 
\end{aligned}
\label{52 Calculate transferred mass every period 1} 
\end{equation}
where coefficients $C_{\dot{m}_1}$ and $C_{\mathrm{\varepsilon n} }$ are derived in Appendix \ref{App: Calculate transferred mass every period}. 
The critical phase $f_0$  satisfying $R_{\mathrm{L}}(f_0)$\,$=$\,$R_{\mathrm{WD}}$, i.e. $\varepsilon(f_0)$\,$=$\,$0$. 
The function $f_0(e,\epsilon)$ in Equation (\ref{App3: 9}) is due to $\varepsilon(f_0)$\,$=$\,$0$ and the variable  $\epsilon$ $=$ $\varepsilon(f$\,$=$\,$0)$.
The definite integral can be expressed in terms of the Appell hypergeometric function, $\mathrm{AF1}(\alpha ; \beta ,\beta ^{\prime}; \gamma ;x,y)$ \citep{DLMF}. 
After the integral processing,  the function $\delta M_{\mathrm{WD}} (\epsilon ,e, M_{\mathrm{WD}},q)$ is  
\begin{align}\label{59 Calculate transferred mass every period 8} 
\delta M_{\mathrm{WD}} \!  \simeq &\frac{4\ C_{\dot{m}_1} r_{\text{p}}^{3/2}}{\sqrt{G m } } \sqrt{\frac{\epsilon }{2e} }
\sum_{n=0}^{\infty }\sum_{m=0}^{\infty  }C_{\mathrm{\varepsilon n} }   \begin{pmatrix}
 n\\
m
\end{pmatrix}\left ( -1 \right ) ^m \left ( 1-\epsilon  \right ) ^m \times  \nonumber \\
&\quad \mathrm{AF1}\left [\frac{1}{2} ;\frac{1}{2} ,m+2;\frac{3}{2}; \frac{1+e}{2e}\epsilon,\epsilon \right ]. 
\end{align}
The  stripping mass is directly related to the outflow parameter $\epsilon$. This simplified yet physically complete model enables efficient computation while maintaining the essential physics of the MT process.

\subsection{Long-term orbital effects in Newtonian elliptical orbits}\label{sec:The long-term orbital effects of MT are calculated using discrete method for classic elliptical orbits}

To evaluate the long-term orbital evolution, we also require the secular change rate of orbital eccentricity. 
For MT, $ \left \langle   \dot{e}^{\mathcal{B}}_{\mathrm{MT} }\right \rangle_{\text{sec}}$ is obtained using Equations (\ref{eq:elements differential equation11 b}) and (\ref{f_MT change_2 form}). 
For gravitational radiation, $\left \langle   \dot{e}_{\mathrm{GW} }\right \rangle_{\text{sec}}$ is taken from  \citet{peters1964gravitational}. They are
\begin{align}\label{5.2 long-term orbital effects of MT discrete method 1} 
& \left \langle   \dot{e}^{\mathcal{B}}_{\mathrm{MT} }\right \rangle_{\text{sec}} =  \left ( -\frac{\delta M_{\mathrm{WD}} }{ M_{\mathrm{WD} }}  \right )\sqrt{\frac{G M_{\mathrm{WD} } }{c^2 a_{\text{r}} } }\frac{c(1+e_{\text{r}})}{\pi a_{\text{r}} } (1-q)\sqrt{\frac{1+q}{q} }\tilde{h} , \nonumber \\
&\left \langle   \dot{e}_{\mathrm{GW} }\right \rangle_{\text{sec}} = - \left (\frac{G M_{\mathrm{WD}}}{c^2 a_{\text{r}}}   \right )^3 \frac{304}{15}\frac{c}{a_{\text{r}}} \frac{1+q}{q^2} \frac{e_{\text{r}}\left ( 1+ \frac{121}{304} e_{\text{r}}^{2}\right )}{(1-e_{\text{r}}^2  )^{5/2} }.
\end{align}
For the pericenter distance, using $\left \langle \dot{r}_{\text{p}} \right \rangle_{\text{sec}}$\,$\simeq $\,$(1-e)\left \langle \dot{a} \right \rangle_{\text{sec}} $\,$-$\,$ a\left \langle \dot{e} \right \rangle_{\text{sec}}$: 
\begin{align} \label{5.2 long-term orbital effects of MT discrete method 1 drp} 
& \left \langle   \dot{r}^{\mathcal{B}}_{\text{pMT}}\right \rangle_{\text{sec}} = 0 , \\
&\left \langle   \dot{r}_{\mathrm{pGW} }\right \rangle_{\text{sec}} = - \left (\frac{G M_{\mathrm{WD}}}{c^2 a_{\text{r}}}   \right )^3 \frac{64}{5}c\frac{1+q}{q^2} \frac{ \left(1-\frac{7 }{12}e_{\text{r}}+\frac{7 }{8}e_{\text{r}}^{2}+\frac{47 }{192}e_{\text{r}}^{3}\right) }{(1-e_{\text{r}}  )^{3/2}(1+e_{\text{r}}  )^{7/2} }. \nonumber 
\end{align}  
The zero pericenter shift from MT arises because our delta-function model concentrates all mass loss at pericenter.

Assuming $\gamma$\,$=$\,$1$, the orbital period is approximately $\left \langle \dot{T}\right \rangle_{\text{sec}}$\,$\simeq$\,$3\pi \sqrt{a/Gm} \left \langle \dot{a} \right \rangle_{\text{sec}}$\,$=$\,$3 T \left \langle \dot{a} \right \rangle_{\text{sec}}/2a$. Hence, we have:
\begin{align}\label{5.2 long-term orbital effects of MT discrete method 0}
& \left \langle   \dot{T}^{\mathcal{B}}_{\text{MT}}\right \rangle_{\text{sec}} = \left ( -\frac{\delta M_{\mathrm{WD}} }{ M_{\mathrm{WD} }}  \right )3 \frac{1+e_{\text{r}}}{1-e_{\text{r}}} (1-q ),\\
&\left \langle   \dot{T}_{\mathrm{GW} }\right \rangle_{\text{sec}} =  - \left (\frac{G M_{\mathrm{WD}}}{c^2 a_{\text{r}}}   \right )^{5/2 } \frac{64}{5}3 \pi  \sqrt{\frac{1+q}{q^3}}  \frac{ 1+\frac{73}{24} e_{\text{r}}^{2}+\frac{37}{96} e_{\text{r}}^{4}}{(1-e_{\text{r}}^2  )^{7/2} }. \nonumber
\end{align}

Now, we can make qualitative predictions about the long-term evolution: 
\begin{enumerate}
\item  If the pericenter distance $r_{\text{p}}$\,$\ge$\,$R_{\text{WD}}/C_{\text{L}}$ (i.e. $\epsilon$\,$\le$\,$0$), the Roche lobe is underfilled, and the MT rate is negligible. 
In this case, we can assume that orbital decay is driven only by gravitational radiation, i.e. $\delta M_{\mathrm{WD}}$\,$=$\,$0$ and $\dot{a}^{\mathcal{B}}{}_{\mathrm{MT} } $\,$=$\,$0$. Consequently, $a$, $e$, $T$ and $r_{\text{p}}$ all decrease.
\item  
When $r_{\text{p}}$\,$<$\,$R_{\text{WD}}/C_{\text{L}}$ (i.e. $\epsilon $\,$ >$\,$ 0$), and $0$\,$<$\,$ \left |\dot{a}^{\mathcal{B}}{}_{\mathrm{MT} } / \dot{a}_{\mathrm{GW}}  \right |_{\text{sec}}$\,$<$\,$1$, gravitational radiation still dominantes.  The Roche lobe overflow.
In this case, $a$, $e$, $T$ and $r_{\text{p}}$ continue to decrease. 
However, $\epsilon$ gradually increases due to the radius-mass relationship of the WD (see Section \ref{sec:Orbital evolution with MT process} for details). 
Thus, according to Equation (\ref{59 Calculate transferred mass every period 8}),  $\left |   \delta M_{\mathrm{WD}}\right |$ and $\left |\dot{a}^{\mathcal{B}}{}_{\mathrm{MT} } / \dot{a}_{\mathrm{GW}}  \right |_{\text{sec}} $ gradually increases. 
\item  
As   $\left |\dot{a}^{\mathcal{B}}{}_{\mathrm{MT} } / \dot{a}_{\mathrm{GW}}  \right |_{\text{sec}}$ $\rightarrow$ $1$, both $a$ and $T$ reach their minimum values simultaneously.    
Similarly, as $\left |\dot{e}^{\mathcal{B}}{}_{\mathrm{MT} } / \dot{e}_{\mathrm{GW}}  \right |_{\text{sec}} \rightarrow 1$, the  eccentricity $e$ reaches its minimum value.  
The times at which $a$, $T$ reach their minima differ slightly from the time when $e$ does, as the change rates due to gravitational radiation follow different formulae.
\item  
Beyond the threshold $\left |\dot{a}^{\mathcal{B}}{}_{\mathrm{MT} } / \dot{a}_{\mathrm{GW}}  \right |_{\text{sec}}=1$,  the parameters $a$, $T$, and $e$ will increase according to the equations in Sections \ref{sec:Comparison of the effects of MT and gravitational radiation on orbit} and \ref{sec:The long-term orbital effects of MT are calculated using discrete method for classic elliptical orbits}. 
At this point, MT begins to dominate.  
However, $\delta M_{\mathrm{WD}}$ continues to rise as described by Equation (\ref{59 Calculate transferred mass every period 8}), and $r_{\text{p}}$ continues to decrease as given by Equation (\ref{5.2 long-term orbital effects of MT discrete method 1 drp}). 
\end{enumerate} 

In the absence of other mechanisms, the MT process remains dominant  without reverting to gravitational radiation dominance.  
Since  $\left \langle \dot{r}_{\text{p}} \right \rangle_{\text{sec}}$\,$=$\,$\left \langle \dot{r}^{\mathcal{B}}_{\text{pMT}} \right \rangle_{\text{sec}}$\,$+$\,$\left \langle \dot{r}_{\mathrm{pGW} } \right \rangle_{\text{sec}} $\,$\le$\,$ 0$  always holds according to Equation (\ref{5.2 long-term orbital effects of MT discrete method 1 drp}), we infer that the MT process of a WD in a bound, high-eccentricity orbit around an IMBH is inherently irreversible.

\begin{figure} 
\includegraphics[width=1.00\linewidth]{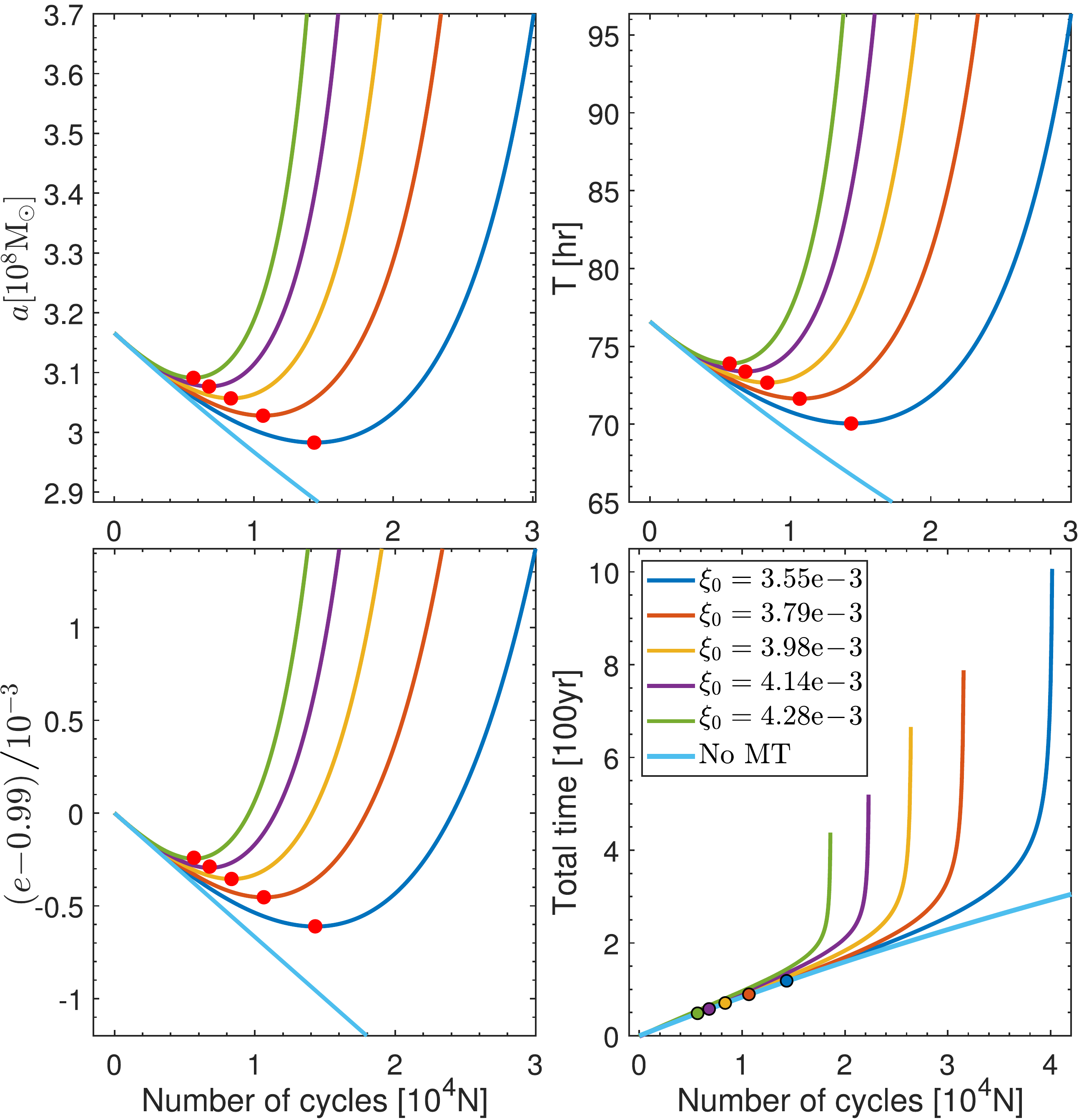}
\begin{minipage}{\linewidth}
\caption{The synchronous  evolution of $a$, $T$, $e$,  and total time respectively.  
Red marks  indicate the points where $a$, $T$, $e$ reach their minimum values, and colored marks  indicate  total time reach its inflection value. 
The sky-blue curves (without markers) show evolution considering   GW emission, they  exhibit monotonic behavior without any minima or inflection points.
}
\label{5.2 Variation of the semi-major axis, eccentricity, WD period}
\end{minipage} 
\end{figure} 

\begin{figure} 
\includegraphics[width=1\linewidth]{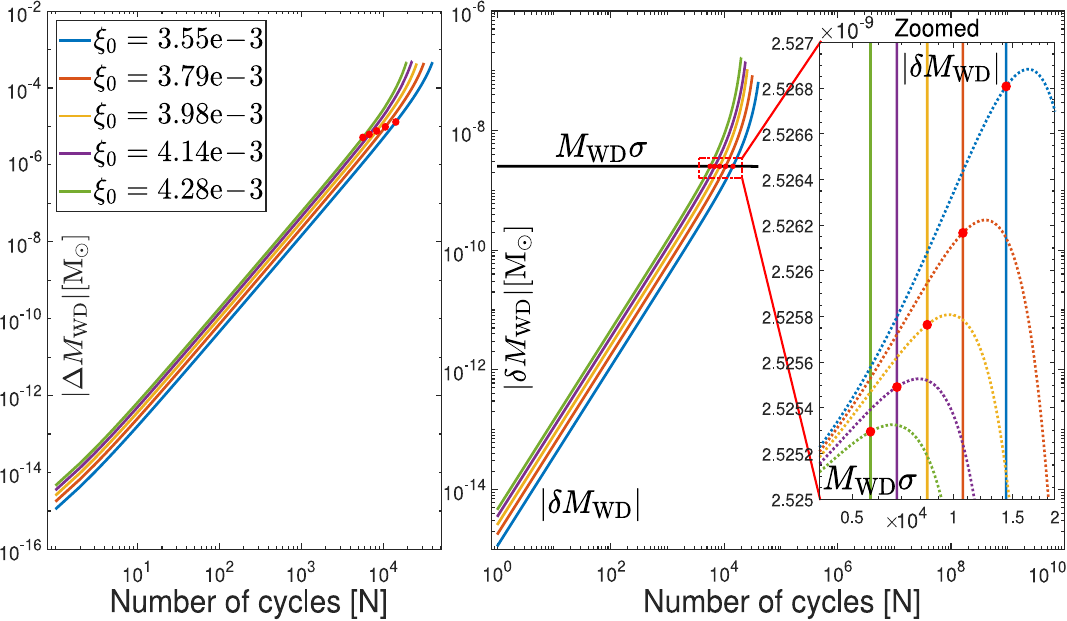}
\begin{minipage}{\linewidth}
\caption{The evolution of  $M_{\text{WD}}$  over time due to the MT process. 
Left panel: Cumulative mass loss  $\left |  \Delta M_{\mathrm{WD} } \right |$ showing the total mass transferred from the WD between the initial state (cycle 0) and cycle N.  Calculated as $ \left | M_{\mathrm{WD} } (\mathrm{N})-M_{\mathrm{WD} }(\mathrm{0}) \right | $.
Right panel: Per-cycle  transferred mass $\left |  \delta M_{\mathrm{WD} } \right |$   showing the mass lost by the WD in each individual cycle.  Calculated as  $\left |  M_{\mathrm{WD} } (\mathrm{N})-M_{\mathrm{WD} }(\mathrm{N}-1) \right |$.
The black curves represent different $\xi_0$ parameter values in the $M_{\mathrm{WD} }\sigma$ function. These curves are nearly indistinguishable at this scale due to their close clustering. 
The zoomed inset provides a magnified view comparing $\left |  \delta M_{\mathrm{WD} } \right |$ (solid line) and $M_{\mathrm{WD} }\sigma $ (dashed line). 
Red markers indicate critical points where both orbital separation $a$ and orbital period  $T$ reach their minimum values. These occur at the intersections of the $\left |\delta M_{\mathrm{WD} }\right |$ and $M_{\mathrm{WD} }\sigma $ curves.
}
\label{5.2 Variation of WD mass}
\end{minipage}  
\end{figure} 

Therefore, the long-term evolution for $\boldsymbol{f}_{\text{MT}}^{\mathcal{B}}$ may lead to two  possible outcomes:  
\begin{enumerate}
\item 
Tidal Disruption Scenario : 
The WD will be tidally disrupted if its pericenter distance $r_{\mathrm{p} }$ shrinks below the critical tidal radius $ r_{\mathrm{t} }$. Following established theory \citep{Rees1988}, this occurs when $r_{\mathrm{p} } $\,$<$\,$ r_{\mathrm{t} }$\,$\simeq$\,$R_{\mathrm{WD}}\left(M_{\mathrm{IMBH}} / M_{\mathrm{WD}}\right)^{1 / 3}$.
In our framework, this condition translates to a critical overflow disruption threshold $\epsilon_{\,\mathrm{t} }(q)$. For systems with extreme mass ratios ($q$\,$\ll $\,$1$), this threshold approaches $\epsilon_{\,\mathrm{t} }$\,$=$\,$1-C_{\mathrm{L}} q^{-1/3}$\,$\sim$\,$ 0.465$.  
\item  
Ejection Scenario : 
The WD may escape the system entirely if its orbital eccentricity reaches or exceeds unity ($e \ge 1$) before reaching the disruption threshold ($\epsilon < \epsilon_{\,\mathrm{t} }$).
\end{enumerate}  

The actual outcome depends sensitively on the initial conditions:
First, systems starting near the critical $\epsilon_{\,\mathrm{t} }$ boundary (initial $\epsilon_0$ is slightly smaller than $\epsilon_{\,\mathrm{t} }$) overwhelmingly favor tidal disruption. WD will be tidally disrupted before the eccentricity reaches 1.
Second, systems beginning with much lower initial overflow parameter $\epsilon_0 \ll \epsilon_{\,\mathrm{t} }$ typically lead to ejection.   
We also note that for $\boldsymbol{f}_{\text{MT}}^{\mathcal{A}}=\boldsymbol{0}$, the case follows the description in \citet{zalamea2010white}, tidal disruption is inevitable.

For numerical simulation, we consider spinless WD and IMBH with initial conditions: $e_0$\,$=$\,$0.99$, $\epsilon_0$\,$=$\,$0$, $m_{10}$\,$=$\,$0.15\, \text{M}_{\sun}$, $m_{20}$\,$=$\,$4\times 10^{5} \text{M}_{\sun}$.  
The dimensionless surface equivalent depth $\xi_0$ in Equations  (\ref{65 fMT}) and (\ref{eq:24 rhoph}) is used to adjust $\delta M_{\mathrm{WD}}$, with values $\xi_0\times10^{3}$\,$=$\,$3.55$, $3.79$, $3.98$, $4.14$, $4.28$. 

In Figure \ref{5.2 Variation of the semi-major axis, eccentricity, WD period},  we show  the evolution of $a$, $T$, $e$, and the total elapsed time, respectively. 
Initially, with $\epsilon=0$, gravitational radiation dominates, causing $a$, $T$, $e$ to decrease. 
After reaching the minimum, the three parameters begin to increase as the MT effect becomes dominant. 
The lower the initial value of $\xi_0$, the longer time it takes to reach the minimum, and the lower the minimum values.
The  evolution of the total elapsed time shows that, in the absence of the MT process, the slope decreases with increasing orbital cycles,  corresponding to the decline in $T$ due to gravitational radiation.  
After the inflection point, where the MT process becomes dominant, the slope increases, corresponding to the rise in $T$ due to the MT effect.  
The red point on each curve marks the transition from GW-dominated evolution to an MT-dominated one. 
Figure \ref{5.2 Variation of the semi-major axis, eccentricity, WD period} is consistent with the evolution we predicted earlier in this section.

In Figure \ref{5.2 Variation of WD mass}, we show the evolution of $M_{\mathrm{WD}}$.  
Smaller $\xi_0$ requires more cycles to achieve the same mass, because it results in a lower MT rate   $\left | \dot{m}_1 \right | $.  
Although $\xi_0$ differs for each line, when $\left |\dot{a}^{\mathcal{B}}_{\mathrm{MT} } / \dot{a}_{\mathrm{GW}}  \right |_{\text{sec}} $\,$\rightarrow  $\,$1$, $\delta M_{\mathrm{WD} }$ remains almost unchanged (i.e., the values of red points in the right panel are nearly the same).
This is because the size of  $\sigma$ in Equation (\ref{44 aMT divided aGW}) is only related to $q$, $e_{\text{r}}$, and $r_{\text{p}}$.  
In the right panel, $M_{\mathrm{WD} } \sigma (\mathrm{N})$ shows little change (The black lines show curves for different values of $\xi_0$ of the function $M_{\mathrm{WD} } \sigma$ which are clustered together, with $\sigma$\,$\approx$\,$1.7\times 10^{-8}$).  
This function is closely tied to the initial conditions, and since all our evolutions use the same initial orbital parameters, the function’s initial value is the same for each dashed line. 



\begin{figure}  
\includegraphics[width=1\linewidth]{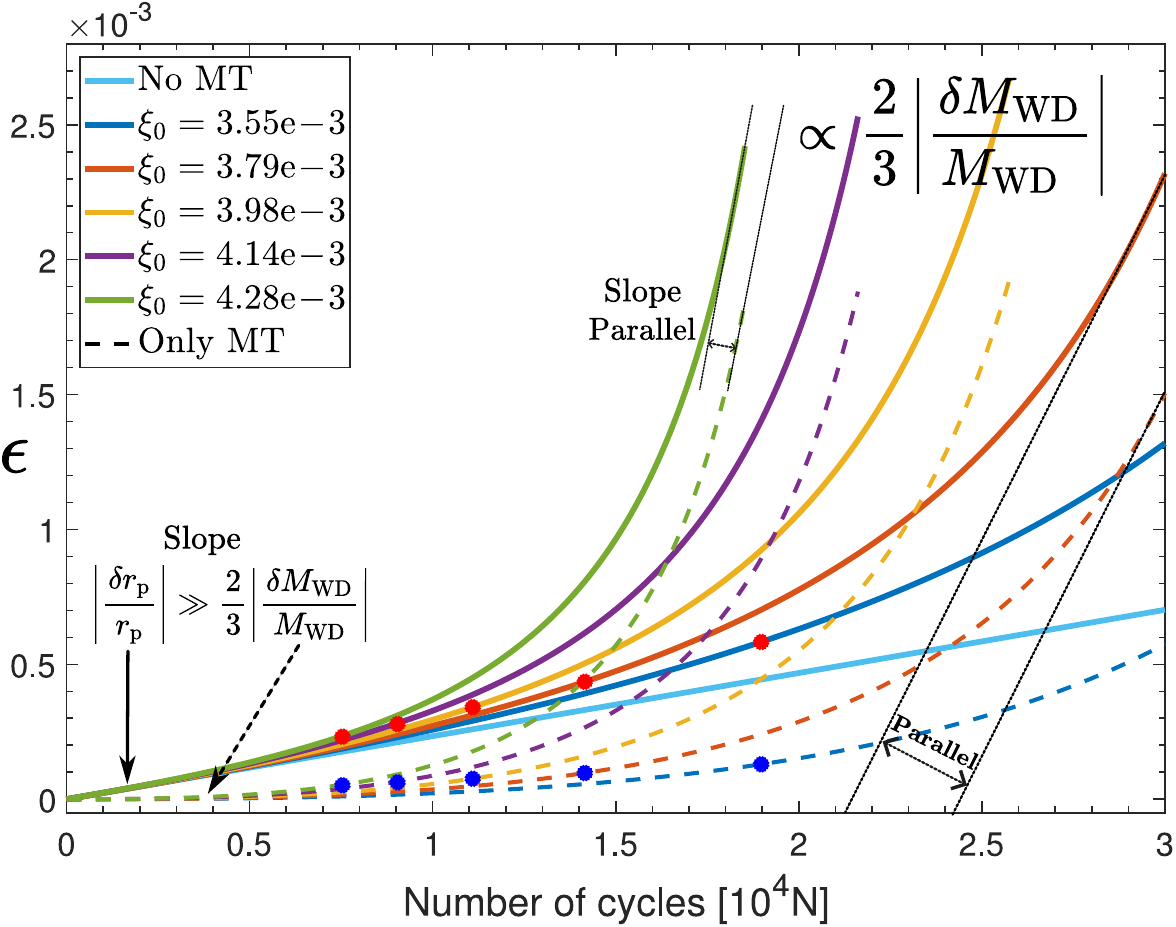}
\begin{minipage}{\linewidth}
\caption{The evolution of  $\epsilon$\,$=$\,$1$\,$-$\,$r_{\mathrm{p}} C_{\mathrm{L} }/R_{\mathrm{WD} } $ on long timescale.  
Sky blue solid line is the evolution of $\epsilon$ without MT. 
Solid lines in other colors are the cases where both gravitational radiation and the MT process are considered (Equation \ref{5.2 long-term orbital effects of epsilon}). 
Dashed lines are the function $1$\,$-$\,$(M_{\mathrm{WD} }(\mathrm{N}) / M_{\mathrm{WD} }(\mathrm{0}))^{(2+\gamma q(\mathrm{N}))/3}$.   
Red and blue  marks indicate the points where semimajor axis $a$  reach the minimum.}
\label{5.2 ratio of WD epsilon1}
\end{minipage}  
\end{figure}

In Figure \ref{5.2 ratio of WD epsilon1}, we show the evolution of overflow parameter $\epsilon(\mathrm{N})$. The presence or absence of MT significantly affects $\epsilon$. 
From Equations (\ref{45 epsilon - rp}) and (\ref{47 CLdRWD - MWDdMWD}), the slope $d\epsilon /d\mathrm{N} $ is about $ \epsilon (\mathrm{N})$\,$-$\,$\epsilon (\mathrm{N}$$-$$1)$ equal to $\left |  \delta r_{\mathrm{p} }/ r_{\mathrm{p} }\right |$\,$+$\,$2/3 \left | \delta M_{\mathrm{WD} }/ M_{\mathrm{WD} }  \right |$.  
The integral for $\epsilon(\mathrm{N})$ with $\epsilon(\mathrm{N}=\mathrm{0})=0$  is 
\begin{equation}
\begin{aligned}
\epsilon(\mathrm{N}) &\simeq 1-\frac{\, r_{\mathrm{p} }(\mathrm{N}) }{\, r_{\mathrm{p} }(\mathrm{0}) } \left ( \frac{M_{\mathrm{WD} }(\mathrm{N}) }{M_{\mathrm{WD} }(\mathrm{0})  } \right )^{\frac{2+\gamma q(\mathrm{N})}{3} } \\
&\simeq -\frac{\Delta r_{\mathrm{p} }(\mathrm{N}) }{r_{\mathrm{p} }(\mathrm{N}) }-\frac{2+\gamma  q(\mathrm{N})}{3}  \frac{\Delta M_{\mathrm{WD} }(\mathrm{N})}{M_{\mathrm{WD} }(\mathrm{N}) }.
\end{aligned}
\label{5.2 long-term orbital effects of epsilon}
\end{equation}
where cumulative quantity $\Delta r_{\mathrm{p} }(\mathrm{N})$ is $ r_{\mathrm{p} }(\mathrm{N})$\,$-$\,$r_{\mathrm{p} }(0) $,  and $\Delta M_{\mathrm{WD} }(\mathrm{N})$ is $  M_{\mathrm{WD} } (\mathrm{N})-M_{\mathrm{WD} }(0)$.

Because the initial orbital parameters are the same,  the initial slopes of solid lines in Figure  \ref{5.2 ratio of WD epsilon1} are similar.  
For comparison, we also plot the function obtained by excluding the change in $r_{\mathrm{p} }$  (dashed line).  This situation corresponds to considering only MT, without considering the change of GW emission for $r_{\mathrm{p} }$. 
When the MT effect is  weak, $\epsilon$ is very small,  and gravitational radiation dominates, causing $\left |  \delta r_{\mathrm{p} }/ r_{\mathrm{p} }\right |$\,$\gg$\,$ \left |  \delta M_{\mathrm{WD} }/  M_{\mathrm{WD} }  \right |$. 
According to Equation (\ref{5.2 long-term orbital effects of MT discrete method 1 drp}), the change in $r_{\mathrm{p} }$ is entirely governed by GW emission.
Therefore, the initial slope of the dashed line is nearly zero, reflecting that the MT effect is very weak at the beginning. 
Correspondingly, when the MT effect dominates, $\left |  \delta r_{\mathrm{p} }/ r_{\mathrm{p} }\right |$\,$\ll$\,$\left |  \delta M_{\mathrm{WD} }/  M_{\mathrm{WD} }  \right |$, this relationship has also been shown in the numerical simulations in Section \ref{sec:Orbital evolution with MT process}. 
Therefore, the slopes for the solid and dashed lines start to converge to  $2/3 \left |  \delta M_{\mathrm{WD} }/  M_{\mathrm{WD} }  \right |$. 
The fundamental reason for this relationship is the radius-mass relationship of the WD shown in Equation (\ref{eq:RWD}).

\subsection{Verification of long-term results}\label{sec:Verification of long-term results}
In Sections \ref{sec:Estimation of mass transfer effects} and \ref{sec:The long-term orbital effects of MT are calculated using discrete method for classic elliptical orbits}, we computed the evolution of Newtonian elliptical orbits. 
In Section \ref{sec:Analyzing orbital results without MT in PN method}, we  include the effects of PN dynamics terms,  
specifically  the contributions of $ \mathcal{A}$, $\mathcal{B}$, and the accelerations $\boldsymbol{a}_{\mathrm{SO}}$, $\boldsymbol{a}_{\mathrm{SS}}$, $\boldsymbol{a}_{\mathrm{SOPN}}$ in Equation (\ref{eq:relative acceleration}). 
In long-term evolution,  we observe  minima in the evolution of orbital parameters (Figure \ref{5.2 Variation of the semi-major axis, eccentricity, WD period}), and the MT rate approaches consistency near these  minima (Figure \ref{5.2 Variation of WD mass}). 
In order to verify the correctness of the long-term results, it is necessary to obtain the same conclusions in short-term calculations.
Orbital parameters (such as $a$, $T$, and $e$) should also reach their minima under relativistic corrections.  
When MT and gravitational radiation are in equilibrium, the transferred mass per cycle, $\delta M_{\mathrm{WD} }$, should remain unaffected by  $\xi_0$, i.e. $\left | \delta M_{\mathrm{WD} } \right | $\,$= $\,$M_{\mathrm{WD} } \sigma $, when $\left |\dot{a}^{\mathcal{B}}{}_{\mathrm{MT} } / \dot{a}_{\mathrm{GW}}  \right |_{\text{sec}}$\,$=$\,$1$.
We aim to investigate the minimum of  $a$ during  short-term evolution using parameters from Section \ref{sec:The long-term orbital effects of MT are calculated using discrete method for classic elliptical orbits}, and analyze  $\delta M_{\mathrm{WD} }$ near $a_{\text{min}}$.
In these short-term calculations, we  do not assume a delta function for the MT rate $\dot{M}_{\mathrm{WD} }$ like  Equation (\ref{42 dMWDdt }) (we use Equations \ref{eq:58 dotm1} and \ref{eq:59 dotm1}), nor do we use the  time-averaged orbital parameter change rate per cycle  (we use Equation \ref{eq:elements differential equation11}). 
As these phenomena occur near the minimum, we focus on the evolution over 100 cycles to verify these conclusions. 
The initial orbital parameters for the short-term calculation are adjusted to correspond to those of the first few cycles before the minimum.

\begin{figure} 
\includegraphics[width=1\linewidth]{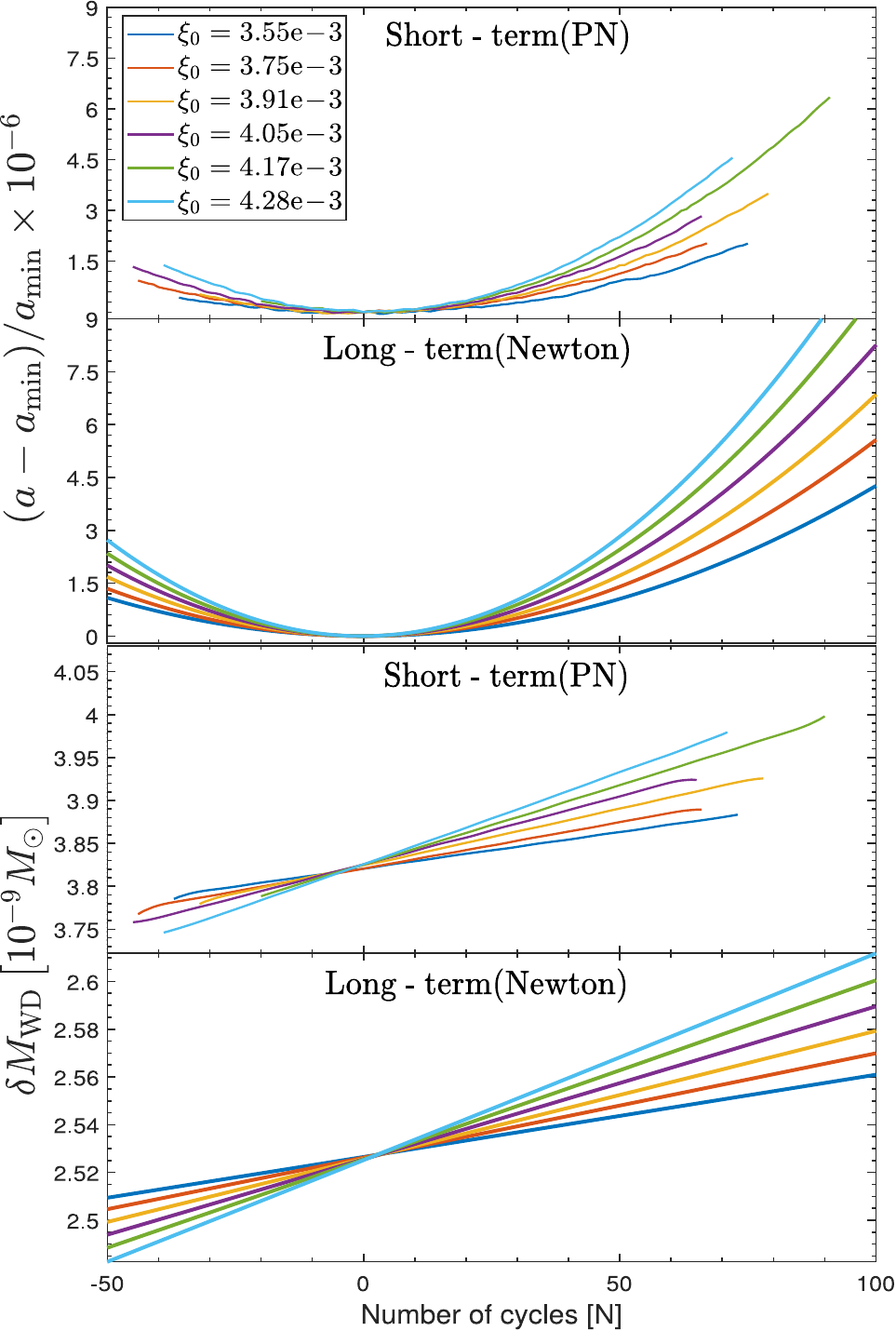}
\begin{minipage}{\linewidth}
\caption{Comparison of the short-term and long-term evolution of $(a-a_{\text{min}})/a_{\text{min}}$ and $\delta M_{\text{WD}}$.  
The short-term evolution  uses the PN results from  Section \ref{sec:Orbital evolution with MT process}. 
The long-term evolution uses the Newtonian elliptical orbit results  from Section \ref{sec:The long-term orbital effects of MT are calculated using discrete method for classic elliptical orbits}. 
The x-axis is set to $\text{N}(a_{\text{min}})=0$.}
\label{5.2 ratio of WD epsilon}
\end{minipage} 
\end{figure}

In Figure \ref{5.2 ratio of WD epsilon}, we compare the short-term and long-term evolution of $(a-a_{\text{min}})/a_{\text{min}}$ and $\delta M_{\text{WD}}$. 
Different surface densities parameter $\xi_0$ lead to varying minimum semi-major axis values $a_{\text{min}}$. 
We normalize the results  by the ratio $(a-a_{\text{min}})/a_{\text{min}}$. 
As $\xi_0$ increases, the rate of change of $a$ accelerates, indicating that larger  surface density of WD correspond to stronger MT effects.

We also show the variation of $\delta M_{\text{WD}}$ near $a_{\text{min}}$ in Figure \ref{5.2 ratio of WD epsilon}. 
According to Equation (\ref{44 aMT divided aGW}), the value of $\delta M_{\text{WD}}$ near $a_{\text{min}}$ is independent on $\xi_0$.  
Therefore, for the same initial values of $r_{\text{p0}}$,  $e_{\text{r0}}$ and $q_0$, the transferred mass  $\left | \delta M_{\mathrm{WD} } \right | $ near $a_{\text{min}}$ should converge into  $ M_{\mathrm{WD} } \sigma$. 
This convergence is well demonstrated in Figure \ref{5.2 ratio of WD epsilon} for both short-term and long-term evolution. 
The reasoning behind this behavior is explained in Figure \ref{5.2 Variation of WD mass} (the quantity $ M_{\mathrm{WD} } \sigma$ hardly changes) for long-term evolution, and it is confirmed from short-term evolution in Figure \ref{5.2 ratio of WD epsilon}.  

Therefore, we confirm that the MT rate can be represented as a delta function at high eccentricity, and the time-averaged form can be used
to calculate the evolution well. 
Furthermore, in the absence of effects other than MT and gravitational radiation, the evolution of orbital parameters inevitably exhibits extremal points.

\section{Impact on GW signal and detection}\label{sec:Impact on GW sigal and detection}  
As shown in Figure \ref{5.2 Variation of the semi-major axis, eccentricity, WD period}, after the same evolution time around the IMBH, the orbital period of the WD with MT and gravitational radiation is longer than that with only gravitational radiation. 
When transferrd mass $\left | \delta M_{\mathrm{WD} } \right | $ exceeds the limit value $ M_{\mathrm{WD} } \sigma$ in Equation (\ref{44 aMT divided aGW}), the period can even increase over time. 
Because of the change of period, the MT process induces a phase shift in the GW signal. 
Over a sufficiently long observational time, the accumulated phase difference due to MT can be detected.  
In the following, we evaluate whether the MT could induce an accumulative phase shift as large as 1 radian over a period of $t_{\text{obs}}=5 \ \text{yr}$. 
Such a phase shift would be detectable by LISA and other space-based GW detectors \citep{kocsis2011observable,10.1093/mnras/stz1026,10.1093/mnras/stac299,zhang2024gravitational1}.

Our orbit is highly eccentric, resulting in a nonlinear evolution of the orbital phase over time, which complicates the calculation of the phase shift. 
In the GW waveform, significant change of the amplitude occurs near pericenter over very short durations, appearing as periodic pulses in the time domain \citep{poisson2014gravity,mccart2021highly,hughes2021adiabatic}. 
Since the period $T$ changes very slowly due to the MT process, and the phase of each cycle always spans $2\pi$, it is unnecessary to compute the exact phase as a function of time during each period for our purposes. 
Instead, we can estimate the period of each cycle, which is sufficient to address the problem.  
We can demonstrate that (Appendix \ref{App: Equivalence of Phase and Periodic Evolution in the Adiabatic Approximation}), for the same number of cycles, the difference in total time is directly proportional to the difference in phase.

By the proof in Appendix \ref{App: Equivalence of Phase and Periodic Evolution in the Adiabatic Approximation}, under the adiabatic approximation, we can simplify the calculation of phase evolution when two key conditions are met: 1. The number of orbital cycles N is significantly large.  
2. The single orbital period $T_{\text{N}}$ is much shorter than the cumulative duration of all previous cycle.  
When these conditions hold, we can approximate the total phase change by replacing continuous time $t$ with the sum of all previous orbital periods.
Here, integer $\text{N}$ represents the number of cycles where the MT effect causes a phase lag of 1 radian.

The integer $\text{N}$ satisfies the following inequality:
\begin{equation}
\begin{aligned}
0\le    \frac{\left |  \Delta u_g\right |}{2\pi }\frac{1-\varsigma _{\text{N}}}{\varsigma _{\text{N}}}-\frac{\sum_{\text{k}=1}^{\text{N}-1}T_{\text{k}}\varsigma _{\text{k}}}{T_{\text{N}} \varsigma  _{\text{N}} } \le 1, 
\end{aligned}
\label{6 1 N equation}
\end{equation}
where $\Delta u_g$ represents the phase difference we want (In our goal, $\Delta u_g$\,$=$\,$-1$), $\varsigma _{\text{k}}$ represents the difference between the period with and  without MT in the k-th cycle, i.e. $\varsigma _{\text{k}}$\,$:=$\,$(T_{\text{k}}-T_{\text{k}}{}^*)/T_{\text{k}}$\,$>$\,$0$. 
Here, the superscript asterisk indicates that MT is not considered.
Inequality (\ref{6 1 N equation}) is equivalent to Inequality (\ref{App4: Delta u inequality equation a}).
For the integer $\text{N}$ is calculated from Inequality (\ref{6 1 N equation}), if the total time for the first $(\text{N}-1)$ cycles   is less than the observation time $t_{\text{obs}}$ (In our goal, $t_{\text{obs}}$\,$=$\,$5 \ \text{yr}$), the phase shift caused by MT is observable within $t_{\text{obs}}$.

 \begin{table}
  \centering
    \caption{The time required for the phase difference to reach 1 radian, for the different initial  $e_0$ and $\epsilon_0$, with same $\xi_{0}$\,$=$\,$3.55\times 10^{-3}$. The time unit is year.
}
 \label{tab: table of phase difference}  
\begin{tabular*}{0.9924\columnwidth}{|@{\hspace*{5.6pt}}c@{\hspace*{5.6pt}}|@{\hspace*{4pt}}c@{\hspace*{4.5pt}}c@{\hspace*{4.5pt}}c@{\hspace*{4.5pt}}c@{\hspace*{4.5pt}}c@{\hspace*{4.5pt}}c@{\hspace*{4.5pt}}c@{\hspace*{4.5pt}}c@{\hspace*{2.5pt}}|}
\hline
         \diagbox{$\epsilon_0$}{$e_0$}& 0.980  & 0.982  & 0.984  & 0.986  & 0.988 & 0.990 & 0.992 & 0.994  \\ \hline
                     0 & 7.188  & 8.170  & 9.427  & 11.087 & 13.367 & 16.674 & 21.846 & 30.934   \\  
        0.30\text{e-4} & 4.267  & 4.785  & 5.436  & 6.280  & 7.413  & 9.013  & 11.437 & 15.521  \\  
        0.60\text{e-4} & 2.945  & 3.286  & 3.713  & 4.263  & 4.999  & 6.032  & 7.587  & 10.186 \\  
        0.90\text{e-4} & 2.266  & 2.524  & 2.847  & 3.262  & 3.816  & 4.593  & 5.760  & 7.707  \\  
        1.20\text{e-4} & 1.858  & 2.068  & 2.330  & 2.667  & 3.117  & 3.748  & 4.694  & 6.272  \\  
        3.51\text{e-4} & 0.847  & 0.941  & 1.059  & 1.210  & 1.412  & 1.695  & 2.119  & 2.826  \\  
        3.73\text{e-4} & 0.809  & 0.899  & 1.011  & 1.156  & 1.349  & 1.619  & 2.024  & 2.699   \\ 
        3.95\text{e-4} & 0.775  & 0.861  & 0.969  & 1.107  & 1.292  & 1.551  & 1.938  & 2.584   \\ \hline
\end{tabular*}  
\end{table}

\begin{table}
  \centering
    \caption{The time required for the semi-major axis $a$ reaches its minimum value due to gravitational radiation and the MT process, for the different initial  $e_0$ and $\epsilon_0$, with same $\xi_{0}$\,$=$\,$3.55\times 10^{-3}$. The time unit is year.}
 \label{tab: table of minimum value}  
 \begin{tabular*}{0.9595\columnwidth}{|@{\hspace*{5.6pt}}c@{\hspace*{5.6pt}}|@{\hspace*{4pt}}c@{\hspace*{4.5pt}}c@{\hspace*{4.5pt}}c@{\hspace*{4.5pt}}c@{\hspace*{4.5pt}}c@{\hspace*{4.5pt}}c@{\hspace*{4.5pt}}c@{\hspace*{4.5pt}}c@{\hspace*{2.5pt}}|}
\hline
       \diagbox{$\epsilon_{0}$}{$e_{0}$} & 0.980  & 0.982  & 0.984  & 0.986   & 0.988   & 0.990 & 0.992 & 0.994  \\ \hline
        3.51\text{e-4} & 4.132  & 4.842  & 5.779  & 7.061  & 8.898  & 11.692  & 16.324  & 25.081  \\  
        3.73\text{e-4} & 2.147  & 2.514  & 3.003  & 3.668  & 4.620  &  6.078  & 8.487   & 13.047  \\  
        3.95\text{e-4} & 0.213  & 0.246  & 0.294  & 0.353  & 0.445  & 0.585   & 0.806   & 1.240   \\ \hline
\end{tabular*} 
\end{table}
 
For a fixed mass combination, such as $m_{1}$\,$=$\,$0.15$\,$\text{M}_{\sun}$, $m_2$\,$=$\,$4 \times 10^{5} \text{M}_{\sun}$, $q$\,$=$\,$3.75\times 10^{-7}$, we use the  long-term evolution of Newtonian elliptical orbits as described in \ref{sec:The long-term orbital effects of MT are calculated using discrete method for  classic elliptical orbits}. 
Therefore, only parameters $e_{0}$, $\epsilon_{0}$, and $\xi_{0}$  need to be adjusted. 
The initial values of $e_{0}$ and $\epsilon_{0}$ reflecting different initial periods and pericenter distances. 
We fix the variable $\xi_{0}$\,$=$\,$3.55\times 10^{-3}$. 
Table \ref{tab: table of phase difference} shows the time required for the phase difference to reach 1 radian. 
Table \ref{tab: table of minimum value} shows the time required for the semi-major axis $a$ to reach its minimum.

In both Tables \ref{tab: table of phase difference} and  \ref{tab: table of minimum value}, the time required to observe the desired phase shift $\Delta u_g$ increases with increasing $e _{0}$ and  decreasing $\epsilon _{0}$.  
These time variations are related to the strength of MT.  
Increasing eccentricity $e$ with a constant pericenter distance $r_{\text{p}}$  reduces the critical phase $f_0$ over which MT occurs (Equation \ref{App3: 9}),  shortening the Roche lobe overflow time in one period and decreasing the transferred mass $\delta M_{\text{WD}}$ (Equation \ref{59 Calculate transferred mass every period 8}). 
This results in a weaker MT effect and a longer required time. 
Conversely, decreasing the pericenter distance $r_{\text{p}}$  at constant eccentricity $e$ increases $\epsilon$, the Roche radius $R_{\text{L}}$ decreases, leading to an increase in $\delta M_{\text{WD}}$. This strengthens the MT effect and reduces the required time.

When the initial overflow parameter $\epsilon_0$ is small, the time required for the phase difference to reach 1 radian  (Table \ref{tab: table of phase difference})  is shorter than the time it takes for the semi-major axis  $a$ to reach its minimum  (Table \ref{tab: table of minimum value}). 
However, for sufficiently large $\epsilon_0$ (as shown in the last row of Tables \ref{tab: table of phase difference} and  \ref{tab: table of minimum value}), this trend reverses: the phase lag takes longer to reach 1 radian than the time needed for the semi-major axis to reach its minimum.
This reversal arises because the phase difference accumulates gradually from an initial value of zero, whereas the MT process begins with a finite $\epsilon_0>0$,  implying an initially strong MT effect. 
Therefore, when $\epsilon_0$ is large, the orbital evolution rapidly becomes MT-dominated even before the phase difference becomes significant. 
In such scenarios, the GW signals could reveal not only the accumulated phase lag due to MT, but also the dynamical transition from GW-dominated to MT-dominated evolution. 

For sufficiently large overflow $\epsilon_0$ and small eccentricity $e_0$, the phase shift can be detected within the observation time $t_{\text{obs}}$. 
Our calculations indicate that for an initial eccentricity $e_0$\,$\simeq $\,$0.99$, the overflow parameter should exceed $10^{-4}$ to allow detection within $t_{\text{obs}}$. 
However, the corresponding  pericenter distance  $r_{\text{p}}$\,$=$\,$(1-\epsilon )R_{\mathrm{WD} } /C_{\mathrm{L} } $ changes only slightly between $\epsilon$\,$=$\,$10^{-4}$ and $\epsilon$\,$=$\,$0$.
This illustrates that even a tiny variation in  pericenter distance $r_{\text{p}}$,  can lead to a significant difference in phase evolution timescales, as reflected in Table \ref{tab: table of phase difference}.
Therefore, the detectability of MT-induced signatures in GW signals is extremely sensitive to the precise value of $\epsilon_0$.

\section{Discussion}\label{sec:Discussion}
\subsection{Implications for QPEs}\label{sec:Discussion 4}
After our analysis, we conclude that the MT process may cause the WD to escape  the gravitational potential of the IMBH (Section \ref{sec:The long-term orbital effects of MT are calculated using discrete method for classic elliptical orbits}), potentially explaining the disappearance of the QPE signal in GSN 069, as reported by \citet{miniutti2023repeating, miniutti2023alive}. 
GSN 069, the first galactic nucleus to exhibit QPEs, was identified in December 2018. These high-amplitude, soft X-ray bursts recur approximately every 9 hours, lasting about 1 hour, with X-ray count rates increasing by up to two orders of magnitude from a stable quiescent level. 
However, following the XMM6 observation, eruptions have not been detected in subsequent exposures from May 2020 to December 2021 (XMM7–XMM11). 
The QPE reappeared in XMM12 observation in July 2022 after disappearing for $2.0$\,$\pm $\,$0.5$ years \citep{miniutti2023alive}. 
The new light curve features differ markedly from the previous ones. 

To estimate the MT rate in the QPE light curve, we use the model of light curve for the partial TDE from the tidal stripping proposed by \citet{Rees1988,ulmer1999flares}.  
The accretion luminosity of the stripped material falling back to the IMBH is $L(t)$\,$=$\,$\tilde{\eta }  \dot{M}_{\text{IMBH}}(t)c^2$, where $\tilde{\eta }$\,$=$\,$0.1$ is the typical radiative efficiency \citep{king2020gsn}.  
The peak rate of mass gain of IMBH is $\delta M_{\text{WD}}/3P_{\text{min}}$, where $P_{\text{min}}$ is the shortest Keplerian orbital period \citep{ulmer1999flares,Wang_2019}. 
Thus, we can give a relationship between the peak of the light curve $L_{\text{peak}}$ and the transferred mass $\delta M_{\text{WD}}$ per cycle  by the WD:
\begin{align}\label{7.1 Lpeak and delta MWD}
\delta M_{\text{WD}} &\simeq 2.5\times 10^{-8} \mathrm{M}_{\sun}\left ( \frac{\tilde{\eta } }{0.1}  \right )^{-1}\left ( \frac{L_{\mathrm{peak}} }{10^{42} \operatorname{erg} \mathrm{s}^{-1}} \right )\left ( \frac{P_{\mathrm{min}}}{1 \mathrm{ks} }  \right ).
\end{align}
Based on the initial parameters provided in Sections \ref{sec:Analyzing orbital results without MT in PN method} and \ref{sec:Analyzing orbit with MT},  if $\delta M_{\text{WD}}$ has the value given by Equation (\ref{7.1 Lpeak and delta MWD}), the MT effect on the orbit becomes dominant over gravitational radiation ($\delta M_{\text{WD}}$ exceeds the value indicated by the red points in the right panel of Figure \ref{5.2 Variation of WD mass}).

In our model, when MT becomes dominant, the orbital period and eccentricity increase (Figure \ref{5.2 Variation of the semi-major axis, eccentricity, WD period}), and the WD finally escapes from the gravitational potential of the IMBH. 
The lifetime of the WD-IMBH system, when MT dominates, is much shorter than the gravitational radiation timescale, i.e. $t_{\text{life}}$\,$\sim $\,$ \left ( 1-e \right )/\left \langle \dot{e} \right \rangle  _{\mathrm{sec} }$\,$ \ll $ $t_{\text{GW}}$\,$\sim$\,$ \left | \left (  r_{\mathrm{p} }-R_{\mathrm{t} } \right ) /\left \langle \dot{r} _{\mathrm{pGW} } \right \rangle  _{\mathrm{sec} }  \right | $ (use Equations \ref{5.2 long-term orbital effects of MT discrete method 1} and \ref{5.2 long-term orbital effects of MT discrete method 1 drp}). 
The escape timescale is approximately $10$ to $10^2$ years.
Higher surface densities parameter  $\xi_0$ results in higher transferred mass $\delta M_{\text{WD}}$, causing the WD to escape the system on a shorter timescale. 
Therefore, for a low-mass WD core that has been stripped of its outer low-density surface, the time required to enter and escape the system will be short in 10 years.

In addition, the MT rate is highly sensitive to the pericenter distance.  
For this escape to occur, the initial pericenter distance must be sufficiently large to avoid tidal disruption,  i.e. $r_{\text{p}}$ satisfies $\epsilon_0$\,$=$\,$1$\,$-$\,$r_{\text{p}}C_{\mathrm{L} }/R_{\mathrm{WD} }$\,$\ll $\,$\epsilon_{\,\mathrm{t} }$\,$\sim $\,$1$\,$-$\,$C_{\mathrm{L}} q^{-1/3}$ (see Section \ref{sec:The long-term orbital effects of MT are calculated using discrete method for classic elliptical orbits} for details).

The mechanism developed in our work accounts only for the dynamical escape of the WD from the gravitational potential of IMBH. 
The observed recurrence or reappearance of QPEs likely involves additional physical processes not captured in our model. 
One plausible explanation is the perturbative influence of external massive bodies—such as nearby high-speed moving stars ($10$\,$-$\,$10^2\, \mathrm{M} _{\sun}$), stellar-mass BHs ($10^2\, \mathrm{M} _{\sun}$), or even intermediate-mass BHs ($10^4\, \mathrm{M} _{\sun}$), which may induce gravitational focusing or orbital re-injection of the escaping WD. 
Some studies have considered the von Zeipel–Lidov–Kozai (ZLK) effect, which is the perturbation of the primary-secondary orbital system by a third party (such as another star or an IMBH) \citep{Bode2013,Lei2022,Kei-ichi2025}. Through the ZLK cycle, the eccentricity and inclination of the orbit can be significantly adjusted, causing the WD to occasionally approach the primary BH, triggering the emission of QPEs.

\subsection{The assumptions for mass transfer}\label{sec:Discussion 1}
Equation (\ref{44 aMT divided aGW}) provides  the MT rate when gravitational radiation and MT are in equilibrium. 
In this subsection, we will show that our MT rate is consistent with the estimates in \citet{king2020gsn,king2022quasi}, which use the gravitational radiation losses $ ( \dot{J} /J   ) _{\mathrm{GW} } $  to drive the MT rate.  

According to \citet{king2022quasi,king2023angular}, the dynamical stability of MT arises from the conservation of orbital angular momentum on a dynamical timescale, which is much shorter than that for gravitational radiation losses. 
The accretion disk reaches an effectively steady state after a few orbital periods. After this, the disk passes almost all its angular momentum back to the orbit. 
This is the reabsorption of the original angular momentum of the mass transferred to the IMBH. 
As a result, conservation of orbital angular momentum forces the binary separation to increase, i.e. $\left \langle \dot{a}^{\mathcal{B}}_{\mathrm{MT} } \right \rangle_{\text{sec}}$\,$>$\,$0$. 
Our analysis of orbital evolution builds upon the orbital angular momentum analysis from \citet{king2022quasi} and the eccentric orbital dynamics with MT from \citet{sepinsky2007interacting}. 
Finally, we present a new possible scenario for orbital evolution.

We reconsider the condition $\left \langle   \dot{R}_{\mathrm{L}}/R_{\mathrm{L}}-\dot{R}_{\mathrm{WD} }/R_{\mathrm{WD} }\right \rangle=0$ in \citet{king2022quasi} and recalculate the expression  using definitions $\varepsilon$\,$=$\,$1-C_{\mathrm{L} } D/R_{\mathrm{WD} }$ and $\epsilon$\,$=$\,$1-C_{\mathrm{L} } r_{\mathrm{p} }/R_{\mathrm{WD} }$:   
\begin{align}\label{64 1} 
\frac{\dot{R}_{\mathrm{L}}}{R_{\mathrm{L}}}-\frac{\dot{R}_{\mathrm{WD} }}{R_{\mathrm{WD} }}&=-\frac{\dot{\varepsilon} }{1-\varepsilon}=\frac{\dot{D} }{D}-\left (   \frac{\dot{\epsilon} }{1-\epsilon}+\frac{\ \dot{r}_{\mathrm{p} } }{\ r_{\mathrm{p} }}\right )\\
&=\frac{\dot{D} }{D}-\frac{\ \dot{m}_{1}}{\ m_{1}}\left [ \frac{m_1}{R_{\mathrm{WD} }}\frac{ \partial R_{\mathrm{WD} }}{\partial m_1}-\frac{q}{ C_{\mathrm{L} } }\frac{ \partial C_{\mathrm{L}}}{\partial q} \left ( 1+\gamma q \right ) \right ]. \nonumber
\end{align}
Thus, we got the relationship between $\dot{\epsilon}$ and  $\dot{r}_{\mathrm{p} }$ from $\dot{J}$:
\begin{subequations}
\begin{align}
\frac{\dot{\epsilon} }{1-\epsilon}=&\frac{\ \dot{m}_{1}}{\ m_{1}} \left [  \frac{ \partial \ln R_{\mathrm{WD} }}{\partial \ln m_1}- \frac{ \partial \ln C_{\mathrm{L}}}{\partial \ln q} \left ( 1+\gamma q \right ) \right ] -\frac{\ \dot{r}_{\mathrm{p} } }{\ r_{\mathrm{p} }}, \label{66 3 a} \\
\frac{\dot{r}_{\mathrm{p} } }{r_{\mathrm{p} }} = -&\frac{\ \dot{m}_{1}}{\ m_{1}}\frac{2(1- \gamma  q^2)+q(1-\gamma )}{1+q}-\frac{\dot{e}}{1+e} +\frac{2 \dot{J}}{J} . \label{66 3 b} 
\end{align}
\label{66 3} 
\end{subequations}
If  taking $R_{\mathrm{WD} }$\,$\propto$\,$m_1^{\zeta_1}$ and  $C_{\mathrm{L} }$\,$\propto$\,$q^{\zeta_2}$,  we obtain $\partial \ln  R_{\mathrm{WD} }/\partial \ln  m_1$\,$=$\,$\zeta_1$ and $\partial \ln  C_{\mathrm{L} }/\partial \ln  q$\,$=$\,$\zeta_2$. Under the conditions $M_{\mathrm{WD}}$\,$\ll$\,$ M_{\mathrm{Ch}}$ and  $q$\,$\ll $\,$1$, we can assume $\zeta_1$\,$\rightarrow$\,$-1/3$ (Equation \ref{eq:RWD}) and $\zeta_2$\,$\rightarrow$\,$ 1/3$ (Equation \ref{coefficient f A,q}). Thus, if we further take $\gamma$\,$=$\,$1$, which represents the conservation of mass, Equation (\ref{66 3 a})  reduces to Equation (12) in \citet{king2022quasi}.  
The last term of Equation (\ref{66 3 b}), the change in angular momentum, is driven only by gravitational radiation, which is much longer than the dynamical scale. 
The second term, the change in eccentricity, has additional MT effect, which further increases the MT rate.

\citet{king2022quasi} sets $\left \langle \dot{\epsilon}_{\text{MT}} \right \rangle_{\text{sec}}$\,$=$\,$0$ in Equation (\ref{64 1}) in order to keep the MT rate stable in the long time, and show  the effect of gravitational radiation losses $\dot{J} / J$ is to drive MT. And based on this, the orbit needs to be expanded (i.e., increases in semi-major axis, period, and eccentricity).
But this setting is not the only possible choice for orbital expansion. 
Assuming orbital evolution starts from $\epsilon_0$\,$=$\,$0$ (as shown in Figures \ref{5.2 Variation of the semi-major axis, eccentricity, WD period}, \ref{5.2 Variation of WD mass} and  \ref{5.2 ratio of WD epsilon1}), 
the transferred mass per cycle $\left |  \delta M_{\text{WD}} \right |$ always increases, and the orbital period also increases due to MT dominance, the change of the period  is faster than the change of $\left |  \delta M_{\text{WD}} \right |$ (When the eccentricity approaches $1$, the period approaches infinity).

The average MT rate $\left | \delta M_{\text{WD}}/T \right |$  may reaches a maximum value and then decreases when the orbit expands. 
This reduction in average MT rate must require the MT process leading to orbit expansion.

Alternatively, the setting that the pericenter distance $r_{\text{p}}$ remains unchanged, i.e. $\left \langle  \dot{r}_{\text{pMT}}\right \rangle_{\text{sec}}$\,$=$\,$0$ in Equation (\ref{5.2 long-term orbital effects of MT discrete method 1 drp}), can also allow to orbital expansion.  
As shown in Section \ref{sec:Dynamics under mass transfer}, the analysis of $J_{\text {orb }}$ has given an expression for the component $\mathcal{S}$ (Equation \ref{39 Conservation of angular momentum for S}).
From Equation (\ref{eq:elements differential equation11}) for $\dot{p}$ and $\dot{e}$, the contribution of $\mathcal{S}$ is much larger than that of $\mathcal{R}$ at high eccentricity. 
Therefore, under different settings for component $\mathcal{R}$, the evolution of $a$, $T$ and $e$ will not be different. 
They still evolve as shown in Figure \ref{5.2 Variation of the semi-major axis, eccentricity, WD period}.

We set $\left \langle  \dot{r}_{\text{pMT}}\right \rangle_{\text{sec}}$\,$=$\,$0$ in this paper. For highly eccentric orbits, as discussed in Section \ref{sec:Dynamics under mass transfer},  the acceleration $\boldsymbol{f}_{\text{MT}}^{\mathcal{B}}$  is primarily significant near pericenter.  
Equation (\ref{5.2 long-term orbital effects of MT discrete method drp}) shows that $\boldsymbol{f}_{\text{MT}}^{\mathcal{B}}$ has almost no contribution to $\left \langle   \dot{r}_{\text{pMT}}\right \rangle_{\text{sec}}$,  
taking $\left \langle   \dot{r}_{\text{pMT}}\right \rangle_{\text{sec}}$\,$ =$\,$0$ at high eccentricity more reasonable than $\left \langle   \dot{\epsilon} _{\text{MT}} \right \rangle_{\text{sec}} $\,$=$\,$ 0$ in \citet{king2022quasi}.

We can explain the orbital difference between the two settings: 
(i) setting Equation (\ref{66 3 a}) to zero corresponds to $\left \langle   \dot{\epsilon}_{\text{MT}} \right \rangle_{\text{sec}}$\,$=$\,$0$, (ii) setting Equation (\ref{66 3 b}) to zero corresponds to  $\left \langle  \dot{r}_{\text{pMT}}\right \rangle_{\text{sec}}$\,$=$\,$0$.
However, only one of Equations (\ref{66 3 a}) or   (\ref{66 3 b}) can be set to zero, since $\dot{m}_{1}$\,$ \ne$\,$ 0$ with MT.   
First, if $\left \langle   \dot{\epsilon}_{\text{MT}} \right \rangle_{\text{sec}} $\,$=$\,$0$, then $\left \langle   \dot{r}_{\text{pMT}}\right \rangle_{\text{sec}} $\,$>$\,$0$; MT leads to an increase in the pericenter distance, the orbital pericenter decay caused by gravitational radiation will be slowed down significantly. 
Second, if $\left \langle   \dot{r}_{\text{pMT}}\right \rangle_{\text{sec}} $\,$=$\,$0$, then  $\left \langle   \dot{\epsilon}_{\text{MT}} \right \rangle_{\text{sec}}$\,$>$\,$0$, we have discussed the evolution of $\epsilon$ in this setting in Figure \ref{5.2 ratio of WD epsilon1}.

To drive the MT rate, we average $\dot{J}_{\text {orb }} /J_{\text {orb }}$ in Equation (\ref{38 Conservation of angular momentum 1}) over one orbital period and apply  Equation (\ref{42 dMWDdt }) to  include $\delta M_{\mathrm{WD}}$, accounting for gravitational radiation losses \citep{peters1964gravitational}:
\begin{align}\label{66 4} 
&\left \langle \frac{\dot{J}_{\text {orb }}}{J_{\text {orb }}} \right \rangle_{\text{MT\,sec}} =\left ( \frac{\delta M_{\mathrm{WD}} }{ M_{\mathrm{WD} }}  \right )\sqrt{\frac{G M_{\mathrm{WD} } }{c^2 a_{\text{r}} } }\frac{1}{2\pi } \frac{c}{a_{\text{r}}} (1-q)\sqrt{\frac{1+q}{q} }  \tilde{h}, \nonumber \\
&\left \langle \frac{\dot{J}_{\text {orb }}}{J_{\text {orb }}} \right \rangle_{\text{GWsec}} =- \left (   \frac{G M_{\mathrm{WD} }}{c^2 a_{\text{r}}} \right )^3 \frac{32}{5}\frac{c}{a_{\text{r}}}  \frac{1+q}{q^2} \frac{ 1+\frac{7}{8} e_{\text{r}}^2 }{(1-e_{\text{r}}^2 )^{5/2} } .
\end{align}
Assuming $\left \langle \dot{J}_{\text {orb }}/J_{\text {orb }}\right \rangle_{\text{MTsec}}$\,$\simeq $\,$ \left \langle \dot{J}_{\text {orb }}/J_{\text {orb }} \right \rangle_{\text{GWsec}}$, we obtain:
\begin{equation} 
\begin{aligned}
&\left | \frac{\delta M_{\mathrm{WD}} }{T}  \right | \simeq  \frac{  M_{\mathrm{WD}} }{T}\tilde{\lambda }  ^{-5/2} q \frac{64}{5} \pi  \frac{ 1+\frac{7}{8} e_{\text{r}}^2 }{(1+e_{\text{r}} )^{5/2} } \frac{\sqrt{1+q} }{(1-q)\tilde{h}} . 
\end{aligned}
\label{66 5} 
\end{equation}
Here, $\left | \delta M_{\mathrm{WD}}/T \right | $ derived using this equation is the same as ($-\dot{M}_2$) in Equation (15) of \citet{ king2022quasi}.  
Equation  (\ref{66 5}) is similar to Equation (\ref{44 aMT divided aGW})  for $\left |\dot{a}^{\mathcal{B}}{}_{\mathrm{MT} } / \dot{a}_{\mathrm{GW}}  \right |_{\text{sec}}$\,$=$\,$1 $. 
Therefore, the MT rate from \citet{king2022quasi} is consistent with our view that gravitational radiation and MT are equilibrium.

\subsection{Mass loss and angular momentum transfer}\label{sec:Discussion 1 2}

In Section \ref{sec:Dynamics under mass transfer}, we derived  the form of MT acceleration $\boldsymbol{f}_{\text{MT}}^{\mathcal{B}}$,  but we assume the conservation of mass and angular momentum in the mathematical treatment in the following sections.
If these assumptions are relaxed, i.e. $\gamma <1$, $\tilde{f} \ne 0 $ and $\tilde{g} \ne 0 $,  the time required for WD ejection maybe change. 
However, we find that the lifetime  $t_{\text{life}}$\,$\sim $\,$ \left ( 1-e \right )/\left \langle \dot{e} \right \rangle  _{\mathrm{sec} }$ for the WD to escape is largely insensitive to the factor $\tilde{h}$.
From Equation (\ref{44 aMT divided aGW}), when MT balances gravitational radiation, the transferred mass scales as $\left | \delta M_{\text{WD}} \right |$\,$ \sim $\,$ M_{\text{WD}}\sigma $\,$\propto $\,$\tilde{h}^{-1} $. 
Meanwhile, the secular rate of change of eccentricity due to MT  (Equation \ref{5.2 long-term orbital effects of MT discrete method 1}) satisfies $\left \langle   \dot{e}^{\mathcal{B}}_{\mathrm{MT} }\right \rangle_{\text{sec}} $\,$ \propto $\,$\left | \delta M_{\text{WD}} /M_{\text{WD}}\right |    \tilde{h}$\,$ \propto  $\,$ \sigma \tilde{h}  $.
Combining these relations reveals that the eccentricity evolution rate is effectively independent of $\tilde{h}$, as the dependence cancels out. Consequently, the escape timescale $t_{\text{life}}$, which is closely linked to the eccentricity evolution, is not significantly affected by variations in factor $\tilde{h}  $.

In addition, the non-conservation of orbital angular momentum may also affect the stripping mass, this helps us understand the evolution of WD.
However, we do not know the exact values of these coefficients. 
We will not discuss the specific motion of MT flow in detail in this paper. We only make assumptions about the coefficients in Equation (\ref{39 Conservation of angular momentum for S}).
For the factor $\tilde{h}$, a zeroth expansion in the small mass ratio limit  ($q\ll  1$) is $\tilde{h}=1-\gamma \tilde{f} +\mathcal{O}  \left (q  \right )$. 
The contribution forom $\tilde{g}$ appears only in combination with
$q$ and can thus be neglected to leading order.  
Thus, we temporarily ignore the second term in Equation  (\ref{38 Conservation of angular momentum 1}).
As an example, we take a typical parameter combination: $\gamma=\tilde{f}=1/2$ (half of the mass escapes from the system, taking with it half of its specific angular momentum), this yields $\tilde{h}=3/4$. 
Under this condition, the transferred mass required to balance GW losses  (Equation \ref{44 aMT divided aGW})  increases relative to the conservative case ($\tilde{h}=1$). Quantitatively, for $\tilde{h}=3/4$, the required $\delta M_{\text{WD}}$ increases by approximately $33\%$. This implies a longer evolutionary path from $\epsilon =0$ to equilibrium, though the timescale remains broadly comparable.

A extreme scenario where $\tilde{h}$\,$\ll $\,$1 $ would require $\gamma$\,$\sim $\,$1$ (most of the mass retained in the system) and $\tilde{f}$\,$\sim$\,$ 1$  (almost no angular momentum is reconverted into orbital angular momentum), this  combination that seems physically implausible. 
In high-eccentricity systems, the accretion disc circularises on a few orbital periods, with a typical circular radius close to the orbital pericentre. 
As the WD continues to perturb the disc during each pericentre passage, the disc structure is periodically destabilised, facilitating the transfer of a non-negligible fraction of angular momentum back to the orbit. 
Such systems are unlikely to exhibit strongly non-conservative angular momentum behaviour.

\subsection{Relativistic orbital correction for the transferred mass}\label{sec:Discussion 2}
In Section \ref{sec:Simplification of mass transfer model}, for Keplerian elliptical orbits, Equation (\ref{59 Calculate transferred mass every period 8}) gives the transferred mass $\delta M_{\mathrm{WD}}$ during each orbital period.
However, for high eccentricity ($e_{\text{r}} $\,$\sim $\,$0.99$) and small pericenter distance ($\tilde{\lambda }$\,$=$\,$r_{\mathrm{p}}/(G m_{2}/c^2)$\,$ \sim$\,$ 8$), this result should be corrected to account for the effects due to GR.
 
Using PN approximation, the $\dot{f}_{\text{N} }$ in Equation  (\ref{52 Calculate transferred mass every period 1}) should be replaced by $df/dt$ in Equation (\ref{eq:elements differential equation11 c}). 
When the companion star  ($q$\,$\ll$\,$1$) has a high velocity near pericenter, the effect of relativistic corrections is significant,  the geodesic equation in curved spacetime becomes more applicable. 
If we use geodesic equations in Schwarzschild spacetime instead, we can adopt $dt/d\psi$ from  \citet{cutler1994gravitational} and \citet{martel2004gravitational}  to replace $\dot{f}_{\text{N} }^{-1}$. 
Here, $\psi$ is a phase parameter that controls radial motion. 
The final result for $\delta M_{\mathrm{WD}}$ is then given by 
\begin{align}\label{65 Calculate transferred mass every period 14} 
 \delta M_{\mathrm{WD}}&  \simeq \sum_{k=0}^{\infty } \left \{ F_{k}(\hat{p}) \cdot (1+e)^k \sqrt{1+\frac{4(1-e^2)}{\hat{p}(\hat{p}-4) }}  \times \right .  \nonumber \\
& \left. \frac{4\ C_{\dot{m}_1} r_{\text{p}}^{3/2}}{\sqrt{G m } } \sqrt{\frac{\epsilon }{2e} }  \sum_{n=0}^{\infty }\sum_{m=0}^{\infty  }C_{\mathrm{\varepsilon n} }\begin{pmatrix}
 n\\
m
\end{pmatrix}\left ( -1 \right ) ^m \left ( 1-\epsilon  \right ) ^m\times  \right . \nonumber \\
&\quad  \left. \mathrm{AF1}\left [\frac{1}{2} ;\frac{1}{2} ,m-k+2;\frac{3}{2}; \frac{1+e}{2e}\epsilon,\epsilon \right ] \  \right \} ,
\end{align}
where the series $F_{k}(\hat{p})$  is derived in Appendix \ref{App: Calculate transferred mass every period}. Here, $\hat{p}$ satisfies  $p=\hat{p}(Gm/c^2)=r_{\text{p}} (1+e) $, with $e=(r_{\text{a}}-r_{\text{p}})/(r_{\text{a}}+r_{\text{p}})$.  
The detailed derivation of Equation (\ref{65 Calculate transferred mass every period 14}) is provided in Appendix \ref{App: Calculate transferred mass every period}.
 
The relativistic correction to the Schwarzschild spacetime geodesic introduces the coefficients $F_k(\hat{p})(1$\,$+$\,$e)^k$ and $1$\,$+$\,$4(1-$\,$e^2)/\hat{p}(\hat{p}-4) $, which reduce to $1$ in the 0PN limit. 
The coefficient  $F_k(\hat{p})(1+e)^k$  is positive for any $k$, and for   bound orbit that won't fall into central BH, i.e. $0$\,$<$\,$e$\,$<$\,$1$, $\hat{p}$\,$>$\,$4$, we have $4(1$\,$-$\,$e^2)/\hat{p}(\hat{p}-4) $\,$>$\,$0$. 
These coefficients collectively enhance $\delta M_{\mathrm{WD}}$ compared to Equation (\ref{59 Calculate transferred mass every period 8}),  which is due to the orbital precession effect caused by GR. 
Since the orbital precession due to the Schwarzschild metric is prograde, the precession angle during one orbital period $\Delta \varphi$ is positive. 
This positive $\Delta \varphi$ increases the time interval during which $R_{\text{WD}}$\,$>$\,$R_{\text{L}}$, i.e. $dt/d\psi$\,$>$\,$\dot{f}_{\text{N} }^{-1}$.
Here, the expression for $dt/d\psi$ is provided in Equation (\ref{App3: 10}). 
Because $f_0$ satisfies $R_{\text{WD}}$\,$=$\,$C_{\text{L}}D(f_0)$ and  we substitute $D$ with  $p/(1+e \cos \psi)$, the value of $f_0$ remains consistent with Equation (\ref{App3: 9}).
The definite integral of $dt/d\psi$ from $-f_0$ to $f_0$ must be greater than that of $\dot{f}_{\text{N} }^{-1}$.

As orbital  eccentricity increases with same pericenter distance, the angular phase evolves more slowly, reflected by an increase in $dt/d\psi$ (see Equation \ref{App3: 10}). 
This results in a longer pericenter interaction time and consequently enhances MT per orbit.
For a quantitative example,  a representative system with $m_1$\,$=$\,$0.15$\,$\text{M}_{\sun}$, $q$\,$=$\,$3.75\times 10^{-7}$,  $\epsilon$\,$=$\,$5$\,$\times$\,$10^{-3}$ and $e$\,$=$\,$0.985$, the relative difference in transferred mass between relativistic and Newtonian calculations is: $(\delta M_{\mathrm{WD}}^{\mathrm{R}}-\delta M_{\mathrm{WD}}^{\mathrm{NR}})/\delta M_{\mathrm{WD}}^{\mathrm{NR}}$\,$\simeq$\,$ 65\%$, where "$\mathrm{R}$" and "$\mathrm{NR}$" denote the relativistic and non-relativistic cases, respectively.
This  discrepancy highlights the importance of relativistic corrections in systems with extreme eccentricity and high velocity, where even subtle changes in orbital dynamics significantly affect MT efficiency.

The transferred mass per cycle, $\delta M_{\mathrm{WD}}$, obtained from  Equation (\ref{65 Calculate transferred mass every period 14}), is consistent with that from our short-term PN orbital simulations (Section \ref{sec:Analyzing orbital results without MT in PN method}). 
These simulations incorporate the MT acceleration $\boldsymbol{f}_{\text{MT}}^{\mathcal{B}}$ (Equation \ref{f_MT change_2 form}) and the mass loss rate $\dot{m}_1$ (Equations \ref{eq:58 dotm1} and \ref{eq:59 dotm1}). The deviation between the two approaches remains below $5\%$, confirming  the accuracy of the relativistic treatment.

\subsection{Contribution of white dwarf spin angular momentum}\label{sec:Discussion 3}
 For a typical spin parameter of WD $\chi_1$\,$ \sim $\,$ 0.1$, the corresponding orbital perturbation scales as $\chi_1 q$, rendering the resulting orbital modulation subdominant. 
The low spin value $\chi_1$\,$ \sim $\,$ 0.1$ is expected, as the structure of WDs is supported by electron degeneracy pressure; excessive spin would lead to centrifugal forces exceeding gravitational binding, causing structural instability or disintegration \citep{poisson2014gravity}. 
Therefore, given the extreme mass ratio $q$\,$\ll$\,$1$, the orbital influence of the WD’s spin—through spin-orbit ($\boldsymbol{a}_{\mathrm{SO}}$), spin-spin ($\boldsymbol{a}_{\mathrm{SS}}$), and spin–post-Newtonian ($\boldsymbol{a}_{\mathrm{SOPN}}$) acceleration terms—is negligible.

In contrast to the PN spin correction, the MT process exerts a more substantial influence on the dimensionless parameter $A$\,$=$\,$\Omega _{1}^2  D^3 / G m$, as introduced in Section \ref{sec:MT rate and the first lagrangian point}. This parameter enters both the effective potential-density (Equation \ref{eq:56 rho}) and the Lagrange point condition (Equation \ref{29 equaiton of X1}), both of which are ultimately used to derive the transferred mass per orbit. These expressions are further expanded into series of $\delta M_{\mathrm{WD}}$ in Equations (\ref{59 Calculate transferred mass every period 8}) and (\ref{65 Calculate transferred mass every period 14}). The impact of $A$ manifests directly through the coefficients of these expansions.
We first estimate the size of $A$ at pericenter: 
\begin{align}\label{5.5 estimate A}
 &A=\frac{ \Omega _{1}^2  r_{\text{p} }^3}{G m}= \left ( \frac{R_{\mathrm{WD}}\Omega _{1}}{c}  \right ) ^2\left ( \frac{R_{\mathrm{WD}} }{r_{\text{p} }}  \right )  ^{-2} \left ( \frac{G m}{c^2 r_{\text{p} } }  \right ) ^{-1} \\
&\ \ \  =  \frac{5}{4896} \frac{q}{ 1+q} \left (  \frac{1-\epsilon }{C_{\mathrm{L} }}  \right )^3\frac{M_{\mathrm{WD} }^{4/3}}{\text{M}_{\sun}M_{\mathrm{Ch}}^{1/3}} \left (  \frac{\chi _{1}}{\tilde{\gamma } }  \right )^2 =\frac{ \varpi}{1+q}(1-\epsilon)^3 , \nonumber
\end{align}
where $\varpi $ and $\tilde{\gamma } $ are given in Equations (\ref{App2: 10}) and (\ref{App2: 10 1}).

For a representative system with $M_{\mathrm{WD} }$\,$=$\,$0.15 $\,$\text{M}_{\sun }$, $q$\,$=$\,$3.75\times 10^{-7}$, $\epsilon$\,$=$\,$5\times 10^{-3}$,  $\chi _{1}$\,$=$\,$0.1$, and $\tilde{\gamma } $\,$=$\,$0.5$, we find $A$\,$=$\,$1.86\times 10^{-5}$\,$\gg $\,$\chi _{1} q$\,$\sim$\,$10^{-8}$. 
For coefficient $D_{\varepsilon 1}$ in Equation (\ref{App2: 14}),  we see that $\varpi C_{\mathrm{L} }/D_{\varepsilon 1}$\,$\gg$\,$ \chi _{1} q$.
Therefore, the orbital effect induced by the spin of the WD is much weaker than the effect due to MT. 
However,  $10^{-5} D_{\varepsilon 1}$ is still very small compared to $ D_{\varepsilon 1}$, making it negligible.

\subsection{ Mass loss through the L2 Lagrange point }\label{sec:Analyzing L2 Mass Loss}
In binary systems where the primary is significantly more massive than the secondary, MT from the secondary to the primary is typically dynamically stable, occurring through the inner Lagrange point L1. 
However, if the system evolves rapidly, mass loss may also occur through the outer Lagrange point L2 \citep{MasslossLinial2017,linial2023unstable}. 
It is essential to evaluate the conditions under which this L2  mass transfer occurs.

Using the results of \citet{sepinsky2007equipotential}, the position of the L2 point can be determined. 
In the limit of an extreme mass ratio ($q \to 0$), its distance from the donor is nearly identical to that of the L1 point.  
Generally, the equipotential surface volume at L2 exceeds that at L1. 
Assuming a small amount of mass is transferred, the WD's radius closely approximates the Roche radius at both L1 and L2. 
Therefore, the Roche radius $R_{\mathrm{L2}}$ at L2 can be estimated via a Taylor expansion around the   calculated Roche radius $R_{\mathrm{L1}}$ at L1  (Equation \ref{coefficient f A,q}). 
We can approximate the volumetric radius by \citep{ritter1988turning,MasslossLinial2017}:
\begin{align}\label{40 Calculation of Lagrange 2 points} 
R_{\mathrm{L2}}-R_{\mathrm{L1}}\approx \frac{\phi _{\mathrm{2}}-\phi_{\mathrm{1}}}{GM_{\mathrm{WD} }/R_{\mathrm{WD}}^2 }>0 ,
\end{align} 
where $\phi _{\mathrm{2}}$ is the potential at point L2, which can be approximated using the potential function
\begin{align}\label{eq:59 phi(X)} 
\phi(D,\tilde{X}_{\mathrm{D}})=-\, \frac{G m_2}{D} \left ( \frac{q}{\tilde{X}_{\mathrm{D}}}+\frac{1}{2}  A(1+q)\tilde{X}_{\mathrm{D}} ^2+\frac{1}{1-\tilde{X}_{\mathrm{D}}} -\tilde{X}_{\mathrm{D}}  \right ),
\end{align} 
where $\tilde{X}_{\mathrm{D}}=\tilde{X}/D$. 
Thus, the approximate value of $\phi _{\mathrm{2}}-\phi_{\mathrm{1}}$ is
\begin{align}\label{40 Calculation of Lagrange potential} 
\phi _{\mathrm{2}}-\phi_{\mathrm{1}}\approx \phi(D,-\tilde{X}_1)-\phi(D,\tilde{X}_1)= 2 \frac{G m_2}{D} \frac{\tilde{X}_1^3}{1-\tilde{X}_1^2}  .
\end{align} 
For $A$\,$\to$\,$0$ at pericenter, the result of $\phi _{\mathrm{2}}-\phi_{\mathrm{1}}$ is $ G M_{\mathrm{WD}}/D$.   
So we can get  $(R_{\mathrm{L2}}-R_{\mathrm{L1}})/R_{\mathrm{WD}} $\,$\approx  $\,$R_{\mathrm{WD}} /D$. 
We define dimensionless overflow parameter $\varepsilon_{1,2}$\,$:=$\,$\left ( R_{\mathrm{WD}}-R_{\mathrm{L{1,2}}} \right )/R_{\mathrm{WD}}$ at L1 and L2,  so  $\varepsilon_{1}-\varepsilon_{2} $\,$\approx  $\,$R_{\mathrm{WD}} /D$.

If MT through L2 is non-negligible, the condition is: 
(i) MT occurs at L2, i.e. $R_{\mathrm{L2}}<R_{\mathrm{WD}}$.  It means $\varepsilon_{2}>0$ and $\varepsilon_{1}$\,$>$\,$R_{\mathrm{WD}} /D$.  
(ii) the magnitude of  $\varepsilon_{1}$ and $\varepsilon_{2}$ must be comparable, i.e. $(\varepsilon_{1}-\varepsilon_{2})/\varepsilon_{1}$\,$\sim$\,$\mathcal{O}(10^{-1}) $. 
Conversely, if $\varepsilon_{1}$\,$<$\,$R_{\mathrm{WD}} /D$ or $\varepsilon_{1}$\,$\sim $\,$R_{\mathrm{WD}} /D$, i.e. $\varepsilon_{2}<0$ or $\varepsilon_{1}$\,$\gg$\,$\varepsilon_{2}$,  MT through L2 can be neglected.

Here we consider the regime  in which the  transferred  mass $\delta M_{\mathrm{WD}}$ per  cycle, is very small—consistent with the amounts required to produce QPEs (see Section \ref{sec:Discussion 4} for a detailed discussion).   
For typical QPEs, $\delta M_{\mathrm{WD}}$\,$\sim $\,$10^{-8}-10^{-6}$\,$ \mathrm{M}_{\sun}$.   
Numerical results in Figures \ref{5.2 Variation of WD mass} and \ref{5.2 ratio of WD epsilon1} show that, for systems where the WD remains bound, $\delta M_{\mathrm{WD}}<2 \times 10^{-7} \mathrm{M}_{\sun}$ and $\varepsilon_{1}<3\times 10^{-3}<R_{\mathrm{WD}} /r_{\mathrm{p}}\sim 3.95\times 10^{-3}$.  
This indicates that  from the onset of MT  ($\varepsilon_{1}>0$) to the time it escapes from the system ($e>1$), the condition $\varepsilon_{2}<0$ or $\varepsilon_{1}$\,$\gg $\,$\varepsilon_{2}$ holds throughout. 
Therefore, in the analysis of this article,  MT from L2 is not considered.

\subsection{Impact of General Relativity on the Roche Radius}\label{sec:GR correction to Roche radius} 
In EMRI systems, the intense gravitational field of the IMBH significantly  alters tidal forces compared to Newtonian expectations. 
This affects the transferred mass of each cycle.
The relativistic Roche radius is smaller than its Newtonian counterpart at the same orbital separation. 
Determining this relativistic Roche radius is complex, necessitating the use of Fermi normal coordinates (FNCs) to account for the curved spacetime's gravitational potential. 
In FNCs, the spacetime geometry of the black hole is expanded in the vicinity of a timelike geodesic that approximately tracks the center of mass motion of the star and debris \citep{fishbone1973relativistic,marck1983solution,ishii2005black,cheng2013relativistic,cheng2014tidal,banerjee2019tidal,maeda2023chaotic}.

Incorporating fully relativistic MT is complex, thus  we compensate from different perspectives.
In our model, the Roche radius is used solely to estimate the  transferred  mass per orbit. 
This is introduced via the overflow parameter $\varepsilon$\,$=$\,$\left ( R_{\mathrm{WD}}-R_{\mathrm{L}} \right )/R_{\mathrm{WD}}$, which quantifies the extent to which the WD overfills its Roche lobe.
Our analysis begins with the observed periodicity of QPEs, requiring the orbital period to match observations. As a result, the pericenter distance and eccentricity are constrained by our previous results (Sections \ref{sec:Analyzing orbital results without MT in PN method} and \ref{sec:Analyzing orbit with MT}). Within this framework, the transferred mass can be modulated not only by adjusting orbital parameters ($r_{\mathrm{p}}$ and $e$), but also by varying the WD mass, which directly affects its radius $R_{\mathrm{WD}}$.

\begin{figure} 
\centering
\includegraphics[width=0.95\linewidth]{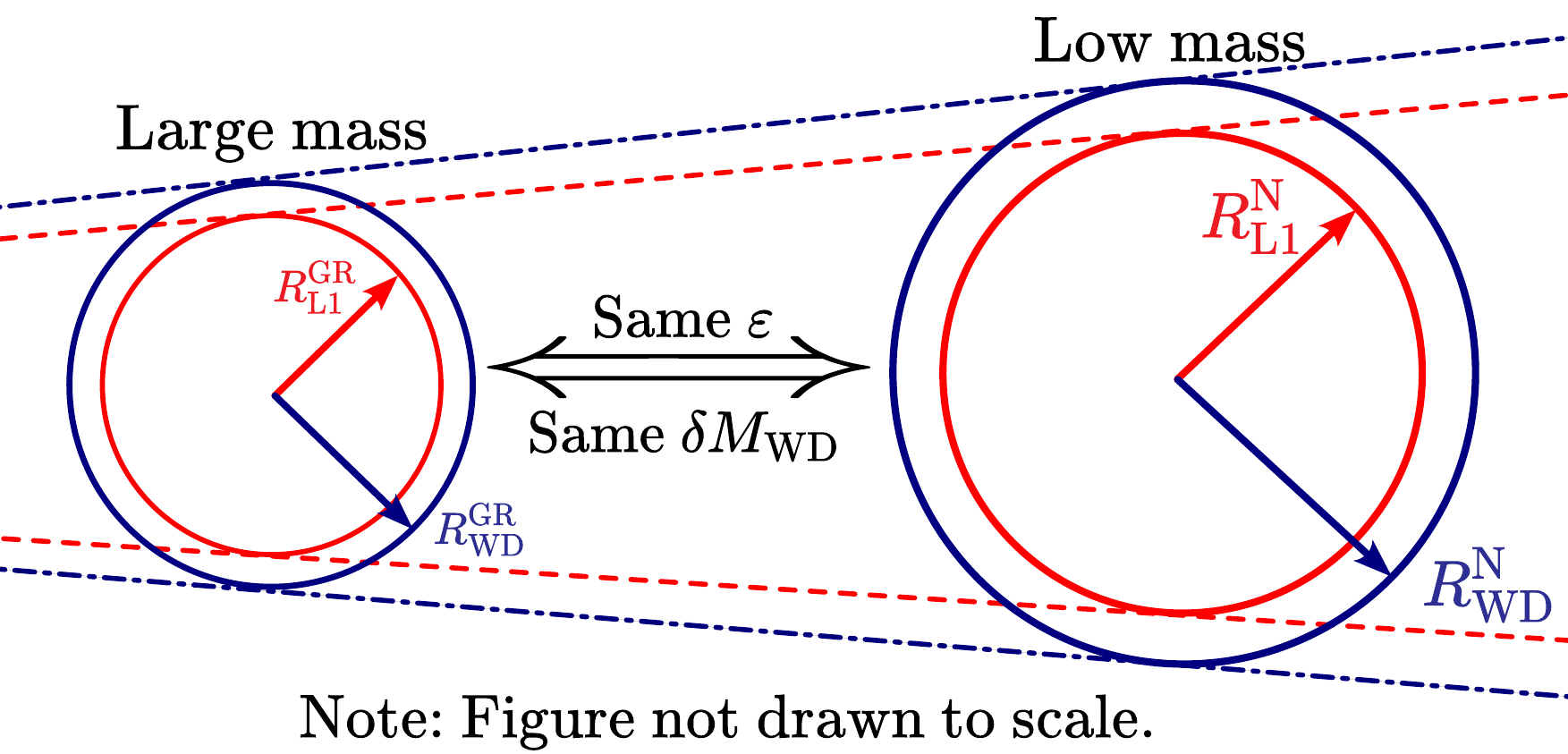}
\begin{minipage}{\linewidth}
\caption{Schematic diagram explaining the mass compensation relation of GR and Newton in Section \ref{sec:GR correction to Roche radius}. GR corrections reduce the Roche radius compared to the Newtonian case at fixed pericentre distance. }
\label{7.6  Schematic diagram of Roche radius correction}
\end{minipage}  
\end{figure}

Importantly, as shown in Figure \ref{7.6  Schematic diagram of Roche radius correction}, at a fixed orbital separation, a more massive WD (with smaller radius) under the relativistically corrected Roche radius yields a similar MT effect (same $\varepsilon$) to a less massive WD (with larger radius) under the Newtonian approximation. 
Thus, the relativistic correction to the Roche radius can be interpreted as effectively increasing the WD mass at a fixed pericentre. We can therefore estimate the additional mass required for the WD to reach the same overflow  $\varepsilon$ as in the Newtonian case.

To achieve the same overflow   $\varepsilon $, the radius of WD should decrease proportionally, i.e. $R_{\mathrm{WD} }^{\mathrm{GR} }/R_{\mathrm{WD} }^{\mathrm{N} }$\,$=$\,$R_{\mathrm{L1} }^{\mathrm{GR} }/R_{\mathrm{L1} }^{\mathrm{N} }$\,$<$\,$1$.  
The relativistic correction to the Roche radius can be estimated by considering the shift in the location of the Lagrange point L1 under relativistic gravity. 
Neglecting the change in the overall shape of the Roche lobe, we assume that the variation in the Roche radius is proportional to the change in the distance between the L1 point and the center of mass of the WD, i.e. $R_{\mathrm{L1}}$\,$\propto$\,$\tilde{X}_1$.  
This approximation offers a practical method to quantify the relativistic modification to the Roche limit without necessitating a comprehensive geometric analysis of the Roche lobe.

By constructing the quadrupole tidal tensor in FNCs, the corresponding gravitational potential can be derived. 
Utilizing the formalism presented by \citet{marck1983solution,ishii2005black,cheng2013relativistic}, the relativistic correction to the position of the Lagrange point L1 in the Schwarzschild metric  is computed by solving  $\nabla \phi^{\mathrm{GR}}$\,$=$\,$0$.  
Here, the corrected potential in FNCs is
\begin{align}\label{68 Calculation of Lagrange potential} 
\phi^{\mathrm{GR}}(D,\tilde{X}_{\mathrm{D}})=-\, \frac{G m_2}{D} \left [ \frac{q}{\tilde{X}_{\mathrm{D}}}+\frac{1}{1-\tilde{X}_{\mathrm{D}}} -\tilde{X}_{\mathrm{D}}+\frac{1}{2} \hat{\mathrm{C}}  \tilde{X}_{\mathrm{D}}^2  \right ],
\end{align} 
where $\hat{\mathrm{C}}$\,$=$\,$3 \hat{L}^2/\hat{D}^2$ (Quadrupole moment correction in GR) and $D$\,$= $\,$\hat{D}G m_2/c^2 $. 
Specifically, the shift in distance between L1 and the WD's centre of mass can be expressed as a series expansion: 
\begin{align}\label{69 Lagrange point relativity correction} 
\tilde{X}_{\mathrm{1} }^{\mathrm{GR} }=\tilde{X}_{\mathrm{1} }^{\mathrm{N} }\left [   1-\mathrm{K}_{\mathrm{1}}\left ( q \right ) \frac{\hat{L}^2}{\hat{r}_{\mathrm{p}}^2 }+\mathrm{K}_{\mathrm{2}}\left ( q \right ) \frac{\hat{L}^4}{\hat{r}_{\mathrm{p}}^4 } +O \left ( \frac{\hat{L}^6}{\hat{r}_{\mathrm{p}}^6 } \right ) \right ],
\end{align} 
where $\hat{r}_{\mathrm{p}}$ is the dimensionless pericenter distance parameter, i.e. $\hat{r}_{\mathrm{p}}$\,$=$\,$\hat{p}/(1+e)$, and $\hat{L}$ is the dimensionless orbital angular momentum. For marginally bound orbits with $\hat{E}\to 1$,  then $\hat{L}^2$\,$\to $\,$\hat{p}^2/(\hat{p}-4)$.  
These dimensionless quantities arise in relativistic geodesic calculations \citep{cheng2013relativistic}.

In the extreme mass-ratio limit ($q$\,$\to$\,$0$),  the coefficients   $\mathrm{K}_{\mathrm{1}}\left ( 0 \right )$\,$=$\,$\mathrm{K}_{\mathrm{2}}\left ( 0 \right )$\,$=$\,$1/2$.  
For pericentre $\hat{r}_{\mathrm{p}}$\,$=$\,$7.875$, the ratio   $R_{\mathrm{L1} }^{\mathrm{GR} }/R_{\mathrm{L1} }^{\mathrm{N} }$\,$\approx$\,$ \tilde{X}_{\mathrm{1} }^{\mathrm{GR} }/\tilde{X}_{\mathrm{1} }^{\mathrm{N} } $\,$\approx$\,$ 0.9$. 
This reflects the fact that the Roche radius in GR is smaller than in the Newtonian case, due to the stronger gravitational field in GR.
From Equation (\ref{eq:RWD}), $R_{\mathrm{WD} }^{\mathrm{GR} }/R_{\mathrm{WD} }^{\mathrm{N} }$\,$=$\,$R_{\mathrm{L1} }^{\mathrm{GR} }/R_{\mathrm{L1} }^{\mathrm{N} }$ implies $M_{\mathrm{WD} }^{\mathrm{GR} }/M_{\mathrm{WD} }^{\mathrm{N} }$\,$=$\,$\left ( R_{\mathrm{WD} }^{\mathrm{GR} }/R_{\mathrm{WD} }^{\mathrm{N} } \right ) ^{-3}$\!$\approx$\,$ 1.43$.
Therefore, a WD with  mass $0.15\mathrm{M} _{\sun }$ (in the Newtonian case) have mass $0.21\mathrm{M} _{\sun }$  in the relativistic case to produce the same transferred mass. 
In this way, relativistic corrections to the tidal field are effectively incorporated.

In addition, the spin of the IMBH modulates the tidal field as a function of latitude, potentially leading to periodic variations in the MT rate near pericenter for inclined (non-equatorial) orbits. 
However, the spin-induced tidal effects are both complex and expected to be subdominant in the regimes considered here. 
For this reason, they are neglected in the present analysis.

\section{Summary and Conclusion}\label{sec:Conclusion}
In this paper, we investigate the orbital evolution of high-eccentricity WD–IMBH systems under the influence of MT.   
We adopt the formulations developed by \citet{sepinsky2007equipotential, sepinsky2007interacting} to derive the governing equations and perform numerical analyses. 
Our treatment of angular momentum follows insights from QPE studies by  \citet{king2020gsn, king2022quasi, king2023angular}.
We give the MT-induced acceleration $\boldsymbol{f}_{\text{MT}}$   using the analysis of the time derivative of orbital angular momentum $J_{\text{orb}}$  (Section \ref{sec:Dynamics under mass transfer}). 
We use  the perturbed Kepler method to calculate the orbital parameter evolution, and use the Roche lobe model to determine the MT rate $\dot{M}_{\text{WD}}$.
Our conclusions come from the analysis of the evolution under these dynamical formulas.

We calculate the evolution where the orbits are relativistically precessed and the IMBH has spin, and analyze the impact of MT on the orbits in this case in Section \ref{sec:Analyzing orbital results without MT in PN method}.  
We find that, despite using the PN approach, the trajectory of the WD closely follows the geodesic motion expected in Kerr spacetime, with only jitter induced by spin-dependent acceleration terms (see Table \ref{tab:average eccentricity} and Figure \ref{Variation of eccentricity er in PN with MT}).
Furthermore, our results suggest that, as predicted,  MT offsets  the effects of gravitational radiation in eccentricity and orbital period,  and greatly promotes the MT process itself. 
The change in overflow parameter far exceeds that from GWs alone,  means the mass stripping process rapidly widens the gap between the WD's radius and its Roche radius.

We find that the effects of gravitational radiation on the semi-major axis $a$, period $T$, and eccentricity $e$ are counteracted by MT. 
When MT and gravitational radiation on the semi-major axis $a$ reach equilibrium, (i.e., the average change rate of $a$ due to the two processes can cancel each other out), the transferred mass $\delta M_{\mathrm{WD}}$ per cycle  in Equation (\ref{44 aMT divided aGW}) is very small, due to the high eccentricity of the orbit.  
Our simulations show that under such conditions, the orbital parameters  $a$, $T$ and $e$ reach minimum values during  evolution,  as shown in Figure \ref{5.2 Variation of the semi-major axis, eccentricity, WD period}. 
When considering only MT and gravitational radiation, and the initial pericenter distance  is far from the tidal radius, the WD will escape  the gravitational potential of the IMBH. 
Our results suggest a qualitatively different picture of long-term orbital evolution compared to previous studies, highlighting the importance of including both MT feedback and relativistic corrections in models of WD–IMBH systems.

We anticipate that the MT process may impart sufficient recoil to cause the WD to escape the gravitational influence of IMBH, which could naturally explain the disappearance of QPE signal  \citep{miniutti2023alive}.
Our analysis suggests that systems with higher per-orbit transferred mass, $\delta M_{\rm WD}$, are more susceptible to rapid escape, shortening the timescale for the WD to leave the system.
We find that the amount of transferred mass is highly sensitive to the difference between the WD radius  and the Roche radius, such that even small variations in this gap can produce large changes in the transferred mass per cycle.

We find that MT effects can imprint observable signatures on the GW signal. 
As MT perturbs the orbital eccentricity and extends the orbital period, a cumulative phase delay emerges when compared with a system driven solely by GW emission.
In particular, we introduce Inequality (\ref{6 1 N equation}), which provides an analytic criterion for estimating the phase shift induced by MT. 
Over a 5-year GW observation period, we compute the accumulated phase shift (Table \ref{tab: table of phase difference}) and the time at which $a$ reaches its minimum (Table \ref{tab: table of minimum value}).
Our results show that, for systems with modest initial eccentricity and sufficiently large pericentre distance, the MT-induced phase shift could be detectable by future GW observatories.

Incorporating MT dynamics into waveform models is thus crucial for accurate source characterization in eccentric EMRIs. 
MT modifies  the phase evolution of the GW signal, affecting key observable parameters such as the component masses, eccentricity, and orbital evolution history. 
It also reveals the intricate interplay between GW-driven inspiral and MT-driven orbital expansion, offering a more complete picture of the lifecycle of WD–IMBH binaries. 
To fully capture the mass transfer signature, joint  electromagnetic and GW observations are essential, with space-based GW detectors triggered alerts playing a critical role in identifying and monitoring these events \citep{sesana2008observing,holley2020getting,torres2024gwnext,Li_2025}.

\section*{Acknowledgements}
We appreciate Yucheng Yin, Han Yan and Di Wang for insightful discussions. 
We also thank the anonymous referee for their constructive and insightful comments, which have improved the clarity and quality of this manuscript.
J.Y. is supported by the National Key Research and Development Program of China (grant No. 2021YFC2203003), the National Natural Science Foundation of China (grant No. 12247101), and the ``111 Center'' under grant No. B20063.
X.C. is supported by the Natural Science Foundation of China (grant No. 12473037).

\section*{Data Availability} 
The data underlying this article will be shared on reasonable request to the corresponding author.





\bibliographystyle{mnras}
\bibliography{example} 




\appendix

\section{A Short Review Of The Perturbed Kepler method}\label{App:A Short Review Of Perturbed Kepler method}
The literature on the Perturbed Kepler Method is extensive
\citep{sterne1960introduction,danbyma,fitzpatrick1970principles,efroimsky2002equations,efroimsky2003gauge,efroimsky2004gauge,fitzpatrick2012introduction,poisson2014gravity,dosopoulou2016orbital1} . 
Here, we provide a brief overview of the formalism \citep{poisson2014gravity}, focusing on approaches that do not employ Lagrangian or Hamiltonian mechanics.

The first time derivative of $\boldsymbol{r}$ and $\boldsymbol{v}$ are:
\begin{equation*} 
\begin{aligned}
 \frac{d \boldsymbol{r}}{d t}=&\frac{\partial \boldsymbol{r}_{\text {Kepler }}}{\partial f} \frac{d f}{d t}+ \sum_{a}\frac{\partial \boldsymbol{r}_{\text {Kepler }}}{\partial \mu^{a}} \frac{d \mu^{a}}{d t}=\boldsymbol{v}_{\text {Kepler }}\left(f(t), \mu^{a}(t)\right),\\
 \frac{d \boldsymbol{v}}{d t}=&\frac{\partial \boldsymbol{v}_{\text {Kepler }}}{\partial f} \frac{d f}{d t}+\sum_{a}\frac{\partial \boldsymbol{v}_{\text {Kepler }}}{\partial \mu^{a}} \frac{d \mu^{a}}{d t}=\boldsymbol{a}, 
\end{aligned}
\end{equation*}
where $\mu^{a}=\left(p , e , \omega , \Omega , \iota \right )$ . 
For Perturbed Kepler method, we further obtain:
\begin{equation*}
\begin{aligned}
&\frac{\partial \boldsymbol{r}_{\text {Kepler }}}{\partial f} \left (  \frac{d f}{d t} -\dot{f} _\mathrm{N} \right )+ \sum_{a}\frac{\partial \boldsymbol{r}_{\text {Kepler }}}{\partial \mu^{a}} \frac{d \mu^{a}}{d t}=\mathbf{0}, \\ 
&\frac{\partial \boldsymbol{v}_{\text {Kepler }}}{\partial f} \left (  \frac{d f}{d t} -\dot{f} _\mathrm{N} \right )+\sum_{a} \frac{\partial \boldsymbol{v}_{\text {Kepler }}}{\partial \mu^{a}} \frac{d \mu^{a}}{d t}=\boldsymbol{f}, 
\end{aligned}
\end{equation*}
where $\dot{f}_\mathrm{N}$ is defined as in Equation  (\ref{fN Wf}). 
The six equations describe the time evolution of six unknown functions $\left(p , e , f, \omega , \Omega , \iota \right )$. 
From a mathematical perspective, each of these functions has solution. 
However,  solving the equations directly is complex and challenging \citep{dosopoulou2016orbital1}.   

The vectors $\boldsymbol{r}_{\text{Kepler}}$ and $\boldsymbol{v}_{\text{Kepler}}$ are not conserved quantities in a classical Kepler orbit,  but they can be combined by various conserved quantities, such as orbital energy $E$ and orbital angular momentum $\boldsymbol{J}_{\text{orb}}$. 
They can yield additional conserved quantities, the orbital period $T$ and the Laplace–Runge–Lenz (LRL) vector $\boldsymbol{A}$. 
By defining an orthogonal frame $(x,y,z)$ with basis vectors $\boldsymbol{e}_x$, $\boldsymbol{e}_y$, $\boldsymbol{e}_z$, where the origin is  at $m_2$, we  align the   $\boldsymbol{e}_x$ with the pericenter and set $\boldsymbol{e}_z$   perpendicular to the orbit plane (different from  basis vectors $\boldsymbol{e}_X$, $\boldsymbol{e}_Y$, $\boldsymbol{e}_Z$). 
In this frame, $\dot{\boldsymbol{e}}_x$\,$ = $\,$\dot{\boldsymbol{e}}_y$\,$ = $\,$\dot{\boldsymbol{e}}_z $\,$= $\,$\boldsymbol{0}$ in a classical Kepler orbit.  
These conserved quantities allow us  to readily derive the $df/dt$ and $d\mu^a/dt$.

We define the energy per unit reduced mass $\mathcal{E} $, the orbital angular momentum per unit reduced mass $\boldsymbol{h}$ and the LRL vector $\boldsymbol{A}$  as follows: 
\begin{align}\label{App: varepsilon h A}
&\mathcal{E} =\frac{E}{\mu}=\frac{v^2}{2} -\frac{G m}{r}, \quad  \boldsymbol{h}=\frac{\boldsymbol{J}_{\text{orb}}}{\mu} =\boldsymbol{r}\times\boldsymbol{v}, \quad
\boldsymbol{A}=\frac{\boldsymbol{v}\times \boldsymbol{h}}{Gm} -\frac{\boldsymbol{r}}{r}.
\end{align}
For Newtonian gravity, we have the following orbital representation:
\begin{align}
r=\frac{h^{2} / G m}{1+\sqrt{1+\frac{2 \mathcal{E}  h^{2}}{G^{2} m^{2}}} \cos f},
\end{align}
where we can define semi-latus rectum $p$ and eccentricity $e$, 
\begin{align}\label{App: p e r}
p=\frac{h^{2}} { G m},\quad e=\sqrt{1+\frac{2 \mathcal{E}  h^{2}}{G^{2} m^{2}}},\quad r=\frac{p}{1+e \cos f}.
\end{align}
We get  $\mathcal{E} $ using the velocity expression $v^2=\dot{r}^{2}+(h /r)^{2}$: 
\begin{align}\label{App: mathcalE}
\mathcal{E} =\frac{\dot{r}^{2}}{2} +\frac{h^{2}}{2 r^{2}}-\frac{G m}{r}.
\end{align}
Therefore, we  obtain $dr/dt$ using Equations (\ref{App: p e r}) and (\ref{App: mathcalE}): 
\begin{align}
\dot{r}= \sqrt{\frac{Gm}{p} }e \sin f.
\end{align}

According to the perturbation theory, the expression of velocity   $\boldsymbol{v}$ remains consistent with Kepler's original form $\boldsymbol{v}_{\text {Kepler }}$:  
\begin{equation}
\begin{aligned}
\boldsymbol{v}=\boldsymbol{v}_{\text {Kepler }}\left(f(t), \mu^{a}(t)\right)=\frac{d }{dt}(r \boldsymbol{n}) =\dot{r} \boldsymbol{n}+ r\dot{f} _\mathrm{N} \boldsymbol{\lambda},
\end{aligned}
\end{equation}
Thus, with $\boldsymbol{r}=r\boldsymbol{n}$, we give the time derivative $d\boldsymbol{n}/dt$: 
\begin{equation}
\begin{aligned}
\frac{d \boldsymbol{n}}{dt} = \dot{f} _\mathrm{N} \boldsymbol{\lambda}. 
\end{aligned}
\label{App: dndt}
\end{equation}
Substituting $\boldsymbol{v}$ into $\boldsymbol{h}$ in Equation (\ref{App: varepsilon h A}), and  using Equation (\ref{App: p e r}), we obtain
\begin{align} \label{App: h l fN}
\boldsymbol{h}=r^2\dot{f} _\mathrm{N}\boldsymbol{\ell},\quad \dot{f} _\mathrm{N} =  \frac{\boldsymbol{h}\cdot\boldsymbol{\ell}}{r^2}=\sqrt{\frac{G m}{p^{3}}}\left (  1+e \cos f\right ) ^{2}.
\end{align}
This is consistent with Equation (\ref{fN Wf}).  In addition,  the vectors $\boldsymbol{n}$, $\boldsymbol{\lambda}$,  $\boldsymbol{\ell}$ can be  expressed in terms of the vectors  $\boldsymbol{e}_x$, $\boldsymbol{e}_y$, $\boldsymbol{e}_z$,
\begin{equation}
\begin{aligned}
\begin{bmatrix}
\boldsymbol{n} \\
\boldsymbol{\lambda}\\
\boldsymbol{\ell}
\end{bmatrix}=\begin{bmatrix}
\ \ \ \cos f  & \sin f & 0\  \\
 -\sin f & \cos f &0\   \\
 0 & 0 &1\  
\end{bmatrix}\begin{bmatrix}
 \boldsymbol{e}_x\\
\boldsymbol{e}_y\\
\boldsymbol{e}_z
\end{bmatrix}.
\end{aligned}
\label{App: matrix e x y z}
\end{equation}
Due to the orthogonality of ($\boldsymbol{n}$, $\boldsymbol{\lambda}$,  $\boldsymbol{\ell}$) and ($\boldsymbol{e}_x$, $\boldsymbol{e}_y$, $\boldsymbol{e}_z$), we can get
\begin{subequations}
\begin{align}
&\left \langle \boldsymbol{n},\dot{\boldsymbol{n} }  \right \rangle  =
\left \langle  \boldsymbol{\lambda},\dot{\boldsymbol{\lambda} } \right \rangle  =
\left \langle  \boldsymbol{\ell},\dot{\boldsymbol{\ell} }  \right \rangle \equiv 0, \\
&\left \langle  \boldsymbol{e}_x,\dot{\boldsymbol{e} }_x  \right \rangle  =
\left \langle  \boldsymbol{e}_y,\dot{\boldsymbol{e} }_y  \right \rangle =
\left \langle  \boldsymbol{e}_z,\dot{\boldsymbol{e} }_z  \right \rangle 
\equiv 0. 
\end{align}
\label{orthonormal bases}
\end{subequations}
The LRL vector $\boldsymbol{A}$ is a eccentricity vector defined by $\boldsymbol{r}$ and $\boldsymbol{v}$:
\begin{align} \label{App: A vectors}
\boldsymbol{A}&=\frac{\boldsymbol{v}\times \boldsymbol{h}}{Gm} -\frac{\boldsymbol{r}}{r}=\left ( \frac{v^2}{Gm} -\frac{1}{r} \right ) \boldsymbol{r}-\frac{\boldsymbol{r}\cdot \boldsymbol{v}}{Gm} \boldsymbol{v}\\
&=\left(\frac{r^{3} \dot{f}_{\mathrm{N}}^{2}}{G m}-1\right) \boldsymbol{n}-\frac{r^{2} \dot{r} \dot{f}_{\mathrm{N}}}{G m} \boldsymbol{\lambda} =e\left (\cos f \boldsymbol{n} -\sin f \boldsymbol{\lambda}  \right )=e \boldsymbol{e}_x.  \nonumber
\end{align}

Therefore, we can establish the relationship between orbital parameters $f$,  $p$,   $e$,   $\omega$,  $\Omega$,  $\iota$, and the conserved quantities $\boldsymbol{h}$, $\boldsymbol{A}$, $\boldsymbol{e}_x$, $\boldsymbol{e}_y$, $\boldsymbol{e}_z$ as follows:
\begin{equation}
\begin{aligned}
p&:=\frac{h^{2}}{G m}, \quad \cos f :=\boldsymbol{n} \cdot \boldsymbol{e}_{x}, \quad  \sin \iota \sin \Omega :=\boldsymbol{e}_{z} \cdot \boldsymbol{e}_{X}, \\
e&:=A, \quad \quad     \cos \iota :=\boldsymbol{e}_{z} \cdot \boldsymbol{e}_{Z}, \quad   \sin \iota \sin \omega  :=\boldsymbol{e}_{x} \cdot \boldsymbol{e}_{Z}. 
\end{aligned}
\end{equation}
Taking the time derivative on both sides of the equations, we get
\begin{equation}
\begin{aligned}
\dot{e}&=\dot{A},\quad \dot{f} =-\frac {\dot{\boldsymbol{n}} \cdot \boldsymbol{e}_{x}+\boldsymbol{n} \cdot \dot{\boldsymbol{e}}_{x}}{ \sin f}, \quad  \dot{\omega}  =\frac{ \left ( \dot{\boldsymbol{e}}_{x} +\cot \iota \sin \omega \ \dot{\boldsymbol{e}}_{z}\right )\cdot \boldsymbol{e}_{Z}}{\sin \iota\cos\omega},\\
\dot{p}&=\frac{2h }{G m}\dot{h},\quad \dot{\iota} =-\frac{\dot{\boldsymbol{e}}_{z} \cdot \boldsymbol{e}_{Z}}{\sin \iota} , \quad \dot{\Omega}=\frac{\dot{\boldsymbol{e}}_{z}\cdot \left ( \boldsymbol{e}_{X} +\cot \iota \sin \Omega \ \boldsymbol{e}_{Z}\right ) }{\sin \iota\cos\Omega}.
\end{aligned}
\label{App:dot p e f}
\end{equation}

Therefore, we now need to relate the external force $\boldsymbol{f}$ to $\dot{\boldsymbol{h}}$, $\dot{\boldsymbol{A}}$, $\dot{h}$, $\dot{A}$, $\dot{\boldsymbol{e}}_z$,  $\dot{\boldsymbol{e}}_x$.  
The form of $\boldsymbol{f}$ is  $\boldsymbol{f}=\mathcal{R } \boldsymbol{n}+ \mathcal{ S } \boldsymbol{\lambda}+\mathcal{ W} \boldsymbol{\ell}$.  
We can calculate $\dot{\boldsymbol{h}}$ and $\dot{\boldsymbol{A}}$ using Equation (\ref{App: varepsilon h A}): 
\begin{subequations}
\begin{align} 
\frac{d \boldsymbol{h}}{dt}&=\frac{d \boldsymbol{r}}{dt} \times  \boldsymbol{v}+\boldsymbol{r} \times  \frac{d \boldsymbol{v}}{dt}=\boldsymbol{r} \times\boldsymbol{f}=-r\left (\mathcal{ S }\boldsymbol{\ell} -\mathcal{ W} \boldsymbol{\lambda} \right )   ,\\
\frac{d \boldsymbol{A}}{dt}&=\frac{\boldsymbol{a} \times(\boldsymbol{r} \times \boldsymbol{v})+\boldsymbol{v} \times(\boldsymbol{r} \times \boldsymbol{a})}{G m}-\left ( \frac{\boldsymbol{v}}{r}-\frac{\dot{r}}{r^{2}} \boldsymbol{r} \right ) \notag\\
&=\frac{\boldsymbol{f} \times\boldsymbol{h}+\boldsymbol{v} \times(\boldsymbol{r} \times \boldsymbol{f})}{G m} \notag\\
&=\frac{1}{Gm} \left [ 2 h  \mathcal{S}\boldsymbol{n}-(h\mathcal{R}+r \dot{r}\mathcal{S})\boldsymbol{\lambda}-r \dot{r}\mathcal{W}\boldsymbol{\ell}  \right ]\\
&=\frac{1}{Gm} \left \{ \left [  2 h  \mathcal{S}\cos f+(h\mathcal{R}+r \dot{r}\mathcal{S})\sin f\right ]\boldsymbol{e}_x\right. \notag\\
&\quad \quad \ \, \, \left.+ \left [  2 h  \mathcal{S}\sin f-(h\mathcal{R}+r \dot{r}\mathcal{S})\cos f\right ]\boldsymbol{e}_y -r \dot{r}\mathcal{W}\boldsymbol{e}_z\right \} .  \notag
\end{align}
\label{App: dvhdt dvAdt}
\end{subequations}
We then  calculate $\dot{h}$ and $\dot{A}$ using Equations (\ref{App: h l fN}), (\ref{App: A vectors}) and (\ref{App: dvhdt dvAdt}), along with the orthogonality condition  (\ref{orthonormal bases}):
\begin{subequations}
\begin{align} 
\frac{d h}{dt}&=\frac{d \boldsymbol{h}}{dt} \cdot \boldsymbol{\ell}+\boldsymbol{h} \cdot  \frac{d \boldsymbol{\ell}}{dt}=\frac{d \boldsymbol{h}}{dt} \cdot \boldsymbol{\ell}=r\mathcal{ S } ,\\
\frac{dA}{dt}&=\frac{d \boldsymbol{A}}{dt} \cdot \boldsymbol{e}_x+\boldsymbol{A} \cdot  \frac{d \boldsymbol{e}_x}{dt}=\frac{d \boldsymbol{A}}{dt} \cdot \boldsymbol{e}_x \notag \\
&=\frac{1}{Gm} \left [2 h  \mathcal{S}\cos f+(h\mathcal{R}+r \dot{r}\mathcal{S})\sin f \right ]. 
\end{align}
\label{App: dhdt dAdt}
\end{subequations}
And calculate $\dot{\boldsymbol{e}}_z$\,$=$\,$d\boldsymbol{\ell}/dt$ (Equation \ref{App: matrix e x y z}) and $\dot{\boldsymbol{e}}_x$ using Equations (\ref{App: dvhdt dvAdt}) and (\ref{App: dhdt dAdt}): 
\begin{subequations}
\begin{align} 
\frac{d \boldsymbol{e}_z}{dt}&=\frac{1}{h} \left ( \frac{d \boldsymbol{h}}{dt}-\frac{d h}{dt}  \boldsymbol{e}_z \right )=-\frac{r\ \mathcal{ W }}{h}\boldsymbol{\lambda},   \label{App: vector change a}\\
\frac{d \boldsymbol{e}_x}{dt}&=\frac{1}{A} \left ( \frac{d \boldsymbol{A}}{dt}-\frac{d A}{dt}  \boldsymbol{e}_x \right )=\frac{1}{GmA}\times \notag\\
&\quad \left \{    \left [  2 h  \mathcal{S}\sin f  -(h\mathcal{R}+r \dot{r}\mathcal{S})\cos f\right ]\boldsymbol{e}_y -r \dot{r}\mathcal{W}\boldsymbol{e}_z\right \}  \label{App: vector change b}.
\end{align}
\label{App: vector change}
\end{subequations}

Therefore,  the Equation (\ref{App:dot p e f}) changes to
\begin{subequations}
\begin{align}
\dot{p} & =\frac{2h }{G m}r \mathcal{ S },  \label{App: dpefdt a}\\
\dot{e}  &=\frac{h}{Gm} \left [ \sin f\mathcal{R}+\left ( 2  \cos f+ \frac{r \dot{r}}{h}\sin f \right ) \mathcal{S}  \right ], \label{App: dpefdt b}\\
\dot{f}  & =\frac{h}{r^2} +\frac{h}{G m A }  \left [ \cos f \mathcal{R}-\left ( 2  \sin f- \frac{ r\dot{r} }{h}\cos f \right )\mathcal{S}  \right ] ,  \label{App: dpefdt c}\\
\dot{\iota}  & =\frac{r}{h}\cos (\omega +f)\mathcal{W} ,  \label{App: dpefdt d}\\
\dot{\Omega}  & =\frac{r}{h}\sin (\omega +f)\csc \iota \mathcal{W}   ,  \label{App: dpefdt e}\\
\dot{\omega}  & =\frac{h}{G m A }  \left [ -\cos f \mathcal{R}+\left ( 2  \sin f- \frac{ r\dot{r} }{h}\cos f \right )\mathcal{S} \right.  \notag\\
 &\quad \quad \ \ \left. -e \cot\iota \frac{Gm r}{h^2  }   \left ( \frac{h \dot{r} \sec \omega}{GmA} +\tan \omega \cos (\omega +f) \right ) \mathcal{W} \right ] . \label{App: dpefdt f}
\end{align}
\label{App: dpefdt}
\end{subequations}
From Equations (\ref{App: dndt}) and (\ref{App: vector change}), we obtain $d\boldsymbol{n} /dt$ and $d\boldsymbol{\ell}/dt$. We can also get  $d\boldsymbol{\lambda}/dt$ in terms of $\boldsymbol{e}_x$\,$=$\,$\cos f \boldsymbol{n}$\,$-$\,$\sin f \boldsymbol{\lambda}$ (Equation \ref{App: matrix e x y z}), using $\dot{\boldsymbol{e} }_x$ (\ref{App: vector change b}), $\dot{\boldsymbol{n}}$ (\ref{App: dndt}) and $\dot{f}$ (\ref{App: dpefdt c}):
\begin{equation}
\begin{aligned}
\frac{d \boldsymbol{\lambda} }{dt} &=-\frac{\dot{\boldsymbol{e} }_x+\dot{f} \left ( \sin f\boldsymbol{n}+\cos f \boldsymbol{\lambda} \right ) - \cos f \dot{\boldsymbol{n}} }{\sin f}\\
&=-\dot{f} _\mathrm{N}\boldsymbol{n}+ \frac{r\dot{r}\mathcal{W} }{GmA}  \csc f  \boldsymbol{\ell}  .
\end{aligned}
\label{App1: dlambdadt}
\end{equation}

We obtain the final form in Equations (\ref{eq:elements differential equation11}) and (\ref{eq:elements differential equation22}) by substituting the following expressions into Equations (\ref{App: dpefdt}), (\ref{App: dndt}), (\ref{App1: dlambdadt}) and (\ref{App: vector change}):
\begin{align}
r &=\frac{p}{1+e \cos f},  \quad \dot{r}=\sqrt{\frac{Gm}{p} }e \sin f,\\
\dot{f} _\mathrm{N}  &=\sqrt{\frac{G m}{p^{3}}}\left (  1+e \cos f\right ) ^{2}, 
\quad h =\sqrt{Gmp}, \quad A=e . \nonumber
\end{align}

\section{Simplification of the mass transfer model}\label{App: Simplification of the mass transfer model}
In this appendix, we simplify Equation (\ref{eq:59 dotm1}) by considering only the process of Roche lobe overflow. 
We   cancel the first and third terms in Equation (\ref{eq:59 dotm1}), beacuse the density at the Roche radius is much larger than the surface density  ($\rho_\text{L}$\,$\gg $\,$ \rho_{\text{ph}}$)   and $ K^{1/\Gamma}/e $\,$\sim $\,$ F(K,\Gamma)^2 $  for $\Gamma$\,$=$\,$5/3$.  
We get  
\begin{equation} 
\begin{aligned}
\dot{m}_1 \approx -H(\varepsilon )\frac{2 \pi }{\sqrt{B C}}F(K,\Gamma)  K^{\frac{3\Gamma -1}{2\Gamma } } \rho_\text{L}^{\frac{3\Gamma -1}{2 } },
\end{aligned}
\label{App2: dotm1}
\end{equation}
where $\rho_\text{L}$ is $\rho(\tilde{Z}=R_{\text{WD}}-R_{\text{L}})$ can be given by Equation (\ref{eq:56 rho}), and  $2 \pi/\sqrt{B C}$ can  be written in series form:
\begin{equation} 
\begin{aligned}
&\frac{2 \pi }{\sqrt{B C}}=\frac{2 \pi D^3}{G m_2}\frac{1}{ \sqrt{\Re\ [\Re -A(1+q)]} } :=\frac{2 \pi D^3}{G m_2}\frac{1}{\sqrt{\mathcal{R}_0 } }  \left(  \sum_{m=0}^{\infty } \mathcal{S}_m \varepsilon ^m\right), 
\end{aligned}
\label{App2: 2 pi BC}
\end{equation}
where $\Re$ is defined as 
\begin{equation} 
\Re:=\frac{q}{\tilde{X}_1^3} +\frac{1}{(1-\tilde{X}_1)^3}.
\label{eq: 60 Re}
\end{equation}
Here, dimensionless parameter $\tilde{X}_1$,  satisfying $0<\tilde{X}_1\ll 1$, represents the fractional distance from the CM of the WD to the first Lagrangian point L1. The position of L1 is determined by the following equation, as presented in \citet{sepinsky2007equipotential}:
\begin{align}\label{29 equaiton of X1}
\frac{q }{ \tilde{X}_{1}{} ^{2}}+\frac{1}{(1-\tilde{X}_1)^2} +1- A(1+q)\tilde{X}_{1}=0.
\end{align}

Equation (\ref{eq:56 rho}) can be organized into the following form: 
\begin{equation} 
\begin{aligned}
\rho_{\text{L}}^{\frac{3\Gamma -1}{2 } }=\left (  \frac{\Gamma-1}{K \Gamma}\frac{G m_{2}}{D}  C_{\text{L}}\right )^{\frac{3\Gamma -1}{2(\Gamma -1) } } \left (  \hat{ \rho}_{\text{ph}}^{\Gamma -1}+\lambda \right )^{\frac{3\Gamma -1}{2(\Gamma -1) } },
\end{aligned}
\label{App2: 3}
\end{equation}
where $\lambda$\,$:=$\,$\lambda_1+\lambda_O$\,$:=$\,$\sum_{n=1}^{\infty} D_{\varepsilon n} \varepsilon^{n}$, specifically $\lambda_1$\,$=$\,$D_{\varepsilon 1} \varepsilon$. From Equation (\ref{65 fMT}),   $\hat{ \rho}_{\text{ph}}^{\Gamma -1}$ can be defined as
\begin{equation} 
\begin{aligned}
\hat{ \rho}_{\text{ph}}^{\Gamma -1}:=\left ( \frac{\Gamma-1}{K \Gamma}\frac{G m_{2}}{D} C_{\text{L}} \right ) ^{-1}\rho_{\text{ph}}^{\Gamma -1}=\frac{q}{C_{\mathrm{L} }^2} \frac{ \xi_{0} }{1- \xi_{0}}\left ( 1-\varepsilon  \right ).
\end{aligned}
\label{App2: 4}
\end{equation}
Define $\alpha$ to re-express the variable $\hat{ \rho}_{\text{ph}}^{\Gamma -1}$, 
\begin{equation} 
\begin{aligned}
\frac{q}{C_{\mathrm{L} }^2} \frac{ \xi_{0} }{1- \xi_{0}} :=\frac{D_{\varepsilon 1} }{1-\alpha}, \ \ \mathrm{where} \ \, \alpha:=1-\left(\frac{q}{C_{\mathrm{L}}^{2}} \frac{\xi_{0}}{1-\xi_{0}}\right)^{-1} D_{\varepsilon 1}.
\end{aligned}
\label{App2: 5}
\end{equation}
We can assume that $\lambda_1+\hat{ \rho}_{\text{ph}}^{\Gamma -1}$\,$\gg$\,$ \lambda_O$ near pericenter. 
Let us define a new series to represent  $\hat{ \rho}_{\text{ph}}^{\Gamma -1}+\lambda$,  
\begin{equation} 
\begin{aligned}
\left (   \lambda_1+\hat{ \rho}_{\text{ph}}^{\Gamma -1}+\lambda_O \right )^{\frac{3\Gamma -1}{2(\Gamma -1) } }:= \left(\sum_{n=0}^{\infty} \mathcal{U}_{\varepsilon n} \varepsilon^{n}\right).
\end{aligned}
\label{App2: 6}
\end{equation}
We can take a first-order expand with respect to  $\lambda_{O}$:
\begin{align}\label{App2: 7}
& \left(\sum_{n=1}^{\infty} \mathcal{U}_{\varepsilon n} \varepsilon^{n}\right) \simeq   \left(D_{\varepsilon 1}\frac{1-\alpha\varepsilon}{1-\alpha} \right)^{\frac{3\Gamma -1}{2(\Gamma -1) } }\left(1+\frac{3\Gamma -1}{2(\Gamma -1) }\frac{  \lambda_{O}}{\lambda_{1}+\hat{\rho}_{\mathrm{ph}}^{\Gamma-1}}\right), \nonumber \\
&  \frac{  \lambda_{O}}{\lambda_{1}+\hat{\rho}_{\mathrm{ph}}^{\Gamma-1}}=\left(D_{\varepsilon 1}\frac{1-\alpha\varepsilon}{1-\alpha} \right)^{-1}\left(\sum_{n=2}^{\infty} D_{\varepsilon n} \varepsilon^{n}\right). 
\end{align}
Therefore, the coefficient $\mathcal{U}_{\varepsilon n}$ are given analytically. 

We define $\mathcal{S}_0$\,$\equiv $\,$1$, so that $\mathcal{R}_0$ is the zeroth-order approximation of $\Re\, [\Re -A(1+q)]$ with respect to  $\varepsilon$. When $q$\,$\ll$\,$ 1$, the simplest solution to Equation (\ref{29 equaiton of X1}) for $\tilde{X}_1$ is 
\begin{equation} 
\begin{aligned}
&\tilde{X}_1 \simeq \left (  \frac{q}{2+A(1+q)} \right ) ^{1/3} \! = \left (  \frac{q}{2+\varpi (1-\varepsilon)^3 } \right ) ^{1/3}  \! . 
\end{aligned}
\label{App2: 8} 
\end{equation}
Substitute $\tilde{X}_1$ into Equation (\ref{eq: 60 Re}),  we obtain   $\mathcal{R}_0$:
\begin{equation} 
\begin{aligned}
\mathcal{R}_0 \simeq  \left [  2+\left(1-\left(\frac{q}{2+\varpi }\right)^{1 / 3}\right)^{-3}\right ] \left [  2+\varpi +\left(1-\left(\frac{q}{2+\varpi }\right)^{1 / 3}\right)^{-3}\right ].
\end{aligned}
\label{App2: 9} 
\end{equation}
Here, dimensionless $\varpi$ is 
\begin{equation} 
\begin{aligned}
&\varpi =C_{\mathrm{L}}^{-3}\left ( \frac{R_{\mathrm{WD}}\Omega _{1}}{c}  \right ) ^2 \left ( \frac{G M_{\mathrm{IMBH} }}{c^2 R_{\mathrm{WD}}}  \right ) ^{-1} \! \simeq \frac{5}{4896}\frac{q}{C_{\mathrm{L}}^3 } \frac{M_{\mathrm{WD} }^{4/3}}{M_{\sun}M_{\mathrm{Ch}}^{1/3}}\left (  \frac{\chi _{1}}{\tilde{\gamma } }  \right )^2 \! , 
\end{aligned}
\label{App2: 10} 
\end{equation}
where $\tilde{\gamma } $ the difference between the WD's moment of inertia and that of a uniform-density sphere. 
It can be related to $\boldsymbol{S}_1$: 
\begin{equation} 
\begin{aligned}
&\left | \boldsymbol{S}_1 \right |:=\frac{G M_{\mathrm{WD} }^2}{c}\chi _{1}  :=\frac{2}{5}\tilde{\gamma } M_{\mathrm{WD}} R_{\mathrm{WD}}^2 \Omega _{1}.
\end{aligned}
\label{App2: 10 1} 
\end{equation}
With $\mathcal{R}_0$ in Equation (\ref{App2: 9}), we can calculate $\mathcal{S}_m$ from Equation (\ref{App2: 2 pi BC}). 
Taking the first-order expand of  $\mathcal{S}_m$ about $\varpi$  gives a  simple expression:
\begin{equation*}  
\begin{aligned} 
&\mathcal{S}_0\simeq 1,\ \mathcal{S}_1\simeq \frac{3}{4} \varpi  Q,\ \mathcal{S}_2\simeq -\frac{3}{4} \varpi  Q,\ \mathcal{S}_3\simeq \frac{1}{4} \varpi  Q,\ \mathcal{S}_4\simeq \mathcal{O}(\varpi ^2),
\end{aligned} 
\label{App2: 11} 
\end{equation*}
where
$$ Q=\frac{4-10 \times 2^{2 / 3} q^{1 / 3}+12 \times 2^{1 / 3} q^{2 / 3}-8 q+2^{2 / 3} q^{4 / 3}}{6-9 \times 2^{2 / 3} q^{1 / 3}+12 \times 2^{1 / 3} q^{2 / 3}-8 q+2^{2 / 3} q^{4 / 3}}\simeq \frac{2}{3}.$$
Thus, $\mathcal{S}_m$  are entirely influenced by the spin of the WD.

Next, we only need to calculate the  $\sum_{n=1}^{\infty} D_{\varepsilon n} \varepsilon^{n}$. Since $\lambda $  is given by Equation  (\ref{eq:56 rho}), we have 
\begin{align}\label{App2: 12} 
\lambda= &\left (  \frac{G m_{2}}{D}  C_{\text{L}}\right )^{-1} \left \{ \frac{G m_1}{R_{\mathrm{WD}}} \left ( \frac{R_{\mathrm{WD}}}{R_{\mathrm{WD}}-\tilde{Z}} -1  \right )-\frac{1}{2} \Omega _{1}^2(2 R_{\mathrm{WD}}-\tilde{Z})  \tilde{Z} \right. \nonumber \\
&\left.+\frac{G m_2}{D}\left [ \frac{\tilde{Z}}{D} +\frac{D}{D-R_{\mathrm{WD}}+\tilde{Z}} -\frac{D}{D-R_{\mathrm{WD}}} \right ]   \right \}.
\end{align}
where $\tilde{Z}=R_{\mathrm{WD}}-C_{\mathrm{L}}D=R_{\mathrm{WD}}\varepsilon$. Thus, $\lambda$ can be expressed as 
\begin{equation} 
\begin{aligned}
\lambda= \sum_{n=1}^{\infty} D_{\varepsilon n} \varepsilon^{n}&= \frac{q}{C_{\mathrm{L} }} \varepsilon -\frac{\varpi C_{\mathrm{L} }}{2}\left (  2-\varepsilon\right )  \left ( 1-\varepsilon \right ) \varepsilon \\
&  +\frac{ \varepsilon }{1- \varepsilon }\left [ 1-\frac{ 1- \varepsilon}{(1- \varepsilon -C_{\mathrm{L} })(1-C_{\mathrm{L} })} \right ].
\end{aligned}
\label{App2: 13} 
\end{equation}
Therefore, $D_{\varepsilon n}$ is given by
\begin{equation} 
\begin{aligned} 
&D_{\varepsilon n}=1-\frac{1}{(1-C_\text{L})^{n+1}}+\frac{q}{C_\text{L}{}^2} \delta _{n1}  -\frac{\varpi C_\text{L}}{2} \left ( 2\delta _{n1} -3\delta _{n2} +\delta _{n3} \right ).
\end{aligned}
\label{App2: 14} 
\end{equation}

\section{Transferred mass for every orbital period}\label{App: Calculate transferred mass every period}
Following  Appendix \ref{App: Simplification of the mass transfer model}, we adopt a simplified form for $\dot{m}_1$:
\begin{equation} 
\begin{aligned}
& \dot{m}_1:=C_{\dot{m}_1} \sum_{n=0}^{\infty } C_{\mathrm{\varepsilon n} }\varepsilon^n, 
\end{aligned}
\label{App3: 1} 
\end{equation}
where $C_{\dot{m}_1}$ and $C_{\mathrm{\varepsilon n} }$ are defined as
\begin{subequations}
\begin{align}
&C_{\dot{m}_1}=- 2  \pi  W(K,\Gamma) \cdot\left ( G M_{\mathrm{IMBH} } \right )^2   C_\text{L}^3/\sqrt{\mathcal{R}_0 },\\ 
&\sum_{n=0}^{\infty } C_{\mathrm{\varepsilon n} } \varepsilon^n =\left ( \sum_{m=0}^{\infty } \mathcal{S}_m \varepsilon ^m    \right )  \left ( \sum_{n=0}^{\infty } \mathcal{U}_{\varepsilon n}\varepsilon ^n     \right ). 
\end{align}
\label{App3: 2} 
\end{subequations}
 
 We can express the integral of $\dot{m}_1$ for a Keplerian elliptical orbit:
\begin{equation} 
\begin{aligned}
& \int_{-\pi}^{\pi}  \dot{m}_{1} \dot{f}_{\text{N} }^{-1}  df= C_{\dot{m}_1} \sqrt{\frac{p^3}{Gm} } \sum_{n=0}^{\infty } C_{\mathrm{\varepsilon n} }  \int_{-f_0}^{f_0}\frac{\varepsilon^n df}{(1+e \cos f)^{2}}. 
\end{aligned}
\label{App3: 3} 
\end{equation}
Here, $\varepsilon^n$  can be rewritten as a function of the angle $f$:
\begin{equation} 
\begin{aligned}
  \varepsilon^n&=\left ( 1- \frac{p C_{\mathrm{L} }}{R_{\mathrm{WD} }}\frac{1}{1+e \cos f}   \right ) ^n\\
&=     \sum_{m=0}^{n } \begin{pmatrix}
 n\\
m
\end{pmatrix}\left ( -\frac{pC_{\mathrm{L} } }{R_{\mathrm{WD} }}  \right ) ^m\frac{1}{(1+e \cos f)^{m }}. 
\end{aligned}
\label{App3: 4} 
\end{equation}

Therefore, $\delta M_{\mathrm{WD}}$ can be written as  
\begin{align}\label{App3: 5}
 \delta M_{\mathrm{WD}}= &\ C_{\dot{m}_1} \sqrt{\frac{p^3}{Gm} } \sum_{n=0}^{\infty } C_{\mathrm{\varepsilon n} }\times \\
 & \left [ \sum_{m=0}^{n } \begin{pmatrix}
 n\\
m
\end{pmatrix}\left ( -\frac{pC_{\mathrm{L} } }{R_{\mathrm{WD} }}  \right ) ^m\int_{-f_0}^{f_0}\frac{df}{(1+e \cos f)^{m+2}} \right ] . \nonumber
\end{align}
The integration result is the Appell Hypergeometric Function $\mathrm{AF1}(\alpha ; \beta ,\beta ^{\prime}; \gamma ;x,y)$ of two variables, with a  series definition \citep{DLMF}:
\begin{equation} 
\begin{aligned}
\mathrm{AF1}(\alpha ; \beta ,\beta ^{\prime}; \gamma ;x,y)=\sum_{m=0}^{\infty} \sum_{n=0}^{\infty} \frac{(\alpha)_{m+n} (\beta)_{m}(\beta^{\prime})_{(n)}}{(\gamma )_{m+n}\, m!\ n!} x^{m} y^{n} ,  
\end{aligned}
\label{App3: 6} 
\end{equation}
where $(q)_{n}$ denotes the Pochhammer symbol. The integral evaluates to 
\begin{align}\label{App3: 7} 
\int_{-f_0}^{f_0}&\frac{df}{(1+e \cos f)^{m+2}}=\frac{4}{(1+e)^{m+2}} \sin \left ( \frac{f_0}{2}  \right )  \ \times \\
&\quad \mathrm{AF1}\left [\frac{1}{2} ;\frac{1}{2} ,m+2;\frac{3}{2}; \sin^2\left ( \frac{f_0}{2}  \right ),\frac{2e}{1+e}\sin^2 \left ( \frac{f_0}{2}  \right )\right ]. \nonumber
\end{align}

Thus,  we require the value of $\sin (f_0/2)$. 
Since we substitute $D$ with  $p/(1+e \cos f)$, $f_0$ satisfies $R_{\text{WD}}$\,$=$\,$C_{\text{L}}D(f_0)$. 
To relate  $p$ and $R_{\text{WD}}$, we use $\epsilon$  as defined in Section \ref{sec:Comparison of the effects of MT and gravitational radiation on orbit}:
\begin{equation} 
\begin{aligned}
r_{\mathrm{p}}=\frac{p}{1+e}:=(1-\epsilon )\frac{R_{\mathrm{WD}} }{C_{\mathrm{L}}} .
\end{aligned}
\label{App3: 8} 
\end{equation}
Therefore, $f_0$ and $\sin (f_0/2)$ can be written as 
\begin{align} \label{App3: 9} 
&f_0= \sin ^{-1}\left [\sqrt{1- \left ( 1- \frac{1+e}{e} \epsilon  \right )^2 } \right ],  \\
 &\sin \left ( \frac{f_0}{2}  \right ) =\sin\left \{ \frac{1}{2}  \sin ^{-1}\left [\sqrt{1- \left ( 1- \frac{1+e}{e} \epsilon  \right )^2 } \right ]\right \}\equiv \sqrt{\frac{ 1+e}{2e}\epsilon }. \nonumber 
\end{align}
Substituting $\sin (f_0/2)$  into Equation (\ref{App3: 7}) yields Equation (\ref{59 Calculate transferred mass every period 8}).

In  Schwarzschild spacetime,  $dt/d\psi$  can be used to substitute $\dot{f}_{\text{N} }^{-1} $ \citep{cutler1994gravitational,martel2004gravitational}. The primary geodesic equation is: 
\begin{equation} 
\begin{aligned}
&\frac{dt}{d\psi } =\frac{Gm}{c^3} \frac{\hat{p} ^{3/2}}{\hat{x}^2}  \frac{\hat{p} \sqrt{\hat{p} -4} }{ (\hat{p} -2 \hat{x})\sqrt{\hat{p} -4-2 \hat{x}}} \sqrt{1+\frac{4(1-e^2)}{\hat{p} (\hat{p} -4) }},
\end{aligned}
\label{App3: 10} 
\end{equation}
where $\hat{p}$\,$:=$\,$p/(Gm/c^2)$ and $\hat{x}$\,$:=$\,$1+e \cos \psi$ are dimensionless quantities. 
The  pericenter $r_{\text{p}}$, apocenter $r_{\text{a}}$,   eccentricity $e$ are defined as:
\begin{align}\label{App3: 11}
&r_{\text{p}}= \frac{G m}{c^2} \frac{\hat{p}}{1+e},\quad r_{\text{a}}= \frac{G m}{c^2} \frac{\hat{p}}{1-e}, \quad  e=\frac{r_{\text{a}}-r_{\text{p}}}{r_{\text{a}}+r_{\text{p}}}.
\end{align} 
Equation (\ref{App3: 8}) remains valid. The integral variable in Equation (\ref{App3: 3}) is changed from $f$ to $\psi$, with $\psi_0$   satisfying Equation (\ref{App3: 9}). For $\hat{x} $\,$\rightarrow$\,$ 0$, a Taylor expansion of Equation (\ref{App3: 10}) is:
\begin{equation} 
\begin{aligned}
\frac{\hat{p} \sqrt{\hat{p} -4} }{ (\hat{p} -2 \hat{x})\sqrt{\hat{p} -4-2 \hat{x}}}:=\sum_{k=0}^{\infty } F_{k} (\hat{p}) \hat{x}^k,\quad \text{where} \ \ F_{0}=1. 
\end{aligned}
\label{App3: 12} 
\end{equation}
Transferred mass $\delta M_{\mathrm{WD}}$ in Equation (\ref{App3: 5}) then becomes
\begin{align}\label{App3: 13} 
\delta M_{\mathrm{WD}} &=  \sum_{k=0}^{\infty }\left [   F_{k} (\hat{p})C_{\dot{m}_1}\frac{Gm}{c^3} \hat{p}\sqrt{\frac{(\hat{p}-2)^2-4e^2}{\hat{p}-4 }} \times \right. \\
&\left. \sum_{n=0}^{\infty } C_{\mathrm{\varepsilon n} } \sum_{m=0}^{n } \begin{pmatrix}
 n\\
m
\end{pmatrix}\left ( -\frac{Gm}{c^2}\frac{\hat{p}C_{\mathrm{L} } }{R_{\mathrm{WD} }}  \right ) ^m\int_{-\psi_0}^{\psi_0}\frac{d\psi}{\hat{x}^{(m-k)+2}} \right ].  \nonumber
\end{align} 
We continue using Equation (\ref{App3: 7}) for the integral, so Equation (\ref{App3: 13}) reduces to Equation (\ref{65 Calculate transferred mass every period 14}).

\section{Equivalence of Phase and Period Evolution in the Adiabatic Approximation}\label{App: Equivalence of Phase and Periodic Evolution in the Adiabatic Approximation}
As described in Section \ref{sec:Impact on GW sigal and detection}, our objective is to detect the cumulative phase shift $\Delta u$ between systems with and without MT over a finite observation time $t_{\text{obs}}$. 
Importantly, an exact integration of the orbital phase evolution is not required; rather, an estimate of the orbital period for each cycle suffices for our analysis.
 
We adopt the standard formalism for eccentric Keplerian orbits. Specifically, we express the true anomaly $f(t)$ as a function of the eccentric anomaly $u$, where the radial distance is given by: $r$\,$=$\,$a(1$\,$-$\,$e\cos u)$, and the time evolution is governed by Kepler’s equation: $2 \pi (t-t_{\mathrm{p} } )/T$\,$=$\,$u-e\sin u$, with $t_{\mathrm{p} }$\,$\le$\,$  t$ denoting the time of the most recent pericentre passage.
The GW waveform $h(f(t))$, initially expressed in terms of the true anomaly $f$, can be reparameterised in terms of $u$. Therefore, evaluating the GW phase shift $\Delta f(t)$ is equivalent to computing $\Delta u(t)$. The relation between $u$ and $t$ can be  expanded using a Fourier series involving Bessel functions of the first kind $J_{n}(x)$ \citep{murray1999solar}:
\begin{equation}
\begin{aligned}
 &u(t,\text{N})\simeq 2 \pi(\text{N}-1)+l_{\text{N}} +\mathcal{J}(l_{\text{N}}, e_{\text{N}}),
\end{aligned}
\label{App4: u l}
\end{equation}
where
\begin{align}\label{App4: l J}
 &l_{\text{N}}:=2 \pi\left ( t-t_{\text{pN-1}} \right )  /T_{\text{N}},\quad t_{\text{pN-1}}:=\sum_{\text{k}=1}^{\text{N}-1}T_{\text{k}},\\
 &\mathcal{J}(l_{\text{N}}, e_{\text{N}}):= \sum_{n=1}^{\infty } \frac{2}{n} J_{n}(ne_{\text{N}} )\sin nl_{\text{N}}.  
\end{align}
Here, $e_{\text{N}}$ and $T_{\text{N}}$ represent the eccentricity  and period  of the N-th cycle, $\mathcal{J}(l_{\text{N}}(t), e_{\text{N}})$ represent the nonlinear part of the phase evolution caused by eccentricity.  
Assuming  $e_{\text{k}}$ and $T_{\text{k}}$ are approximately constant within the k-th cycle, the Bessel functions show  the difference between elliptical and circular orbits.

However, the phase difference $\Delta u(t)$ is not increase monotonically according to the Bessel function. 
It suddenly increases near pericenter and decreases slowly near apocenter, but satisfies  $u(t_{\text{pN}})-u(t_{\text{pN-1}})$\,$=$\,$2\pi$. 
The phase difference $\Delta u(t)$ does not capture the overall effect of the phase change. 
When $\left [u(t,\text{N})-u(t_{\text{pN}},\text{N})\right ]/u(t_{\text{pN}},\text{N})$\,$\ll$\,$ 1$ (That is, the phase difference near the pericenter), we disregard  the series of Bessel functions, yielding: 
\begin{equation}
\begin{aligned}
 &\frac{u(t,\text{N})}{2 \pi}  \simeq \text{N}-1 +\left ( t-t_{\text{pN-1}} \right )  /T_{\text{N}}  .
\end{aligned}
\label{App4: u l 2}
\end{equation}
The phase difference with and without the MT process is
\begin{equation}
\begin{aligned}
\frac{\Delta u}{2 \pi} \simeq  t\left ( \frac{1}{T_{\text{N}}} -\frac{1}{T^*_{\text{N}}}  \right )-\left ( \frac{  t_{\text{pN-1}}  }{T_{\text{N}}} -\frac{  t_{\text{pN-1}}^* }{T_{\text{N}}^*}  \right )<0,
\end{aligned}
\label{App4: Delta u}
\end{equation}
 where the superscript asterisk indicates that MT process is not considered. 
 Due to the $\left \langle   \dot{T}^{\mathcal{B}}_{\text{MT}}\right \rangle_{\text{sec}}$ in Equation (\ref{5.2 long-term orbital effects of MT discrete method 0}) is positive, the MT process leads to a cumulative increase in the orbital period over successive cycles,  $T_{\text{N}}\ge T_{\text{N}}^*$ and $t_{\text{pN}}\ge t_{\text{pN}}^*$ always holds. 
 
Therefore, the inequality about the  phase difference $\Delta u_g$\,$<$\,$0$ and equation  about required time $t_g$ for our goal (i.e. $\Delta u(t_g)$\,$=$\,$\Delta u_g$)  are
\begin{subequations}
\begin{align}
& \frac{t_{\text{pN-1}} -t_{\text{pN-1}}^*}{T_{\text{N}}^*}  \le \frac{\left |  \Delta u_g\right |}{2 \pi} \le  \frac{t_{\text{pN}} -t_{\text{pN}}^*}{T_{\text{N+1}}^*}, \label{App4: Delta u inequality equation a}\\
& t_g\left ( \frac{1}{T_{\text{N}}} -\frac{1}{T_{\text{N}}^*}  \right )-\left ( \frac{  t_{\text{pN-1}}  }{T_{\text{N}}} -\frac{  t_{\text{pN-1}}^* }{T_{\text{N}}^*}  \right )=-\frac{\left |  \Delta u_g\right |}{2 \pi}.\label{App4: Delta u inequality equation b}
\end{align}
\label{App4: Delta u inequality equation}
\end{subequations} 
For the orbits discussed in Section \ref{sec:The long-term orbital effects of MT are calculated using discrete method for  classic elliptical orbits}, Inequality (\ref{App4: Delta u inequality equation a}) yields a unique integer value for N.  
Substituting this into Equation (\ref{App4: Delta u inequality equation b}) gives the required time $t_{g}$. 
For $\text{N}$\,$\gg$\,$ 1$, $t_{g}$\,$\simeq$\,$ t_{ \text{p}\text{N}-1 }$\,$\simeq$\,$ t_{ \text{p}\text{N}-1 }^*$. This is the equivalence of phase and periodic changes that we want.

In order to better clarify  the meaning of Equation (\ref{App4: Delta u inequality equation}), we define $T_{\text{k}}^*$\,$ :=$\,$T_{\text{k}}(1-\varsigma _{\text{k}} )$\,$<$\,$T_{\text{k}}$ and $t_{\text{pN-1}}$\,$\le$\,$ t_g$\,$:=$\,$(1+\lambda_g  )t_{\text{pN-1}}$\,$\le$\,$ t_{\text{pN}}$. 
Here, $\lambda_g$  represents the difference between $t_{\text{pN-1}}$ and $t_g$. 
From Equation (\ref{App4: Delta u inequality equation b}) and the condition  $t_{\text{pN-1}}$\,$\le$\,$t_g\le t_{\text{pN}}$, we obtain an inequality for $\lambda_g $:
\begin{equation}
\begin{aligned}
 &0\le  \lambda_g= \left (  \frac{\left |  \Delta u_g\right |}{2\pi }\frac{1-\varsigma _{\text{N}}}{\varsigma _{\text{N}}}-\frac{\sum_{\text{k}=1}^{\text{N}-1}T_{\text{k}}\varsigma _{\text{k}}}{T_{\text{N}} \varsigma _{\text{N}} } \right ) \frac{T_{\text{N}}}{t_{\text{pN-1}}} <\frac{T_{\text{N}}}{t_{\text{pN-1}}} .
\end{aligned}
\label{App4: Delta u inequality equation 2}
\end{equation}
From Inequality (\ref{App4: Delta u inequality equation 2}), a unique integer $\text{N}$ can be determined,  
which  corresponds to the closest value satisfying  $\sum_{\text{k}=1}^{\text{N}-1}T_{\text{k}}\varsigma _{\text{k}}  /\left [  T_{\text{N}} (1-\varsigma _{\text{N}} )  \right ]  \simeq  \left |  \Delta u_g\right |/2 \pi$.  
If $T_{\text{N}} $\,$\ll$\,$  t_{\text{pN-1}}$, the factor $\lambda_g$ can be neglected, and $t_g$ can be replaced with $t_{\text{pN-1}}$ for the calculation. 
Solving for $\text{N}$ then reduces to finding the unique value of $\text{N}$ in Inequality (\ref{6 1 N equation}).

\nocite{ZhangZi-hanGW24Jun}


\bsp	
\label{lastpage}

\end{document}